\documentclass[fleqn,usenatbib]{mnras}

\usepackage{newtxtext,newtxmath}

\usepackage[T1]{fontenc}

\DeclareRobustCommand{\VAN}[3]{#2}
\let\VANthebibliography\thebibliography
\def\thebibliography{\DeclareRobustCommand{\VAN}[3]{##3}\VANthebibliography}

\newcommand{\referee}[1]{{\color{black} #1}}

\usepackage{graphicx}	
\usepackage{amsmath}	
\usepackage{xcolor}
\usepackage{ragged2e}

\title[BEARS III: Emission line properties]{Bright Extragalactic ALMA Redshift Survey (BEARS) III: Detailed study of emission lines from 71 \textit{Herschel} targets}

\author[Hagimoto, Bakx et al.]{M. Hagimoto$^{1}$\thanks{E-mail: hagimoto@a.phys.nagoya-u.ac.jp},
T. J. L. C. Bakx$^{1,2}$,
S. Serjeant$^{3}$,
G. J. Bendo$^{4}$,
S. A. Urquhart$^{3}$,
S. Eales$^{5}$,
\newauthor
K. C. Harrington$^{6}$,
Y. Tamura$^{1}$,
H. Umehata$^{7,1}$,
S. Berta$^{8}$,
A. R. Cooray$^{9}$,
P. Cox$^{10}$,
G. De Zotti$^{11}$,
\newauthor
M. D. Lehnert$^{12}$,
D. A. Riechers$^{13}$,
D. Scott$^{14}$,
P. Temi$^{15}$,
P. P. van der Werf$^{16}$,
C. Yang$^{17}$,
\newauthor
A. Amvrosiadis$^{18}$,
P. M. Andreani$^{19}$,
A. J. Baker$^{20,21}$,
A. Beelen$^{22}$,
E. Borsato$^{23}$,
V. Buat$^{24}$,
K. M. Butler$^{16}$,
\newauthor
H. Dannerbauer$^{25,26}$,
L. Dunne$^{5}$,
S. Dye$^{27}$,
A. F. M. Enia$^{28,29}$,
L. Fan$^{30,31,32}$,
R. Gavazzi$^{10,33}$,
\newauthor
J. González-Nuevo$^{34,35}$,
A. I. Harris$^{36}$,
C. N. Herrera$^{8}$,
D. H. Hughes$^{37}$,
D. Ismail$^{24}$,
R. J. Ivison$^{19}$,
\newauthor
B. Jones$^{\referee{13}}$,
K. Kohno$^{\referee{38,39}}$,
M. Krips$^{8}$,
G. Lagache$^{24}$,
L. Marchetti$^{\referee{40,41}}$,
M. Massardi$^{\referee{42,41}}$,
H. Messias$^{6,\referee{43}}$,
\newauthor
M. Negrello$^{5}$,
R. Neri$^{8}$,
A. Omont$^{10}$,
I. Perez-Fournon$^{\referee{44},25}$,
C. Sedgwick$^{3}$,
M. W. L. Smith$^{5}$,
\newauthor
F. Stanley$^{10}$,
A. Verma$^{\referee{45}}$,
C. Vlahakis$^{\referee{46}}$,
B. Ward$^{5}$,
C. Weiner$^{3}$,
A. Wei\ss$^{\referee{47}}$,
and A. J. Young$^{20}$\\
\\
\\
Affiliations are listed at the end of the paper
}

\date{Accepted 2023 March 04. Received 2023 January 26; in original form 2022 December 07} 

\pubyear{2023}

\begin{document}
\label{firstpage}
\pagerange{\pageref{firstpage}--\pageref{lastpage}}
\maketitle

\begin{abstract}
We analyse the molecular and atomic emission lines of 71 bright \textit{Herschel}-selected galaxies between redshifts 1.4 to 4.6 detected by the Atacama Large Millimetre/submillimetre Array. These lines include a total of 156 CO, \textsc{[C\,i]}, and H$_2$O emission lines. For 46 galaxies, we detect two transitions of CO lines, and for these galaxies we find gas properties similar to those of other dusty star-forming galaxy (DSFG) samples.
A comparison to photo-dissociation models suggests that most of \textit{Herschel}-selected galaxies have similar interstellar medium conditions as local infrared-luminous galaxies and high-redshift DSFGs, although with denser gas and more intense far-ultraviolet radiation fields than normal star-forming galaxies.
The line luminosities agree with the luminosity scaling relations across five orders of magnitude, although the star-formation and gas surface density distributions (i.e., Schmidt-Kennicutt relation) suggest a different star-formation phase in our galaxies (and other DSFGs) compared to local and low-redshift gas-rich, normal star-forming systems. 
The gas-to-dust ratios of these galaxies are similar to Milky Way values, with no apparent redshift evolution.
Four of 46 sources appear to have CO line ratios in excess of the expected maximum (thermalized) profile, suggesting a rare phase in the evolution of DSFGs.
Finally, we create a deep stacked spectrum over a wide rest-frame frequency (220--890~GHz) that reveals faint transitions from HCN and CH, in line with previous stacking experiments.
\end{abstract}




\section{Introduction}
\begin{figure*}
\includegraphics[width=\textwidth]{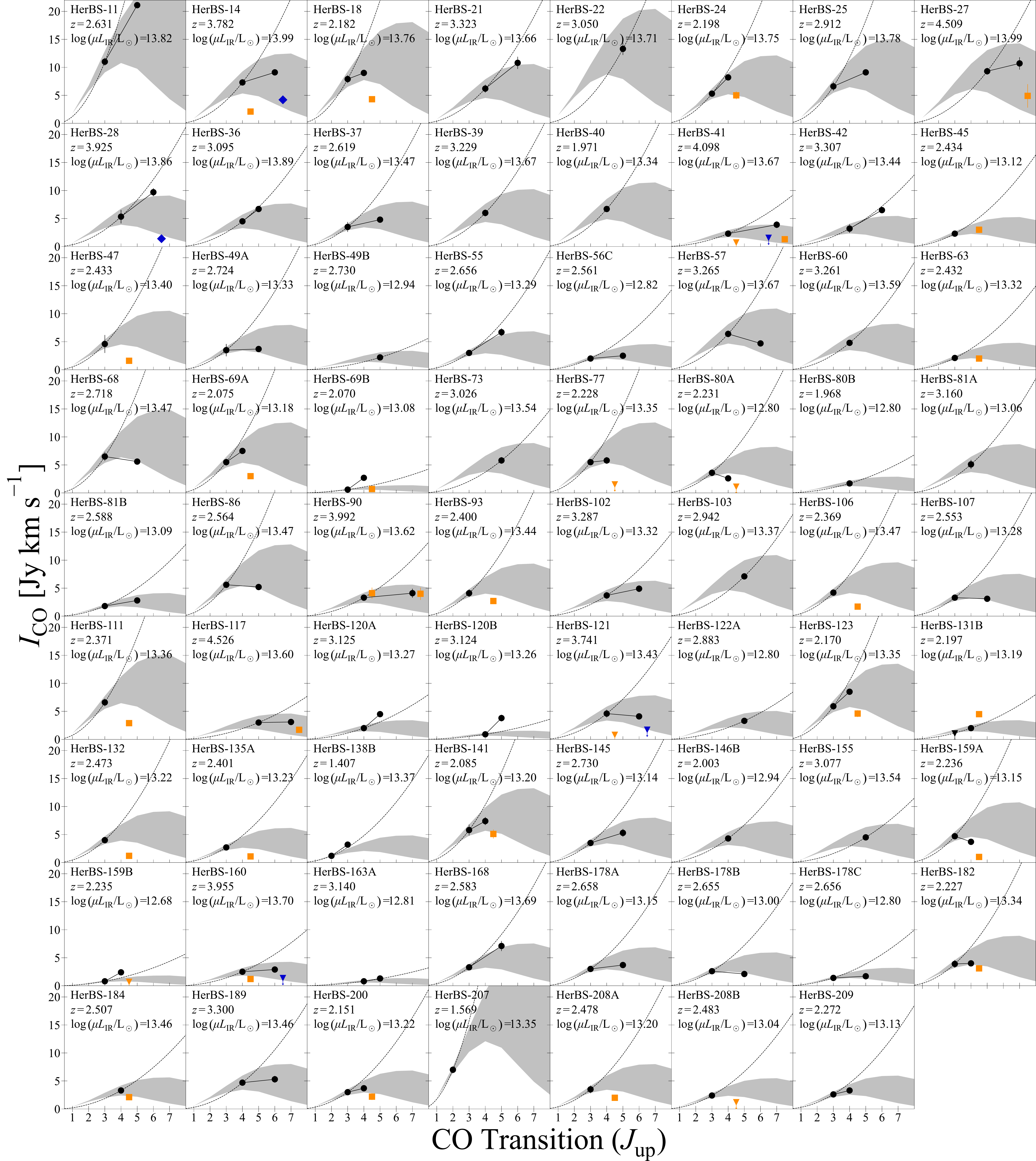}
\caption{Integrated line intensities of the spectral lines for each of the galaxies with robust spectroscopic redshifts (see \citealt{Urquhart2022}). We show the CO-line transitions as {\it connected black points}, and include [C\,{\sc i}]~($^3P_1$--$^3P_0$) and [C\,{\sc i}]~($^3P_2$--$^3P_1$) as {\it unconnected orange squares} at $J_{\rm up} = 4.5$ and 7.5, respectively. The {\it unconnected blue diamond} at $J_{\rm up}= 6.5$ refers to the H$_2$O~(2$_{11}$--$2_{02}$) line. The positions of the [C\,{\sc i}] and H$_2$O lines are chosen based on their frequency relative to the CO lines. We show 3$\sigma$ upper limits as {\it downward triangles}. The {\it dashed black line} shows the thermalized SLED profile fitted to the lowest-$J_{\rm up}$ CO transition, and the {\it filled grey region} shows the mean SLED obtained from the turbulent model for \textit{Planck}-selected galaxies by \citet{Harrington2021} with associated $1\sigma$ spread propagated in quadrature, also fitted to the lowest $J_{\rm up}$ CO transition for each galaxy. We also include the source ID, spectroscopic redshift, and apparent infrared luminosity for each panel.
}
\label{fig:individualSLEDs}
\end{figure*}
Since the advent of wide-field sub-millimetre instrumentation, we have been aware of a population of high-redshift dusty star-forming galaxies (DSFGs), which are subdominant at rest-frame optical wavelengths but contribute 
to the comoving volume-averaged star formation density
\citep[e.g.,][]{Smail1997,Hughes1998,blain2002,Casey2014} \referee{of 20 to 80 percent} over most of the history of the Universe \citep[e.g.,][]{Zavala2021}, particularly at the {\it peak} in star formation density at $z \sim 2.5$ \citep{madau2014}. The \textit{Herschel Space Observatory}, the \textit{Planck} satellite, the South-Pole Telescope (SPT) and the Atacama Cosmology Telescope (ACT) each mapped one to several thousands of square degrees and discovered many high-redshift galaxies apparently forming stars at rates beyond 1000\,$\mathrm{M}_{\odot}$~yr$^{-1}$ \citep{RowanRobinson2016}. However, hydrodynamical models of galaxies initially failed to reproduce these extreme star-forming events \citep[e.g.,][]{narayanan2015}. A further complication is that \referee{high-resolution} follow-up observations of individual sources reveal a heterogeneous population, from smooth disks to clumpy merger-like systems \referee{\citep[e.g.,][]{bussmann13,Bussmann2015,dye2015,Dye2018,Hodge2016,Gullberg2018,Lelli2021,Chen2022}}. Accurate characterization of these diverse sources across larger samples is thus needed, although so far such detailed studies of this population have been limited by the lack of angular resolving power of single-dish telescopes and the intrinsic uncertainty in the photometric redshifts estimated from the far-infrared regime($\Delta z / (1+z) \simeq 0.13$; \citealt{Lapi2011,GN2012,pearson13,ivison16,bakx18}).

The large areas of the DSFG surveys allowed the selection of rare types of galaxies. In fact, simply selecting galaxies with $500$-$\mu$m flux densities $S_{500} > 100$\,mJy has proven to be a reliable selector of strongly gravitationally lensed systems \citep{negrello2007,negrello2010,Negrello2014,negrello2017,wardlow2013,nayyeri2016}. In an effort to better understand bright, distant \textit{Herschel} galaxies, \citet{bakx18} selected sources with $S_{500} \geq 80~\mathrm{mJy}$, as well as a photometric redshift $z_\mathrm{phot} \geq 2$, based on the \textit{Herschel}-Spectral and Photometric Imaging Receiver (SPIRE) fluxes, in order to decrease the contamination rate (see also \citealt{Bakx2020Erratum}). They constructed a sample of 209 sources, known collectively as the \textit{Herschel} Bright Sources (HerBS) sample. A cross-correlation analysis in \cite{bakx2020} has shown it is likely that lower flux criterion ($100~\mathrm{mJy} > S_{500} > 80~\mathrm{mJy}$) selects towards more intrinsically-bright sources with lower magnification factors. Thus, this sample likely not only includes rare strongly lensed sources, but also intrinsically very bright sources, such as hyper-luminous infrared galaxies \citep[HyLIRGS; $L_\mathrm{IR}>10^{13}~\mathrm{L_\odot}$,][]{fu2013,ivison2013,Riechers2013,Riechers2017,Oteo2016}.

The angular magnification in the strongly gravitationally lensed sources enables resolved observations of the gas down to 100-$\mathrm{pc}$ scales \citep[e.g.,][]{dye2015,Tamura2015,Egami2018} with follow-up imaging. Such observations are well-suited to illuminate the physical processes driving the star-formation activity within the galaxies, particularly through resolved kinematics of spectral lines. Similarly, resolved gas morphologies and kinematics will hold the key to understanding the physical origin of HyLIRGs. However, determining the spectroscopic redshift of individual sources is an essential first step, and therefore several redshift campaigns have targeted submm- and mm-wave-selected galaxies. In the northern hemisphere, the Institut de Radio Astronomie Millimetrique (IRAM) 30-m telescope and NOrthern Extended Millimeter Array (NOEMA) have been used to search for CO emission lines for determining spectroscopic redshifts \citep{neri2020,Bakx2020IRAM}. In particular, the NOEMA project $z$-GAL (P.I.s: P. Cox, H. Dannerbauer, and T. Bakx) aims to obtain the spectroscopic redshifts of 137 sources in northern and equatorial fields (Cox et al. in prep.). However, a large fraction of the \textit{Herschel} sources in the southern hemisphere had not yet been done.

Therefore, \citet{Urquhart2022} and \referee{\cite{Bendo2023}} presented the result of the Bright Extragalactic Atacama Large Millimetre/submillimetre Array (ALMA) Redshift Survey (BEARS). They selected the 85 sources from HerBS which are located in the South Galactic Pole field and previously lacked spectroscopic redshifts. Using a series of meticulously-chosen spectral windows, they reported 71 spectroscopic redshifts by targeting the brightest mm-wavelength lines, i.e., the rotational transitions of CO~($J, J-1$) (see \citealt{Bakx2022}). These observations have angular resolutions of $\sim 2$--3 arcsecs for the 2 and 3~mm spectral windows, respectively.
Since higher-$J$ transitions are sensitive to denser and warmer gas, the line luminosity ratios between different transitions can provide insight into the gas conditions within the galaxies such as gas densities and kinetic temperatures (e.g., \citealt{Weiss2007, bothwell2013,yang2017,Canameras2018,dannerbauer2019,  Harrington2021}). 
Furthermore, lines with a high critical density and temperature are excited for the strongly lensed, high-redshift \textit{Planck} selected sources, indicating molecular gas is typically warmer for these dusty star-forming systems with high intrinsic infrared luminosities ($\sim \log_{10} L_\mathrm{IR} = 12.5$--$13.7~\mathrm{L_\odot}$; \citealt{Harrington2021}). Similar findings have been reported for the SPT-selected sources \citep{reuter20} and studies of bright \textit{Herschel} galaxies \citep{yang2017,Bakx2020IRAM,neri2020}. 

The CO lines from this redshift search provide us with a sensitive probe of the interstellar medium (ISM) conditions of a large sample of \textit{Herschel} galaxies.
Specifically, the Schmidt-Kennicutt (SK) relation between the star-formation surface density and molecular gas mass ($\Sigma_\mathrm{SFR}$--$\Sigma_\mathrm{H_2}$; \citealt{Schmidt1959,Kennicutt1998}) is diagnostic of the star-formation mode, i.e., ``main-sequence'' as opposed to ``starburst''. High-redshift dusty starbursts typically have higher star-formation rates (SFRs) relative to their molecular gas compared to normal star-forming galaxies (e.g., \citealt{Casey2014}), which results in a shorter depletion time scale ($\sim 100$~Myr). The reasons for this boost in star formation are often hypothesised to be connected with recent merger events \citep{Sanders1988,Barnes1991,Hopkins2008}, although several other theories have been posited \citep{Cai2013,Gullberg2019,Hodge2019,Rizzo2020,Fraternali2021}. CO is the second-most abundant molecule in the ISM, and the $J=$1--0 transition to the ground state has traditionally been considered as the best tracer of the molecular gas mass. However it is difficult to observe in high-redshift galaxies due to its relative faintness. Alternatively, other low-$J$ CO lines, such as CO~(2--1) or (3--2), can be used as molecular gas tracers, although to use these we have to assume a gas excitation. In addition, we could miss the molecular gas mass behind the ``photosphere'' due to the optically thick nature of $^{12}$CO, particularly for low-$J$ CO lines. In this situation, the \textsc{[C\,i]}~($^3P_1$--$^3P_0$) line is a useful alternative tracer, which has been discussed from both theoretical and observational perspectives (e.g., \citealt{Papadopoulos2004theo,Papadopoulos2004obs}). \textsc{[C\,i]}~($^3P_1$--$^3P_0$) is typically optically thin and therefore probes the full molecular gas mass, although in the local Universe there exists not much difference between molecular gas estimates from CO lines and \textsc{[C\,i]}~($^3P_1$--$^3P_0$) \citep{Israel2020}. In addition, it is a useful probe of the physical conditions of the photo-dissociation regions (PDRs) inside these dusty starbursts (e.g., \citealt{Alaghband-Zadeh2013,Bothwell2017,Jiao2017,Valentino2020CI}). 

In this paper, we exploit the 156 detections and upper-limits of CO rotational transitions, as well as \textsc{[C\,i]} atomic Carbon and the H$_2$O~(2$_{11}$--2$_{02}$) water transition within the BEARS project to estimate the ISM conditions of 71 bright \textit{Herschel}-selected galaxies. We briefly describe the sample and observations in Section\,\ref{sec:ch2}, and show the initial sample results in Section\,\ref{sec:ch3}. We discuss these results in Section\,\ref{sec:ch4}, Section~\ref{sec:ch5} discusses four sources with strange spectral features, and Section~\ref{sec:ch6} discusses the composite spectrum.
We provide our main conclusions in Section\,\ref{sec:ch7}. 
Throughout this paper, we adopted a spatially flat $\Lambda$CDM cosmology with the best-fit parameters derived from the \textit{Planck} results \citep{Planck2015}, which are $\Omega_\mathrm{m} = 0.307$, $\Omega_\mathrm{\Lambda} = 0.693$ and $h = 0.693$. 

\section{Sample and Observations}
\label{sec:ch2}
Our sources are selected from the \textit{Herschel} Bright Sources sample (HerBS; \citealt{bakx18,Bakx2020Erratum}), which contains the 209 brightest, high-redshift sources in the $616.4~\mathrm{deg}^2$, H-ATLAS survey \citep{eales2010}. The H-ATLAS survey used the PACS \citep{Poglitsch2010} and SPIRE \citep{griffin2010} instruments on \textit{Herschel} to observe the North and South Galactic Pole Fields, as well as three equatorial fields, to a $1\sigma$ sensitivity of 5.2 mJy at $250~\mu$m to 6.8 mJy at $500~\mu$m\footnote{These sensitivities are derived from the background-subtracted, matched-filtered maps without accounting for confusion noise} \citep{eales2010,valiante2016}. The sources are selected with a photometric redshift, $z_{\rm{phot}}$, greater than 2 and a $500$-$\mu$m flux density, $S_{500}$, greater than 80 mJy. Blazar contaminants were removed with radio catalogues and verified with $850$-$\mu$m SCUBA-2 observations \citep{bakx18}. \referee{The infrared photometric redshift of each source was initially calculated through the fitting of a two-temperature modified blackbody (MBB) SED template from \cite{pearson13} to the 250, 350 and 500-$\mu$m flux densities \citep{bakx18}. This MBB was derived from the \textit{Herschel}-SPIRE photometry of 40 lensed H-ATLAS sources with spectroscopic redshifts from the H-ATLAS survey, and assumed both a cold (23.9~K) and warm (46.9~K) component of the dust. These sources are similar to the ones in this sample, and would thus provide a good initial photometric redshift estimate, with an estimated error of $\Delta z / (1+z) \approx 0.12$. Improved photometric fits of the data, including the recent ALMA data, are reported in \cite{Bendo2023}.}

Concerted efforts have been made to measure the redshifts of these sources using both single-dish telescopes (e.g., IRAM; \citealt{Bakx2020IRAM}) and interferometers (e.g., NOEMA; \citealt{neri2020}). However most of the sources in the southern hemisphere have been out of reach for earlier redshift search attempts with CARMA, NOEMA and IRAM\ 30-m. As described by \cite{Urquhart2022}, our targeted sources are all located in the South Galactic Pole field \citep{eales2010}. 
The initial line searches were conducted by the Atacama Compact Array (ACA; also called the Morita Array) for 11 sources in the 3-mm band (Band 3). This has been complemented by a larger-scale search using the 12-m~Array that observed another 74 sources at 3~mm, and all 85 targets at 2~mm, in an effort to measure robust redshifts.
In total, the redshifts of 71 galaxies in 62 out of 85 \textit{Herschel} fields were identified. The redshift identification is described in more detail in \cite{Urquhart2022}, with the continuum and source-multiplicity analysis presented in \referee{\cite{Bendo2023}}.

\section{Line fluxes}
\label{sec:ch3}
We use the line fluxes reported in \cite{Urquhart2022} for all detected lines, which we complement with upper-limit extractions using the following method. Briefly, for the detected lines, circular apertures were centred on the peaks of the corresponding continuum emission, and the radii of the apertures were manually adjusted for each source in each image to include as much line emission as possible while still measuring that emission at higher than the $5\sigma$ level. These apertures were several times larger than the beam size, removing potential bias when comparing lines with different signal-to-noise ratios or with different beam sizes. Similarly, the frequency channels were manually selected such that as much of the line flux as possible was included while still yielding measurements above the $5\sigma$ level.
For non-detected lines, the same aperture and line velocity width as the detected lines were used. We then fit a Gaussian profile to the frequency of the undetected spectral lines with a fixed velocity width. The resulting uncertainty on this fit then provided us with the $1\sigma$ error. 

Figure~\ref{fig:individualSLEDs} shows an overview of all the detected spectral lines and the $3\sigma$ upper-limits of undetected lines of all sources. Most of the detected spectral lines are from CO ranging from $J=2$--1 to 7--6, and in addition, we detect the [C\,{\sc i}]~($^3P_1$--$^3P_0$) line for 23 sources, a water line (H$_{\rm 2}$O~(2$_{\rm 11}$--2$_{\rm 02}$)) for two sources, and the [C\,{\sc i}]~($^3P_2$--$^3P_1$) line for four sources. 
We also include the upper limits of a single CO, six [C\,{\sc i}]~($^3P_1$--$^3P_0$) and three H$_{\rm 2}$O~(2$_{\rm 11}$--2$_{\rm 02}$) lines. Most of these lines had already been reported in \cite{Urquhart2022}, however we extracted three extra [C\,{\sc i}]~($^3P_1$--$^3P_0$), three extra [C\,{\sc i}]~($^3P_2$--$^3P_1$), and one extra H$_{\rm 2}$O~(2$_{\rm 11}$--2$_{\rm 02}$) lines. These additional velocity-integrated line fluxes and upper limits are listed in Appendix Table~\ref{tab:upperlimitsAdditional}.
We find velocity-integrated fluxes, $I_{\rm CO}$, ranging from 0.6 to $21.1~\mathrm{Jy~km~s^{-1}}$. We do not correct for cosmic microwave background (CMB) effects (i.e., the observational contrast against the CMB and the enhanced excitation of the CO or other line transitions), since this requires extensive modelling of the ISM conditions and exceeds the scope of this paper, particularly since it is not a large correction at these redshifts \referee{\citep{daCunha2013,Zhang2016,Tunnard2017}}. The velocity-width FWHM of our sources ranges between 110 and $1290~\mathrm{km~s^{-1}}$. This observed diversity in line profiles is also seen in the Figure~\ref{fig:JD_graph}, where the velocity distribution of the BEARS systems are compared against one another.

\section{Discussion}
\label{sec:ch4}
\subsection{CO spectral line energy distributions}
\label{sec:COSLEDs}
\begin{figure}
\includegraphics[width=\linewidth]{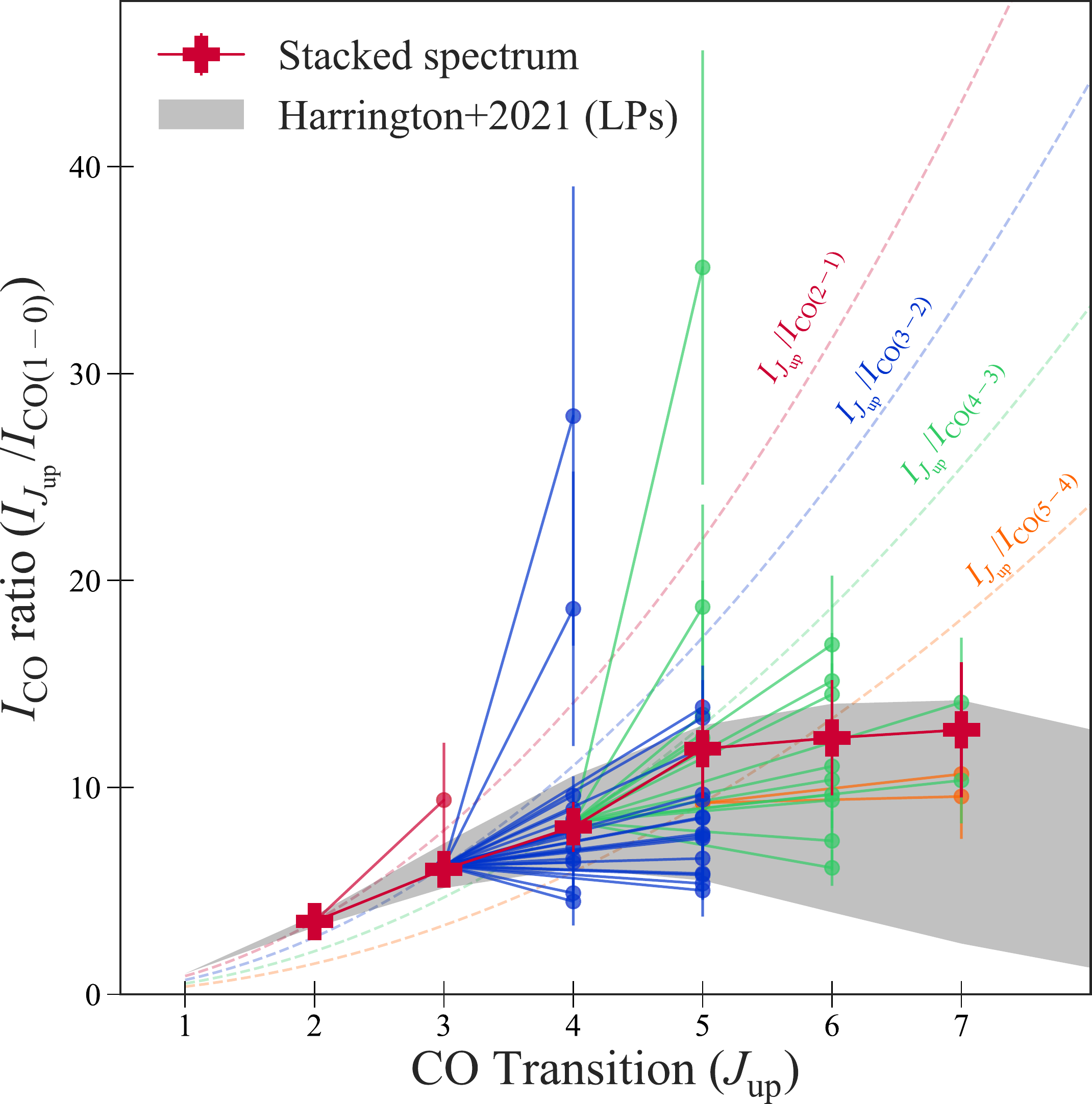}
\caption{
Normalized line intensities of BEARS galaxies against the observed CO line transition. The lines are normalized to the CO~(1--0) emission calculated using the mean SLED obtained from the turbulent model for strongly lensed galaxies identified by the \textit{Planck} satelite (LPs) in \citet{Harrington2021}. The colour of the marker reflects the lowest $J_{\rm up}$ transition observed to the galaxy. Similarly, the {\it coloured dashed lines} indicate the thermalized profiles for the lowest transitions ($\propto J^2_{\rm up}$). The {\it grey filled region} shows the mean SLED with $\pm1\sigma$ standard deviation from \citet{Harrington2021}. The {\it connected red plus points} show the SLED from the stacked spectrum (Section~\ref{sec:ch6}), and this is consistent with the mean SLED of \textit{Planck}-selected DSFGs. We note that several sources appear to exceed the maximum (i.e., thermalized) line ratio profile, and we discuss the implications of this in Section \ref{sec:ch5}.
}
\label{fig:togetherSLED}
\end{figure}

Figure~\ref{fig:togetherSLED} compares the normalized integrated line intensities for each of the sources. In this figure, the sources are normalized to the expected CO~(1--0) line velocity-integrated flux, derived from the mean spectral line energy distribution (SLED) of the turbulent model in \cite{Harrington2021}, using the lowest $J_{\rm up}$ CO line with $\pm1\sigma$ standard deviation (shown as the grey shaded region). The colour of the lines indicates the lowest $J_{\rm up}$ transition, starting from $J_{\rm up} = 2$ to 5 shown in {\it red, blue, green} and {\it orange}. The {\it dashed lines} show the associated thermalized profiles for each normalized transition ($I_\mathrm{CO} \propto J_{\rm up}^2$, with the instant of proportionality determined by the SLED from \citealt{Harrington2021}). The {\it connected red plus points} indicate the SLED from the stacked spectrum (see Section~\ref{sec:ch6} for details). 

The CO SLEDs reflect a diverse population that broadly follows the mean ratios reported by \cite{Harrington2021}, which has a slightly smaller median redshift ($z_{\rm med} = 2.4$) relative to our median redshift ($z_{\rm med} = 2.7$). Several sources already show a downward trend in velocity-integrated intensity at $J_\mathrm{up}<4$, which is more in line with  non-starburst galaxies \citep{Dannerbauer2009,Boogaard2020} such as the Milky Way \citep{Fixsen1999}. It is important to note that the general shape of CO SLEDs consists of multiple components (e.g., \citealt{Daddi2015,yang2017,Canameras2018}).
The stacked SLED straddles the top of the expected ratios from \citet{Harrington2021}, but it is normalized only to two CO~(2--1) detections and reflects the average line ratios across a wide range in redshifts. The stacked SLED agrees within the $1\sigma$ error with the model from \cite{Harrington2021}, although it appears higher for $J_{\rm up} > 4$.
While the majority of sources show sub-thermalized profiles (i.e., below their respective {\it dashed lines}), several sources (HerBS-69B, -120A, -120B, -159B) have SLEDs above the typically-assumed theoretical maximum excitation, based on the equipartition distribution of the CO excitation states; this will be discussed separately in Section~\ref{sec:ch5}.

\begin{figure}
\centering
\includegraphics[width=0.9\linewidth]{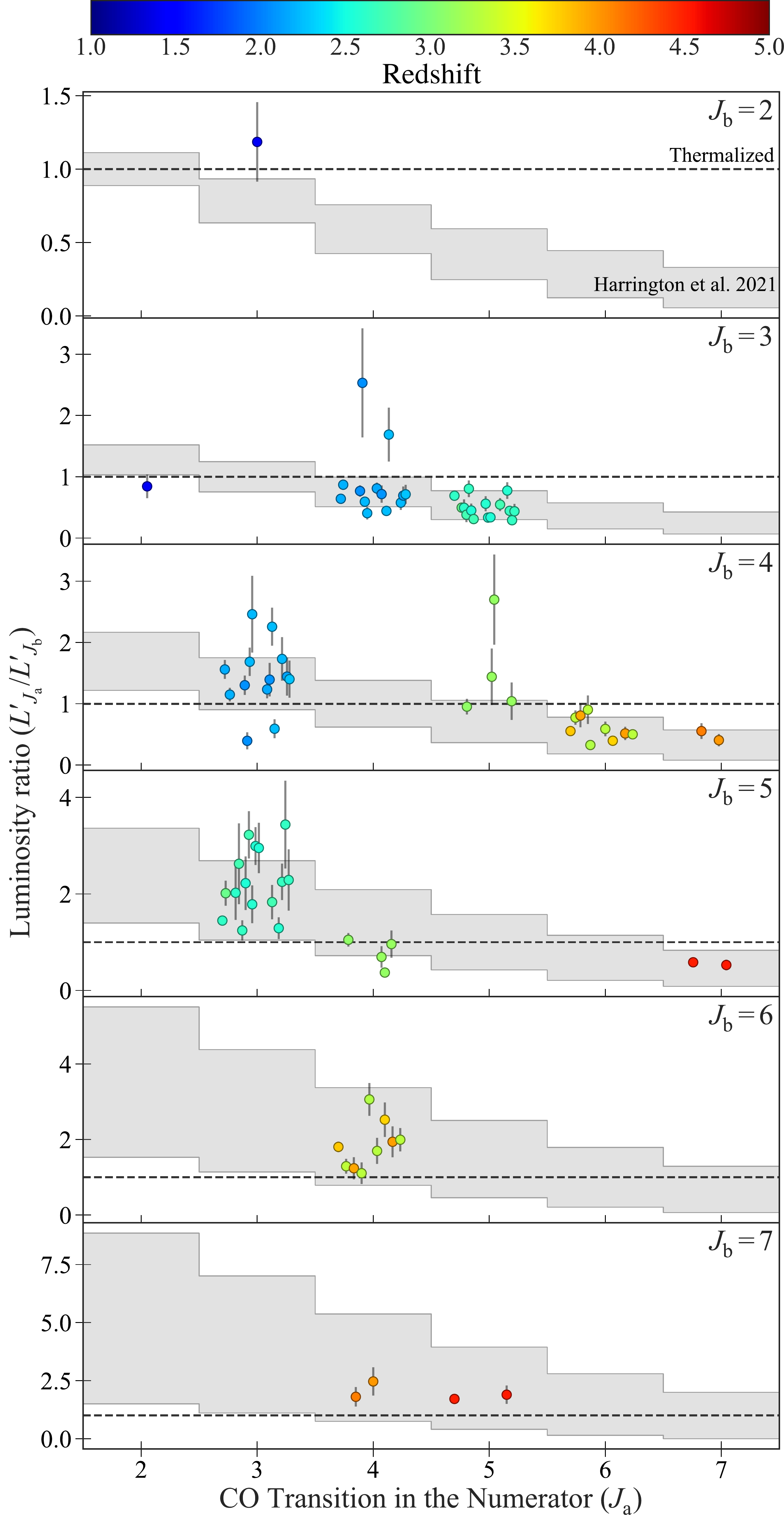}
\caption{Line luminosity ratios between the different lines plotted against the CO transition in the denominator (i.e., the {\it normalizing} CO line transition) using equation~\ref{eq:JaJb}. We compare the ratios to the expected ratios of the compilation of \citet{Harrington2021} (\textit{grey filled region}), and the constant brightness profile (\textit{flat dashed line}). The colour of the data markers reflects their redshift ({\it blue to red} equalling {\it low to high redshift}). {\it Each panel} indicates a different CO-transition in the denominator of the luminosity ratio, while the {\it x-axis} refers to the CO transition in the numerator. We manually shift the points between $J_\mathrm{a}-0.5$ and $J_\mathrm{a}+0.5$ so as not to overlap. The constant brightness profile assumes equal excitation among all CO transitions, and thus has a uniform value of $1.0$ for all CO transitions. The lines detected for each source depends strongly on their spectroscopic redshift, and is reflected in the clustering of colours at specific CO line transitions. For non-thermalized CO lines, the points lie above $1$ at $x$-values lower than $J_{\rm b}$, and below $1$ at $x$-values larger than $J_{\rm b}$.}
\label{fig:luminosityRatioStack}
\end{figure}

We provide an alternative perspective on the CO SLEDs of our sources in Figure~\ref{fig:luminosityRatioStack}, where we compare the luminosity ratios of all detected lines. We calculate the line luminosity for the spectral lines using the equation from \cite{solomon1997}:
\begin{equation}
L^\prime =3.25\times10^7 S_{\rm CO}\Delta v f_{\rm obs}^{-2} D_{\rm L}^2(1+z)^{-3}~[\mathrm{K~km~s^{-1}~pc^2}]. \label{eq:lprime}
\end{equation}
with the integrated flux, $S_{\rm CO}\Delta v$, in $\mathrm{Jy~km~s^{-1}}$, the observed frequency, $f_{\rm obs}$, in GHz, and the luminosity distance, $D_{\rm L}$, in Mpc. We derive the line luminosity ratios, $r_{J_\mathrm{a}, J_\mathrm{b}}$, for each of our sources, using
\begin{equation}
r_{J_\mathrm{a}, J_\mathrm{b}} = \frac{L^\prime_\mathrm{a}}{L^\prime_\mathrm{b}} = \frac{I_\mathrm{a}}{I_\mathrm{b}} \frac{J_\mathrm{b}^2}{J_\mathrm{a}^2}.
\label{eq:JaJb} 
\end{equation}
Here $L^\prime_\mathrm{a}$ and $L^\prime_\mathrm{b}$ refer to the luminosity of lines a and b, $I_\mathrm{a}$ and $I_\mathrm{b}$ refer to the integrated line flux, and $J_\mathrm{a}$ and $J_\mathrm{b}$ refer to the upper rotational quantum level, commonly referred to as $J_{\rm up}$. We use error-propagation rules to obtain the total error of each data point. The {\it dashed black line} at $r_{J_\mathrm{a}, J_\mathrm{b}} = 1$ indicates the constant-brightness or thermalized profile ($I_\mathrm{CO} \propto J_{\rm up}^2$), typically assumed to be the maximum excitation SLED of a galaxy. 

Similar to Figure~\ref{fig:togetherSLED}, the majority of sources fall within the mean SLED from \cite{Harrington2021} ({\it grey filled region}). A handful of sources already appear to peak at or lower than $J_{\rm up} = 3$, with the majority of sources showing a peak in their CO SLED between $J_{\rm up} = 4$--$6$ or beyond.
Several sources, however, lie on the {\it unexpected} side of the thermalization curve, indicating super-thermal excitation within our sources (see Section~\ref{sec:ch5}).
We note that there is a strong redshift-dependence for what lines are detected for each source. This is due to the fixed spectral windows in the ALMA observations described in \cite{Urquhart2022}, where sources at higher redshifts had higher-$J$ CO lines redshifted into the observing windows.

\subsection{Spectral line comparison}
\label{sec:SLcomparisonAndDustLuminosities}
\begin{figure*}
    \centering
    \includegraphics[width=0.8\linewidth]{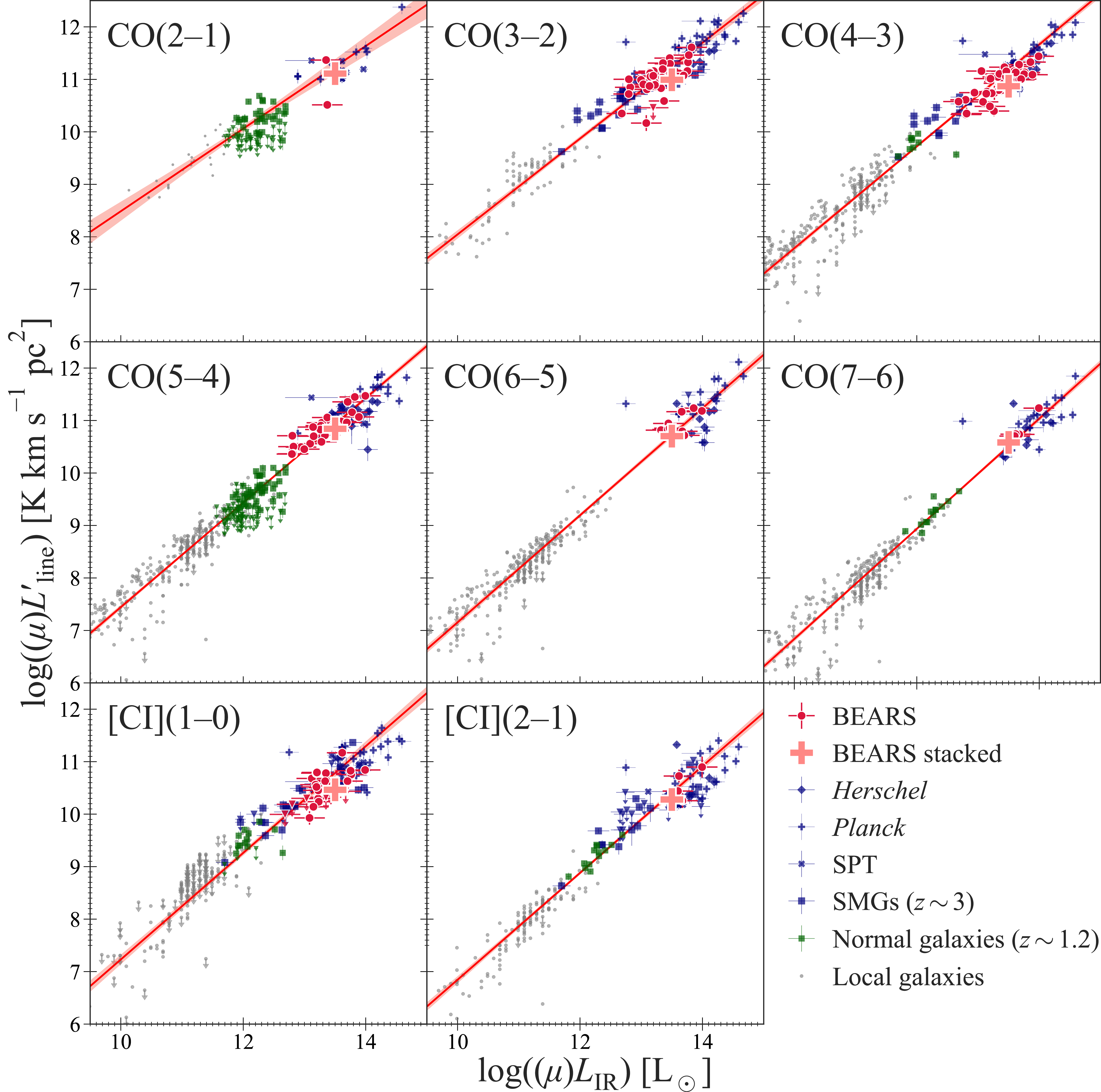}
    \caption{Detected CO and \textsc{[C\,i]} emission lines versus the infrared luminosity. We compare the line luminosity of the BEARS sources against reference samples of low-redshift star-forming galaxies, including infrared-luminous galaxies \referee{(\citealt{Greve2014,Rosenberg2015,Liu2015, Kamenetzky2016, yang2017}}; \textit{grey points}), high-redshift star-forming galaxies (\citealt{Valentino2020CO}; \textit{green squares}), and DSFGs (\citealt{Walter2011, Alaghband-Zadeh2013, yang2017, Bothwell2017, Harrington2021}; \textit{blue squares, diamonds, crosses, and plus points}). We use a linear-regression-fitting technique to estimate the scaling relations, accounting for errors in both the line and infrared luminosities. The red solid line and background pink shade mean the best-fit result and $\pm2\sigma$ uncertainty. The best-fit parameters of the scaling relations are reported in Table~\ref{tab:scalingrelations}. The {\it large plus sign} indicates the result from the stacked spectrum (see Section~\ref{sec:ch6}). Our sources, as well as the stacked result, are consistent with previous works.}
    \label{fig:allCOscalingRelations}
\end{figure*}

In this subsection, we compare several spectral lines against the bolometric infrared luminosity ($8$--$1000$~\micron{}) for the BEARS galaxies. 
\referee{  
We choose to derive the infrared luminosity using only the 151~GHz flux density from the Band~4 observations with ALMA, since all galaxies are identified through their 151~GHz flux density, and over half of all sources discussed in this paper (40 out of 71) have multiples that make direct comparison to the \textit{Herschel} and SCUBA-2 photometry unreliable. For all sources in our sample, the observed 151~GHz emission probes the infrared emission, as all sources lie beyond $z > 1$. Although \cite{Bendo2023} provide estimates of the photometric properties of individual sources and sources with multiples at the same confirmed redshift, a comprehensive study across our sample is still elusive, as it requires extensive work using deblending techniques on the \textit{Herschel} and SCUBA-2 photometry for sources with multiples. The scope of this paper is to provide a comprehensive study of the gas properties of our sample, and we therefore take a simplified approach using the single data point all our sources share (i.e., the 151~GHz detection) and leave a deblending study for future work.
The infrared luminosity is calculated assuming a modified black body with an average dust temperature of $35~\mathrm{K}$ and a dust-emissivity index ($\beta$) of 2. This is in line with the photometric study of individual sources and sources with multiples at the same redshifts from \cite{Bendo2023}, which finds an average dust temperature of $32.3 \pm 5.5$ and a $\beta$ of $2.00 \pm 0.35$, and also agrees with previous studies of dusty galaxies in the distant Universe \citep[e.g.,][]{Eales2000,Dunne2001}.
A single-temperature modified black-body is the most basic representation of the emission of a dusty galaxy, and allows for an easy comparison or future translation to other studies, unlike a direct fit to an observed spectrum such as the Cosmic Eyelash \citep[e.g.,][]{swinbank2010,ivison2010}. The 151~GHz emission traces the Rayleigh-Jeans tail of the emission, and we are thus sensitive to the larger but colder reservoirs of dust. However, we note that this data point provides little constraint on any warmer dust that could be present inside a small portion of these galaxies. 
We account for the CMB using the appropriate equations from \cite{daCunha2013}, which correct both for the dust emission in contrast to the CMB, as well as additional dust heating by the CMB. 
For single-component sources, we validate our method against the luminosities from \cite{Bakx2020Erratum} and \cite{Bendo2023},
and find similar infrared luminosities within the typical uncertainty of the 151~GHz flux density. 
}
Throughout this paper, we calculate the SFR assuming the \cite{Kennicutt2012} relation of \referee{$\mathrm{SFR~[M_\odot~yr^{-1}]}= 1.47 \times{} 10^{-10} L_\mathrm{IR}~[\mathrm{L_\odot}]$}.

Figure~\ref{fig:allCOscalingRelations} shows the line luminosity of the CO and \textsc{[C\,i]} lines against the infrared luminosity of the BEARS sources. These sources are compared against the relationships for low- \referee{(\citealt{Greve2014,Rosenberg2015,Liu2015, Kamenetzky2016, yang2017})} and high-redshift (\citealt{Valentino2020CO, Walter2011, Alaghband-Zadeh2013, yang2017, Bothwell2017, Harrington2021}) galaxies. In order to estimate the scaling relations accounting for errors in both the line and infrared luminosity, we use a linear regression fitting technique of SciPy \citep{Scipy2020} ODR (Orthogonal Distance Regression) package, which is an implementation of the Fortran ODRPACK package \citep{Boggs1992}. The best-fit parameters in the scaling relations are reported in Table~\ref{tab:scalingrelations} according to the perscription
\begin{equation}
    \log_{10} \left(L^\prime_{\rm line}~\left[\mathrm{K~km~s^{-1}~pc^2}\right]\right)=a \log_{10} \left(L_{\rm IR}~\left[\mathrm{L_\odot}\right]\right)+b.
\end{equation}

\referee{The line luminosity scalings of BEARS galaxies are consistent with the reference samples, spanning over four orders of magnitude for all lines.} The best-fit results of the CO and \textsc{[C\,i]} lines (except CO~(2--1) and perhaps CO~(3--2)) agree with a linear scaling relation across our sample. While the line fits are well-constrained, the source-to-source variation across the reference samples and our data is on the order of $\pm 0.5~$dex. This means that we can convert the luminosities of these emission lines into SFR estimates. 

\subsection{Inferred Schmidt-Kennicutt relation of \textit{Herschel} sources}
\label{sec:SchmidtKennicutt}
\begin{figure*}
\includegraphics[width=0.9\linewidth]{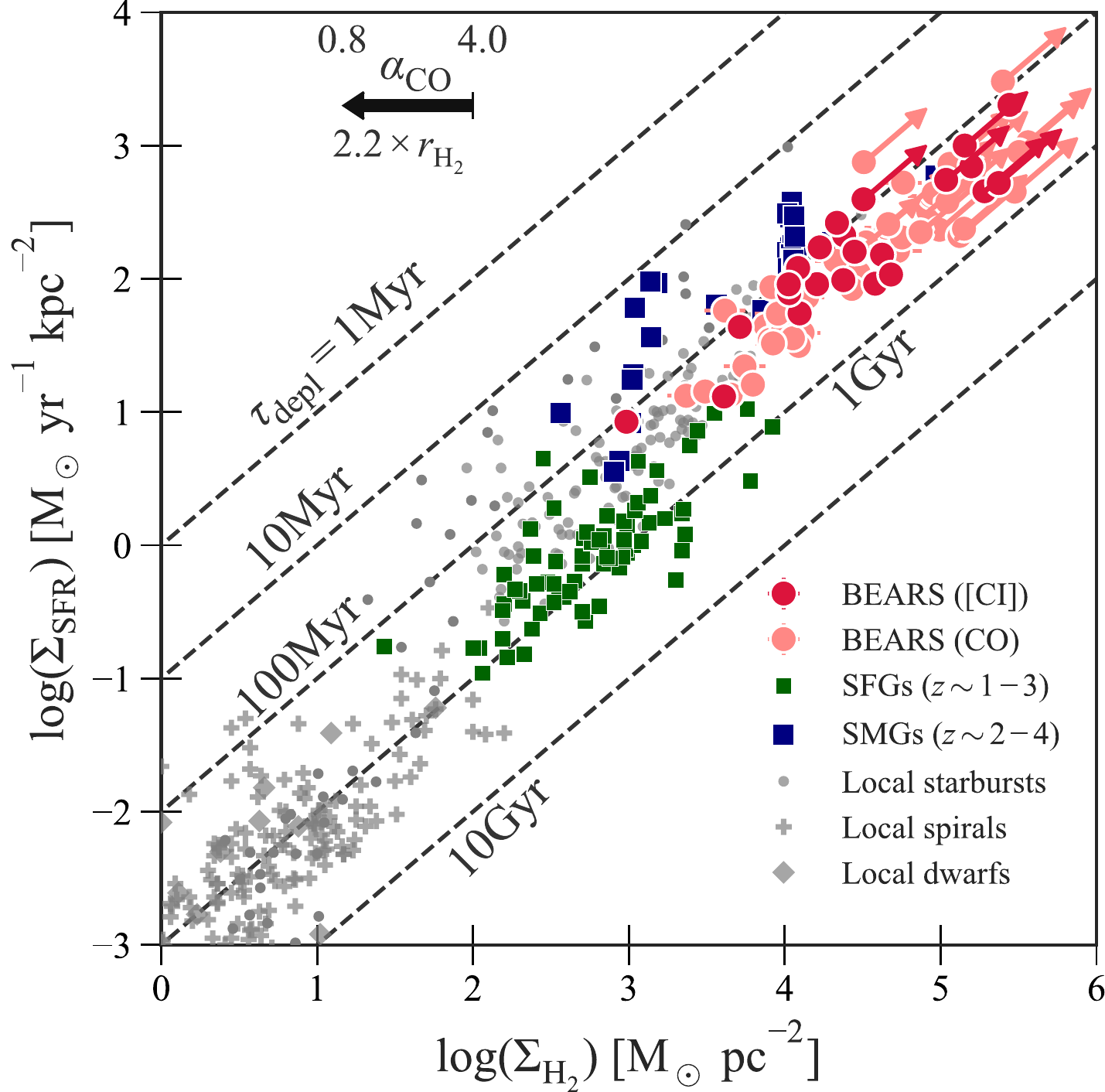}
\caption{Schmidt-Kennicutt scaling relation, i.e., star-formation surface density against molecular gas mass surface density (\citealt{Schmidt1959,Kennicutt1989}) for the BEARS galaxies detected either in \textsc{[C\,i]} ({\it red}) or CO ({\it light red}). We use the deconvolved \texttt{IMFIT} result discussed in Section~\ref{sec:dynamicalmassandsize} to estimate the surface densities, where {\it circles} indicate sources with size estimates and {\it circles with diagonal arrows} indicate sources with upper limits on their size estimates. They are compared against reference samples of low-redshift galaxies (\citealt{Kennicutt1989,delosReyes2019,delosReyes2019b,Kennicutt2021}; {\it grey points, plus points, and diamonds}), high-redshift star-forming galaxies (\citealt{Tacconi2013}; \textit{green squares}) and high-redshift DSFGs (\citealt{Hatsukade2015,C.C.Chen2017} and references therein; {\it blue squares}). We use an $\alpha_{\rm [CI]}$ value of $17.0$ and an $\alpha_{\rm CO}$ value of $4.0$ from the recent study of \citet{Dunne2021,Dunne2022}. We adjust the reference sample of DSFGs ({\it blue squares}) from \citet{C.C.Chen2017} from the initially-assumed $\alpha_{\rm CO} = 0.8$ to $4.0$ for a fair comparison, where the {\it black arrow} in the top-left part of the graph indicates the effect of this change in $\alpha_{\rm CO}$. Recent studies have shown that the size of the molecular gas reservoir extends beyond that of the bright star-forming region. The arrow can thus also be used to indicate the effect of a 2.2 times larger radius of the molecular reservoir relative to the star-formation region.
The {\it diagonal dashed lines} indicate the depletion timescales of 1 Myr to 10 Gyrs. 
The BEARS sources appear to have slightly longer gas depletion times than DSFGs from \citet[][and references therein]{C.C.Chen2017}; however, they do appear to have a similar slope. The SK slope for the BEARS sample is approximately unity.
}
\label{fig:SchmidtKennicutt}
\end{figure*}
Figure~\ref{fig:SchmidtKennicutt} shows the Schmidt-Kennicutt (SK) scaling relation--star-formation surface density against molecular gas mass surface density (\citealt{Schmidt1959,Kennicutt1989}) for all 71 BEARS sources detected in either \textsc{[C\,i]} or CO. 
The molecular hydrogen mass in the galaxy is calculated using
\begin{equation}
M_{\rm H_2} = \alpha  L^\prime_{\rm line}.
\end{equation}
When available, we use the [\textsc{C\,i}] ($^3P_1$--$^3P_0$) line to calculate the molecular gas mass, otherwise we use the mean line luminosity ratios obtained from the turbulence model in \citet{Harrington2021} to calculate the CO~(1--0) line luminosity, specifically $r_{2,1} = 0.88\pm0.07$, $r_{3,1} = 0.69\pm0.12$, $r_{4,1} = 0.52\pm0.14$, $r_{5,1} = 0.37\pm0.15$. 
We use each conversion factor from the recent study of \citet{Dunne2021,Dunne2022};
\begin{eqnarray}
    \alpha_{\rm [CI]} &=& 17.0~\left[\mathrm{M}_\odot~\left(\mathrm{K~km~s^{-1}~pc^2}\right)^{-1}\right] \\
    \alpha_{\rm CO} &=& 4.0~\left[\mathrm{M}_\odot~\left(\mathrm{K~km~s^{-1}~pc^2}\right)^{-1}\right].
\end{eqnarray}
They estimate this value through a self-consistent cross-calibration between three important gas-mass tracers (i.e., [\textsc{C\,i}], CO and submm dust continuum). In total, they use 407 galaxies from low to high-redshift ($z \approx 6$), and fail to find any evidence for a bi-modality in the gas conversion factors between different galaxy types. As a sanity check, our sources also have a good agreement in their estimates of the molecular gas mass between [\textsc{C\,i}] and CO. 
The $\alpha_\mathrm{CO} = 4.0~\mathrm{M}_{\odot}~(\mathrm{K~km~s}^{-1}~\mathrm{pc}^2)^{-1}$ value includes the additional helium correction (see \citealt{Bolatto2013} for a review) of 1.36. This $\alpha_{\rm CO}$ value is larger than the typically-assumed 0.8 for star-bursting galaxies (see e.g., \citealt{Casey2014}); however, this is in line with recent studies of \textit{Planck}-selected galaxies \citep{Harrington2021}. We note that the lower $\alpha_{\rm CO}$ values from previous studies are better able to align the observed dynamical and molecular gas masses (see Section~\ref{sec:moleculargasmass}).
We use the deconvolved \texttt{IMFIT} result discussed in Section~\ref{sec:dynamicalmassandsize} to estimate the surface densities, where {\it circles} indicate sources with size estimates and {\it circles with diagonal arrows} indicate sources with upper limits on their size estimates. We use the same (continuum) size estimate for both the molecular gas and star-formation surface densities. These are compared against reference samples at low- (\citealt{Kennicutt1989,delosReyes2019,delosReyes2019b,Kennicutt2021}) and high-redshift (\citealt{Tacconi2013,Hatsukade2015,C.C.Chen2017} and references therein).

We adjust the reference sample of DSFGs ({\it blue squares}) from \citet[and references therein]{C.C.Chen2017} from the initially-assumed $\alpha_{\rm CO} = 0.8$ to $4.0$ for a fair comparison, where the {\it black arrow} in the top-left part of the graph indicates the effect of this change in $\alpha_{\rm CO}$. Recent studies have shown that the size of the molecular gas reservoir extends beyond the bright star-forming region (e.g., \citealt{C.C.Chen2017}). Therefore, the arrow can also be used to indicate the effect of a 2.2 times ($= \sqrt{4/0.8}$) larger radius of the molecular reservoir relative to the star-forming region.
The diagonal dashed lines indicate the depletion timescales of $1~\mathrm{Myr}$ to $10~\mathrm{Gyrs}$, defined as the molecular gas mass divided by the star-formation rate ($t_{\rm dep}=\mu M_{\rm mol}$/$\mu$SFR).

The BEARS sources appear to have shorter depletion times than local spirals and dwarf galaxies, as well as $z \approx 1$--$3$ star-forming galaxies, suggesting accelerated star formation in these systems and hence implying that these systems are not simply scaled-up versions of gas-rich, normal star-forming systems \citep{Cibinel2017,Kaasinen2020}. The BEARS sources have, on average, longer depletion times than the DSFGs from \citet{Hatsukade2015}, \citet[and references therein]{C.C.Chen2017}.
\citet{Harrington2021} also report the longer depletion timescale for \textit{Planck}-selected DSFGs, suggesting that DSFGs are not necessarily consuming their gas faster than other active galaxies. 
However, they do appear to have a similar slope that is on the order of unity or slightly steeper. This is in contrast to the slopes reported in early studies (e.g., $1.25$--$1.44$ by \citealt{Gao2004}) and in agreement with current studies (e.g., $0.9$--$1.4$ by \citealt{Tacconi2013,Tacconi2018,Tacconi2020} and $1.13$ by \citealt{Wang2022}).
There exists, however, an ambiguity in the measured sizes of these kind of studies. As noted in \cite{C.C.Chen2017}, if the dust size is used instead of the CO-based size estimate, the gas surface density of the sources could increase by over an order of magnitude. 
High-resolution imaging of DSFGs have shown these sources to be compact dusty systems (e.g., \citealt{Ikarashi2015, Barro2016,Hodge2016,Gullberg2019,Pantoni2021}) with sizes on the order of a single kiloparsec. Instead, relative to the $z \sim 2$--4 SMGs, the BEARS systems are likely hosting more extended star formation in their systems seen through an increase in their depletion timescales. Here we note an intrinsic bias in our sample, since all galaxies have spectroscopic redshifts based on CO line measurements. Low gas-surface-density galaxies might remain without a spectroscopic redshift, which could bias our sample towards higher gas surface densities.

The size estimates from \texttt{IMFIT} are on the same order as the beam size of the current ALMA observations ($\approx 2$~arcsec), and could be affected by the magnification of gravitational lensing. These effects, however, would only move the data points along the {\it diagonal lines} of constant depletion timescales (if we exclude differential lensing), and would not affect our estimates of the SK-slope; however it is difficult to exclude any effects from differential lensing at the current resolution \citep{serjeant2012}.

\subsection{Dynamical properties of BEARS galaxies}
\label{sec:dynamicalmassandsize}
The moderate resolution of our observations ($\approx 2~\mathrm{arcsec}$) and velocity width of the lines provides a potential window on the dynamical nature of these high-redshift galaxies. We calculate (apparent) dynamical virial and rotational mass following earlier studies (\citealt{Neri2003,Tacconi2006,Bouche2007,Engel2010,bothwell2013,bussmann13,Wang2013,Willot2015,Venemans2016,yang2017}) as
\begin{eqnarray}
M_{\rm dyn,vir} &=& 1.56\times10^6\left(\frac{\sigma}{\mathrm{km~s^{-1}}}\right)^2\left(\frac{r}{\mathrm{kpc}}\right)~\mathrm{M}_\odot \label{eq:virialdynamical}, \\
M_{\rm dyn,rot} &=& 2.32\times10^5\left(\frac{v_{\rm circ}}{\mathrm{km~s^{-1}}}\right)^2\left(\frac{r}{\mathrm{kpc}}\right)~\mathrm{M}_\odot \label{eq:rotationaldynamical},
\end{eqnarray}
in which $\sigma=\Delta V_\mathrm{CO}/(2\sqrt{2\ln{2}})$, $\Delta V_\mathrm{CO}$ is the full width at half maximum (FWHM) of the line, $v_{\rm circ}$ is the circular velocity (using $v_{\rm circ} = 0.75 \Delta V_\mathrm{CO}/\sin i$, with the inclination angle set as the average of $i = 55 \deg$ following \citealt{Wang2013}), and $r$ is the effective radius. 
Here we note that a wrong estimate of $i$ may lead to a significant systematic shift for individual objects, but that using the average value is useful for statistical estimates of $M_{\rm dyn}$ across the population, even though individual $M_{\rm dyn}$ estimates are not necessarily reliable. We calculate the effective radius from the Band~4 continuum image using the \texttt{CASA} tool \texttt{IMFIT}. Here we note that the size estimation from Band~4 continuum might not equal the source size probed by the CO emission lines \cite[e.g.,][]{C.C.Chen2017}. The source size is calculated from the deconvolved major and minor axes, added in quadrature. Any measurement errors are also added in quadrature. 

\begin{figure}
    \centering
    \includegraphics[width=\linewidth]{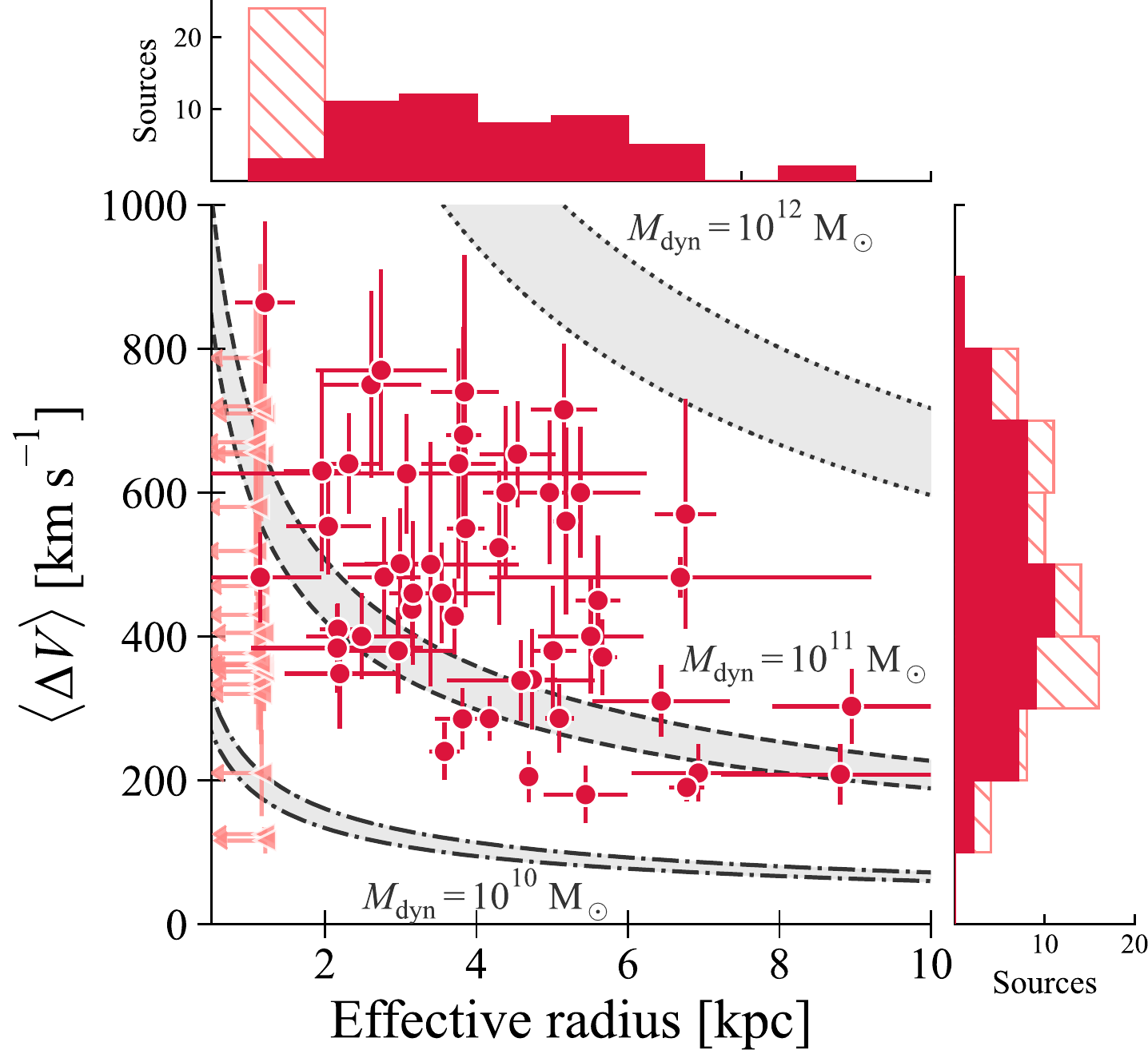}
    \caption{Velocity widths versus effective radii for sources with detected line emission. {\it Dash-dotted, dashed and dotted lines} show constant dynamical masses, which are $10^{10}~\mathrm{M}_\odot$, $10^{11}~\mathrm{M}_\odot$ and $10^{12}~\mathrm{M}_\odot$, respectively. {\it Grey shaded regions} are the variations in which equations of rotational (upper) or virial (lower) estimates shown in \citet{yang2017} (i.e., equations \ref{eq:virialdynamical} and \ref{eq:rotationaldynamical}) are selected to calculate dynamical mass. The {\it pink data points} represent the upper limits of the size estimation from \texttt{IMFIT}. These sources are unresolved so we set the size of them to the minimum size of detected sources ({\it red points}). We show the histograms of the velocity and radius distributions on the right and top sides of the graph respectively; the {\it red filled and pink hatched histograms} correspond to the data points with the same colour. }
    \label{fig:DeltaV_Re}
\end{figure}

In Figure~\ref{fig:DeltaV_Re} we show the velocity distribution of the sources as a function of their effective radius. Twenty-one~galaxies do not have reliable source sizes from the \texttt{IMFIT} routine, and are shown in {\it light red}. For these 21~galaxies, we set their sizes to $0.14~\mathrm{arcsec}$, the lower end of the reliable \texttt{IMFIT} size estimates. There are three likely interpretations of these source sizes. The first and most straightforward is that the source is unlensed, and therefore that the sizes represent the physical sizes of the galaxies. The second option is that the sources are strongly gravitationally lensed, and that the spatial extent is that of a single gravitationally-lensed arc resulting from a foreground galaxy cluster or group. In that case, the observed angular size in the tangential direction would be the magnification $\mu$ multiplied by the intrinsic size, while in the radial direction there would be no magnification. The third option is that the system is a galaxy-galaxy gravitational lens system, in which case the angular extent is likely to reflect that of the Einstein radius of the lensing geometry rather than the size of the BEARS galaxy.

We compare the effective radii and velocities with fixed-mass solutions to the dynamical mass equations, on the simplest assumption that the systems are unlensed. Most galaxies have dynamical masses between $10^{11}~\mathrm{M}_{\odot}$ and $10^{12}~\mathrm{M}_{\odot}$, and the variation is only minor between rotational (upper) and virial (lower) mass limits. There are several caveats to this method, however. Firstly, we do not have any direct reason to assume that these systems are either virialized or stably rotating. Secondly, the systems may be gravitational lenses, in which case the radii in equations~\ref{eq:virialdynamical} and \ref{eq:rotationaldynamical} are overestimates of the underlying source sizes. 

These (apparent) dynamical galaxy masses are at the most massive end of the star-formation main sequence, and appear to indicate that these systems are some of the most massive galaxies in the Universe. Indeed, DSFGs have often been suggested as progenitors of red-and-dead giant elliptical galaxies at $z = 0$ \referee{\citep{Swinbank2006,Coppin2008,Toft2014,Ikarashi2015,Simpson2017,Stach2017}} given (1) their high stellar masses \citep{Hainline2011,Aravena2016}, (2) their high specific star-formation rates \citep{Straatman2014,Spilker2016,Glazebrook2017,Schreiber2018,Merlin2019} and (3) their location in overdense regions (\citealt{Blain2004,weiss2009,Hickox2012}).

However, there exist some important additional caveats to the dynamical mass estimates. A critical underlying assumption is that the systems are self-gravitating and relaxed, as discussed in \citet{Dunne2022}. If submm galaxies are dynamically complex, for example if their gas kinematics is dominated by a major merger, then this could yield apparent anomalously large dynamical masses. \citet{Dunne2022} show that a very wide range of star-forming galaxies can be interpreted consistently as having a constant gas mass conversion factor of $\alpha_\mathrm{CO}=4.0~\mathrm{M}_\odot~(\mathrm{K~km~s^{-1}~pc^2})^{-1}$, and attribute the previous roughly 5 times lower estimates to the unrelaxed dynamical states of submm galaxies (e.g., their equation 9). On the contrary, galaxies undergoing rapid collapse triggering bursts of star formation through violent disc instabilities -- as for example seen in SDP.81 \citep{Dye2018} -- could cause us to underestimate the dynamical masses. Similarly, differential lensing \citep{serjeant2012} of low-dispersion star-forming regions could cause us to underestimate the total velocity widths and thus the dynamical masses.


\subsection{Molecular gas mass}
\label{sec:moleculargasmass}
\begin{figure}
\includegraphics[width=\linewidth]{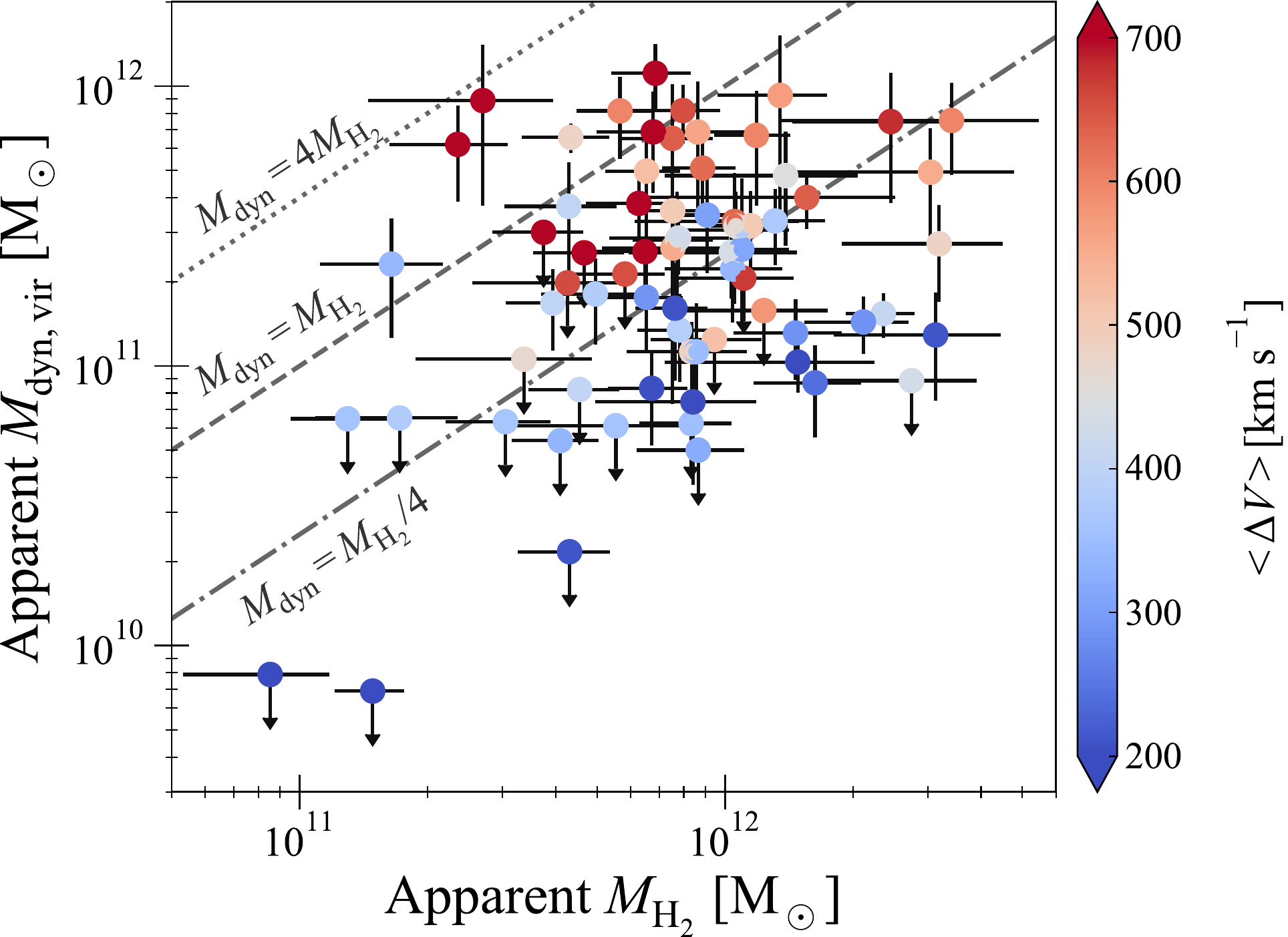}
\caption{
Comparison of the molecular gas masses to dynamical masses for the BEARS sources. The colours indicate the average velocity width of the emission lines, and upper limits indicate sources without accurate size estimates from the \texttt{IMFIT} analysis. The majority of sources lie below the $M_{\rm dyn} = M_{\rm gas}$ relation (\textit{dashed line}), suggesting that we may be underestimating the dynamical masses of these systems. This can be caused by gravitational lensing, which boosts the observed gas mass, or by the compact nature of the dust emission.
}
\label{fig:VirializedDynamicalmass}
\end{figure}

Figure~\ref{fig:VirializedDynamicalmass} shows the (apparent) dynamical virial mass of each galaxy against the estimated molecular gas mass, on the assumptions of no gravitational lensing (see above) and dynamically relaxed (i.e., no major merging) states in the submm galaxies. The figure also shows the 1:4, 1:1 and 4:1 scaling relations between the molecular and dynamical masses.
Remarkably, the majority of our galaxies have a molecular gas mass above that of the dynamical mass. 
We find a scatter of half an order of magnitude around the 4:1 scaling relation, with galaxies having large velocity widths ($\Delta V$) scattering towards higher dynamical masses and vice versa.

Gravitational lensing does not resolve this apparently unphysical pair of mass constraints. Lens magnification corrections would reduce $M_{\rm gas}$ estimates, because this scales with the line luminosity, but lensing would also reduce $M_{\rm dyn}$. The latter correction may even be stronger than that for $M_{\rm gas}$, especially if the physical size estimates from our marginally-resolved ALMA data in equations~\ref{eq:virialdynamical} and \ref{eq:rotationaldynamical} are dominated by Einstein radii, however, the positions of galaxies above the ``main sequence'' in the SK plane in Fig.~\ref{fig:SchmidtKennicutt} would be insensitive to magnification effects. 

An alternative reading of this apparently unphysical result in Fig.~\ref{fig:VirializedDynamicalmass} is that the underlying assumption of the dynamical mass estimates is false, i.e., that the systems are not dynamically relaxed \citep[e.g.,][]{Dye2018} or that the line velocity does not represent the bulk of the galaxy \citep[e.g.,][]{Hezaveh2012,serjeant2012}. 
We argue therefore that our results may still be consistent with the interpretation of a constant $\alpha_{\rm CO}$ conversion factor, in agreement with the merger hypothesis for submm galaxies \citep{Sanders1988,Hopkins2008}. 




Nevertheless, the gas mass alone (modulo the gas fraction) indicates that these systems are at the massive end of the galaxy mass function. The discovery of massive, quenched systems around the first billion years of the Universe \citep{Straatman2014,Glazebrook2017,Schreiber2018} indicates the need for galaxies that rapidly build-up mass and, more importantly, rapidly quench afterwards. At gas masses over $10^{11}~\mathrm{M}_{\odot}$, DSFGs are likely progenitors of the quenched population, although the quenching timescale of $0.1$ to $1~\mathrm{Gyr}$ from Fig.~\ref{fig:SchmidtKennicutt} does not appear to be rapid enough to quench the systems adequately. 

\subsection{Dust-to-gas mass ratio}
\label{sec:G2D}
The dust-to-gas mass ratio is the dust mass divided by the molecular gas mass, and as the dust locks up metals produced through bursts of star-formation, this ratio forms an important evolutionary probe across time \citep{Peroux2020,Zabel2022} sensitive to the gas-phase metallicity in a system \citep{james2002, DraineLi2007,galliano2008,Leroy2011,RemyRuyer2014,Shapley2020,Granato2021}. Meanwhile, the dust destruction mechanisms require the conditions for dust formation to be recent ($\sim 0.3$--$0.5~\mathrm{Gyr}$, \citealt{Hu2019,Hou2019,Osman2020}). Models further suggest that these high star-formation rates also correlate with in/outflows \citep{Triani2020,Triani2021}, ubiquitously observed for DSFGs \citep{Butler2021,berta2021,Riechers2021,Spilker2020a,Spilker2020}.

\begin{figure}
\includegraphics[width=\linewidth]{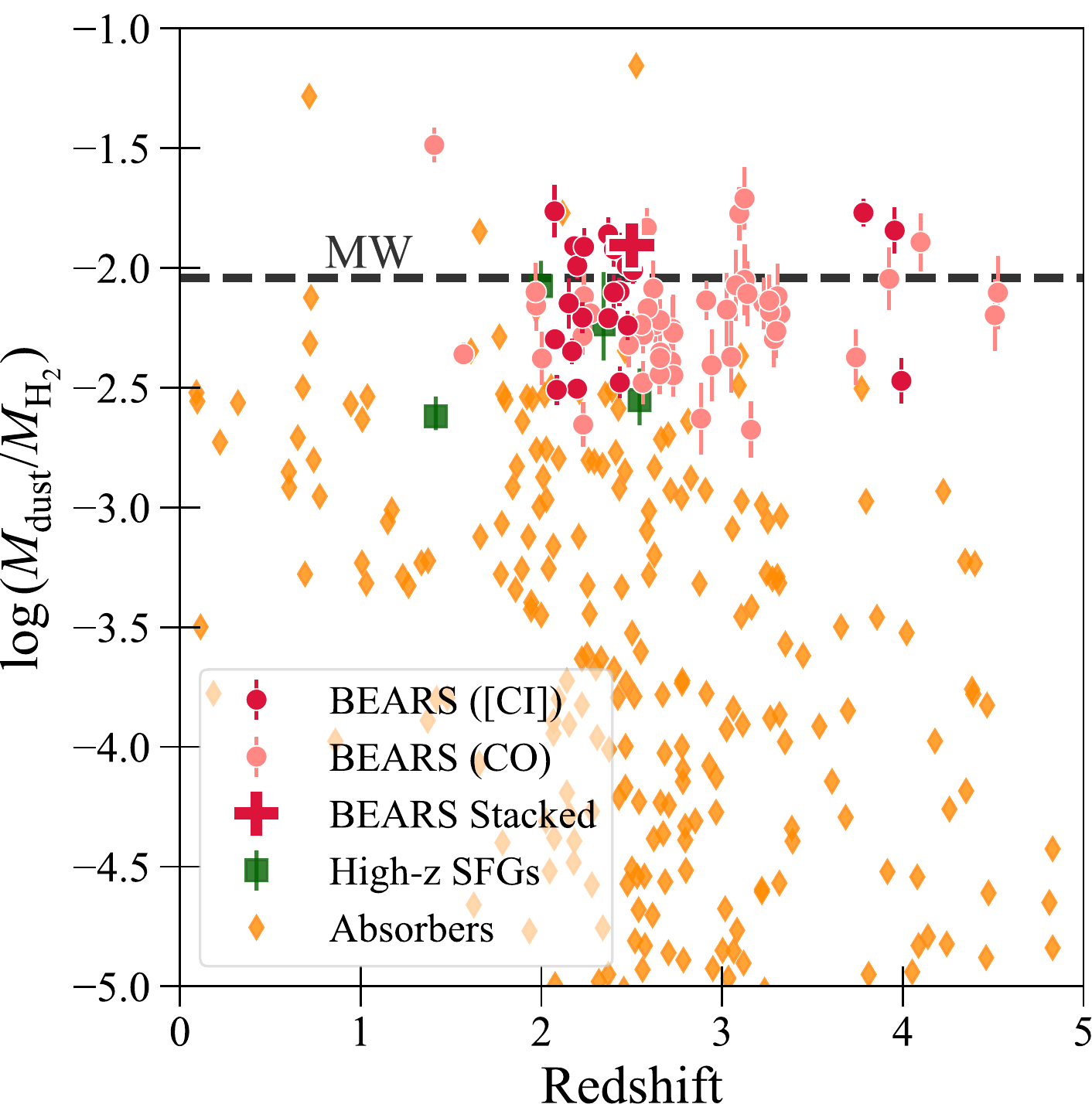}
\caption{Redshift evolution of dust-to-gas ratio of our galaxies, as well as high-redshift star-forming galaxies from \citet{Shapley2020} (\textit{green squares}), and damped Lyman-$\alpha$ absorber systems from \citet{DeCia2018}, \citet[and references therein]{Peroux2020} (\textit{orange points}). The lack of redshift evolution and low scatter across all redshifts suggests we are witnessing an approximately single galaxy type with (recent) intense star formation in line with a metallicity of close to the Milky Way \citep[$\approx0.45~\mathrm{Z_\odot}$;][]{Draine2007,Draine2014}. }
\label{fig:Gas-to-dustratio}
\end{figure}

Figure~\ref{fig:Gas-to-dustratio} shows the dust-to-gas ratio for the BEARS targets, as well as for other high-redshift star-forming galaxies from \cite{Shapley2020} and damped Lyman-$\alpha$ absorber systems from \citet{DeCia2018}, \citet[and references therein]{Peroux2020}. \referee{We use the 151~GHz flux density from \cite{Bendo2023}, and convert this to a dust mass assuming $S_{\nu} = \kappa_{\nu} B_{\nu}(T = 35~{\rm K}) M_{\rm dust} D_\mathrm{L}^{-2}$, where $\kappa_{\nu}$ is the dust mass absorption coefficient, $B_{\nu}$ is the Planck function at temperature $T = 35$~K, and $D_\mathrm{L}$ is the luminosity distance. Here we assume a $\beta = 2$, and approximate the dust mass absorption coefficient ($\kappa_{\rm \nu}$) as $\kappa_{\rm \star} \left( \nu / \nu_{\star}\right)^{\beta}$, with ($\kappa_{\rm \star}, \nu_{\star})$ as (10.41~cm$^2~$g$^{-1}$, 1900~GHz) from \cite{Draine03}.}
The BEARS sources have ratios of the order $10^{-2}$ to $10^{-3}$, similar to what is seen in local dusty galaxies and high-redshift DSFGs. Absorber systems are typically selected solely by their bulk atomic gas and are thus sensitive to the metallicity-evolution of galaxies across time. Unlike absorber systems, BEARS and other SFGs \citep{Shapley2020} do not show any redshift evolution, in agreement with a metallicity close to the Milky Way \citep{Draine2007,Draine2014}. The low scatter and lack of trend with redshift suggests that we are witnessing a single star-forming phase, and the high dust-to-gas ratio suggests that this phase occurred relatively recently. 

\subsection{Photo-dissociation regions inside DSFGs}
\label{sec:PDR}
\begin{figure*}
\includegraphics[width=\linewidth]{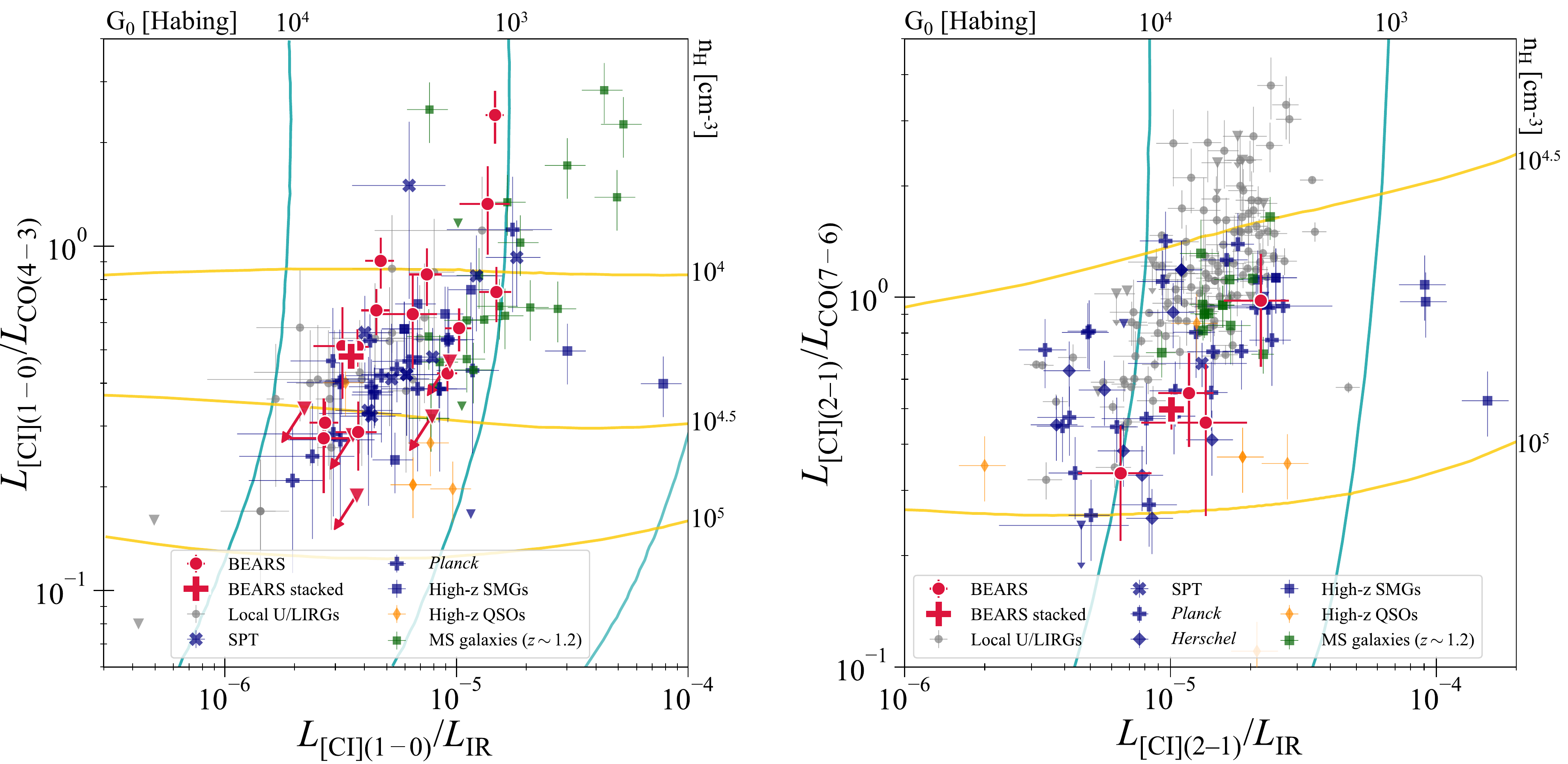}
\caption{{\it Left panel}: Ratio of [C\,{\sc i}]~($^3P_1$--$^3P_0$) to CO~(4--3) plotted versus [C\,{\sc i}]~($^3P_1$--$^3P_0$) luminosity over infrared luminosity for 19 BEARS sources with both [C\,{\sc i}]~($^3P_1$--$^3P_0$) and CO~(4--3) observations ({\it red circles and downward triangles} for detections and upper limits, respectively) as well as reference samples of local U/LIRGs (\citealt{Michiyama2021}; \textit{grey points}), high-redshift SPT-sources (\citealt{Bothwell2017}; \textit{blue crosses}) and \textit{Planck}-sources (\citealt{Harrington2021}: \textit{blue plus points}), classical SMGs (\textit{blue squares}), QSOs at $z\simeq2$--$4$ (\textit{orange diamonds}), and main-sequence galaxies at $z\simeq1.2$ (\textit{green squares}) from \citet[and references therein]{Valentino2020CI}. We also include the result from stacking ({\it red plus sign}). 
{\it Right panel}: Ratio of [C\,{\sc i}]~($^3P_2$--$^3P_1$) to CO~(7--6) plotted versus [C\,{\sc i}]~($^3P_2$--$^3P_1$) luminosity over infrared luminosity for 4 BEARS sources with both [C\,{\sc i}]~($^3P_2$--$^3P_1$) and CO~(7--6) observations ({\it red circles} for detections) as well as reference samples of local U/LIRGs (\citealt{Lu2017}; \textit{grey points}), high-redshift SPT-sources (\citealt{Jarugula2021}; \textit{blue crosses}) and \textit{Planck}-sources (\citealt{Harrington2021}: \textit{blue plus points}), \textit{Herschel} DSFGs (\textit{blue diamonds}), classical SMGs (\textit{blue squares}), QSOs at $z\simeq2$--$4$ (\textit{orange diamonds}) from \citet[and references therein]{Valentino2020CI}, and main-sequence galaxies at $z\simeq1.2$ (\textit{green squares}) from \citet{Valentino2020CO}. We also include the result from stacking ({\it red plus sign}).
The PDR models of \citet{Kaufman2006} are overplotted in both panels, showing the theoretical relations for constant hydrogen densities ({\it near-horizontal yellow solid lines}) and FUV radiation intensity fields ({\it near-vertical cyan solid lines}). }
\label{fig:PDRplot}
\end{figure*}
Figure~\ref{fig:PDRplot} shows the luminosity ratio of [C\,{\sc i}]~($^3P_1$--$^3P_0$) to CO~(4--3) ({\it left panel}) and [C\,{\sc i}]~($^3P_2$--$^3P_1$) to CO~(7--6) ({\it right panel}) against their respective [C\,{\sc i}] luminosity over infrared luminosity. In total, 19 BEARS sources have both [C\,{\sc i}]~($^3P_1$--$^3P_0$) and CO~(4--3) observations, and four BEARS sources have [C\,{\sc i}]~($^3P_2$--$^3P_1$) and CO~(7--6) observations. Here, we calculate the line luminosity using the typical equation from \cite{solomon1997},
\begin{equation}
    L = 1.04 \times{} 10^{-3} S \Delta v f_{\rm obs} D_\mathrm{L}^2~[\mathrm{L_\odot}].
\end{equation}
We also include the reference samples of local U/LIRGs (\citealt{Michiyama2021} in the left panel and \citealt{Lu2017} in the right panel), DSFGs and QSOs in the range of redshift $2$--$4$, main-sequence (MS) galaxies at $z\simeq1.2$ (\citealt{Valentino2020CO,Valentino2020CI} and references therein), high-redshift SPT sources (\citealt{Bothwell2017} in the left panel and \citealt{Jarugula2021} in the right panel) and \textit{Planck} sources (\citealt{Harrington2021}). In this plane, the physical conditions of PDRs \citep[e.g.,][]{Hollenbach1999} can be constrained \citep[e.g.,][]{Umehata2020,Valentino2020CI,Michiyama2021}. We investigate the differences in PDR conditions among our and reference samples using \texttt{PDRToolbox}\footnote{\url{https://dustem.astro.umd.edu/index.html}} \citep{Kaufman2006,Pound2008,Pound2011}, which provides the line intensities for each combination of hydrogen density ($n_\mathrm{H}$) and far-ultraviolet (FUV) radiation intensity fields ($G_0=1.6\times10^{-3}~\mathrm{erg~s^{-1}~cm^{-2}}$; \citealt{Habing1968} units, i.e., the incident FUV field between $6~\mathrm{eV}< h \nu < 13.6~\mathrm{eV}$) assuming plane parallel model \referee{originally by \cite{Kaufman1999}}. We show the theoretical tracks for constant $n_\mathrm{H}$ and $G_0$ using {\it yellow} and {\it cyan solid lines} in Fig.~\ref{fig:PDRplot}.  

Most of our sources are located within $10^4\leq n_\mathrm{H}\leq 10^5~[\mathrm{cm^{-3}}]$ and $10^3\leq G_0\leq 10^4~[\mathrm{Habing}]$. Our sources exhibit denser and more intense radiation environments than MS galaxies at $z\simeq1.2$, but similar properties to local U/LIRGs and other DSFGs. This is consistent with previous works (\citealt{Valentino2020CI,Michiyama2021}), with the exception of the sources HerBS-90 and -131B, where $L_\mathrm{[CI](1-0)}/L_\mathrm{CO(4-3)}\geq1$. This suggests that the gas in these two sources is more diffuse, similar to MS galaxies, in line with photo-ionization modeling to dwarf galaxies by \cite{Madden2020}. We note that the offset could also be caused by the different observed line profiles between CO and [\textsc{C\,i}], i.e., the larger estimation of $\Delta V$ in [\textsc{C\,i}] line emission might have resulted in an overestimate of the line luminosity. 

The ISM properties derived from the different atomic carbon lines, [C\,{\sc i}]~($^3P_1$--$^3P_0$) and [C\,{\sc i}]~($^3P_2$--$^3P_1$), vary slightly in the derived gas densities and FUV intensity fields. This could be due to a change in the internal properties of DSFGs in the early Universe or due to observational biases in selecting our distant galaxies.  
The observational biases could result from a selection towards higher redshift, since all galaxies with CO~(7--6) and \textsc{[C\,i]}~($^3P_2$--$^3P_1$) are detected at higher redshift, where the lines shift into more favourable parts of the atmospheric windows (and the spectral windows used in \citealt{Urquhart2022}). More observations are needed to conclusively test the modest discrepancy between the ISM properties derived from the different atomic carbon lines.

\referee{We use the PDR model from \cite{Kaufman1999} that supposes the simple 1D geometry assuming three discrete layers, i.e., the [C\,\textsc{i}] emission comes only from the thin layer between two gas components traced by CO and singly-ionized carbon emission ([C\,\textsc{ii}]). Meanwhile, spatially-resolved observations of local giant molecular clouds \citep[e.g.,][]{Ojha2001,Ikeda2002} and active star-forming region in local galaxies \citep[e.g.,][]{Israel2015} suggest that the CO- and [C\,\textsc{i}]-bright gas are well mixed. This gas property has been explained more successfully assuming more complicated gas conditions (clumpy geometry: e.g., \citealt{Stutzki1998,Shimajiri2013}, mixing within highly turbulent clouds: e.g., \citealt{Xie1995,Glover2015}, and more complicated 3D geometric models: e.g., \citealt{Bisbas2012}) or additional excitation origins (cosmic rays, CRs: e.g., \citealt{Papadopoulos2004theo,Papadopoulos2018}, shocks: e.g., \citealt{Lee2019}). However, such models require a lot of observational data that trace multi-phase ISM. For example, \cite{Bothwell2017} required multi-transitions of CO, [C\,\textsc{i}] and [C\,\textsc{ii}] lines to constrain gas density and FUV intensity. } 

\referee{In particular, recent works show that CRs can dissociate CO molecules more effectively than FUV radiation in molecular clouds because they are not strongly attenuated by dust \citep[e.g.,][]{Bisbas2017}. Highly star-forming environments such as the ones expected in our sample result in a large amount of CRs (as seen in the stacked spectrum in Section~\ref{sec:ch6}), which will likely affect the properties of the ISM. \cite{Bothwell2017} compare the gas density and FUV intensity obtained from \texttt{PDRToolbox} with those from \texttt{3D-PDR} \citep{Bisbas2012} -- a model that includes the effect of CRs -- for SPT-selected strongly lensed DSFGs. This is a fair comparison, as SPT-selected DSFGs have similar infrared luminosities as our sample. As a result, they find consistency between the two models for the FUV strength, however, a relatively higher gas density ($\langle\log n_\mathrm{H}\rangle = 5.2\pm0.6~\mathrm{cm}^{-3}$) compared to the 1D model from \cite{Kaufman1999} ($\langle\log n_\mathrm{H}\rangle = 4.4\pm0.4~\mathrm{cm}^{-3}$). Since $L_\mathrm{[CI](1-0)}/L_\mathrm{CO(4-3)}$ ratio is sensitive to gas density, CRs can cause us to underestimate the gas density of our sources since we do not consider CRs in our current work. At the moment, it is hard to constrain the physical properties through only two emission lines and infrared luminosity, possibly leading us to underestimate the gas density by nearly one order of magnitude (0.8~dex).}

\subsection{Water lines from BEARS galaxies}
\label{sec:WaterSection}
\begin{figure}
\includegraphics[width=\linewidth]{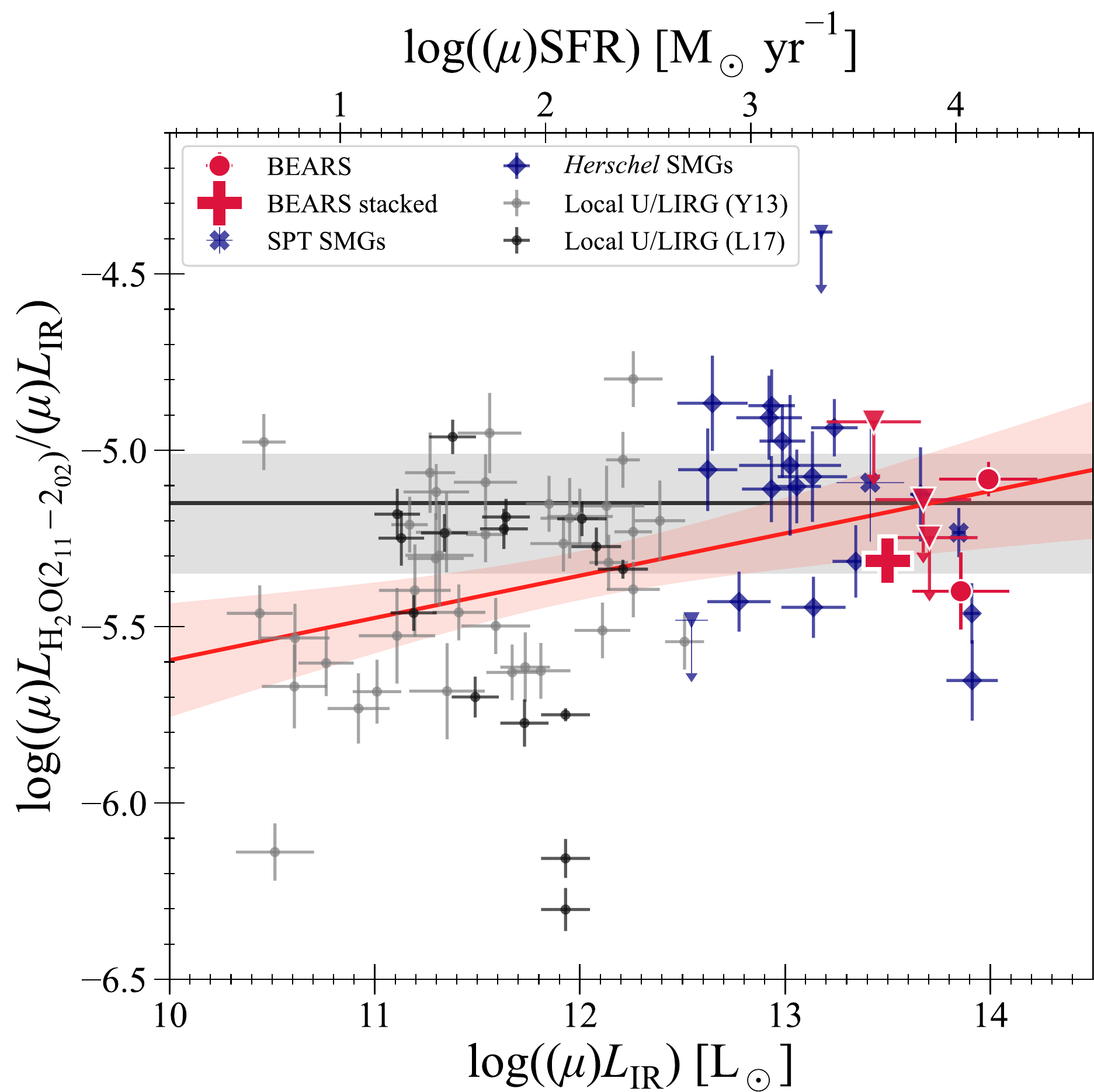}
\caption{Infrared luminosity plotted against the luminosity ratio of H$_2$O~(2$_{11}$--2$_{02}$) emission to $L_{\rm IR}$ of BEARS galaxies ({\it red circles, downward triangles, and plus point} for detections, upper limits, and stacked spectrum, respectively) is shown against a comparison of local infrared-bright galaxies ({\it grey points}; \citealt{Yang2013}, {\it black points}; \citealt{Lu2017}), \textit{Herschel}-selected DSFGs ({\it blue diamonds}; \citealt{Yang2016,Bakx2020IRAM,neri2020}) and SPT-selected DSFGs ({\it blue crosses}; \citealt{Apostolovski2019,Jarugula2021}). The {\it red line and filled region} show the linear regression fit of $\log_{10}\left(L_{\rm H_2O}/L_{\rm IR}\right)=a\log_{10}\left(L_{\rm IR}\right)+b$ with the $\pm2\sigma$ uncertainty. As a result of fitting, we find $a=0.12\pm0.04$ and $b=-6.8\pm0.5$. The {\it black line and grey filled region} corresponds to the best fit result with the slope fixed to zero from \citet{Jarugula2021}, where we use a factor of 1.55 to convert from far-infrared luminosity to infrared luminosity and assume the \citet{Kennicutt2012} relation of SFR~$= 1.47 \times{} 10^{-10} L_{\rm IR}$. Partial collisional excitation of the H$_2$O line could explain our super-linear fit to the luminosity ratio \citep{GonzalezAlfonso2022}.
}
\label{fig:H2O_IR}
\end{figure}
Figure~\ref{fig:H2O_IR} provides a comparison of $L_{\rm IR}$ against the luminosity ratio of H$_2$O~(2$_{11}$--2$_{02}$) emission to $L_{\rm IR}$ for five BEARS sources, three of which are not detected above $3\sigma$. We compare the galaxies against reference samples from \citet{Yang2013, Yang2016}, \citet{Lu2017}, \citet{Apostolovski2019}, \citet{Bakx2020IRAM}, \citet{neri2020} and \citet{Jarugula2021}.
We confirm tight correlations between them and derive a scaling relation for the H$_2$O emission using linear regression to all the samples shown in Fig.~\ref{fig:H2O_IR} in order to account for errors in both the infrared luminosity and line emission, minimising the fit of
\begin{equation}
    \log_{10}\left(\frac{L_{\rm H_2O}}{L_{\rm IR}}\right)=a\log_{10}\left(L_{\rm IR }[\mathrm{L}_{\odot}]\right)+b,
\end{equation}
where both $a$ and $b$ are left as fitting parameters. 

A super-linear relation seems to better fit all the observed data across four orders of magnitude, with $a=0.12\pm0.04$ and $b=-6.8\pm0.5$, which is consistent with previous studies (e.g., \citealt{Omont2013, Yang2013,Yang2016}). Our linear regression fitting favours a super-linear relation at the $3\sigma$ level, although we note that our observations only add two H$_2$O detections, and three upper limits. Contrary to \cite{Jarugula2021}, we also include sources from \cite{Lu2017} that are not included in \cite{Yang2013}. The improved fitting constraints from the additional sources since \cite{Yang2016} -- who find ($a = 1.16 \pm 0.13$) -- likely cause an increase in the significance of the super-linear fit result from $\sim 1$ to $\sim 3 \sigma$.

Both observations (e.g., \citealt{Riechers2013,Yang2013,Yang2016,Apostolovski2019,Jarugula2021}) and modelling (e.g., \citealt{GA2010,GA2014}) find a strong correlation between the infrared luminosity and the H$_2$O line emission. The moderately-higher transitions of H$_2$O (above $\approx 100~\mathrm{K}$) are not excited through collisions, but instead infrared pumping is expected to be their dominant excitation mechanism \citep{GonzalezAlfonso2022}. Meanwhile, the $J = 2$ lines are likely still partially collisionally excited, which could explain the super-linear scaling relation \referee{\citep{Yang2013,Liu2017}}, although the super-linear trend appears to go away in resolved observations \citep{Yang2019}, and could be due to optical depth effects \citep{GonzalezAlfonso2022}. The origins of our observed super-linear relation will require detailed radiative transfer modeling across multiple water transitions. We briefly note that this result is in contrast to the picture of H$_2$O line luminosity being proportional to SFR, even down to the resolved scale of individual GMCs \citep[e.g.,][]{Jarugula2021}.

\section{Companion sources with Unusually-Bright lineS: The BEARS CUBS}
\label{sec:ch5}
\begin{figure*}
\includegraphics[width=\textwidth]{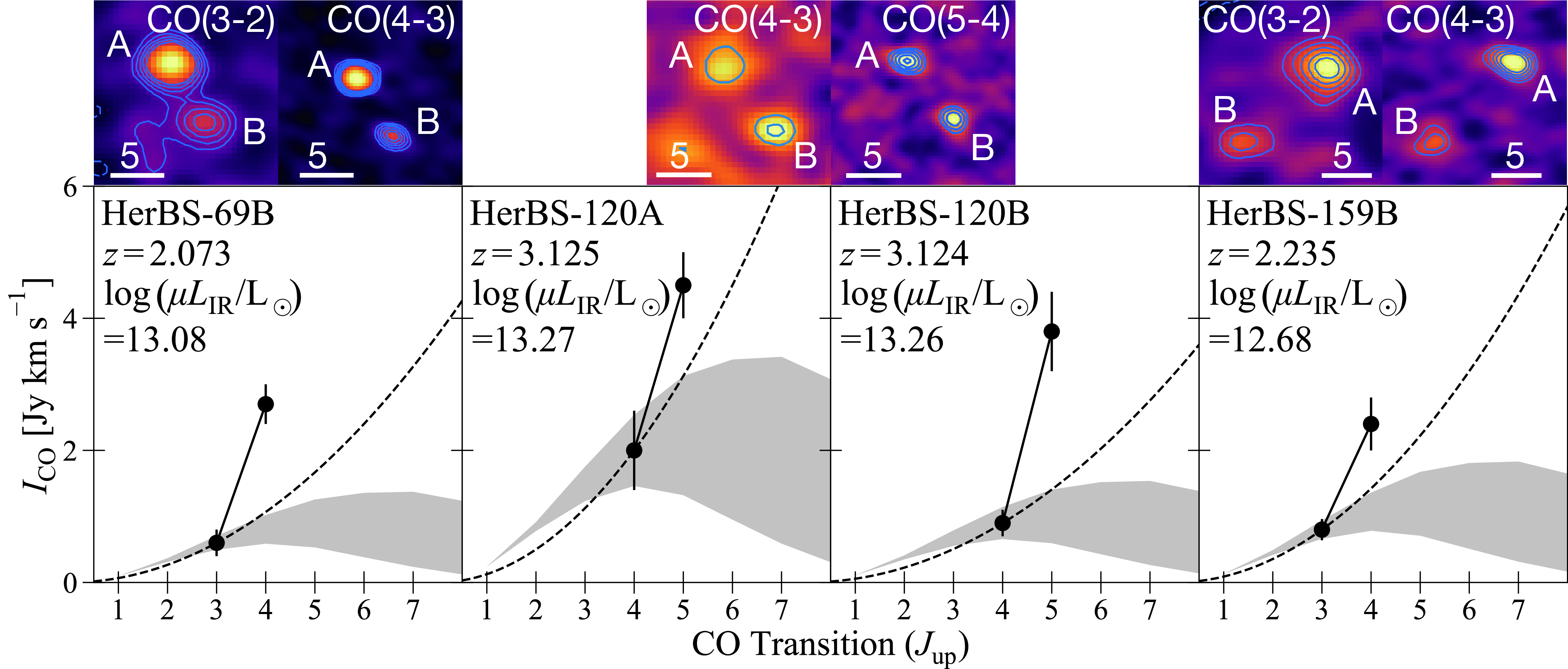}
\caption{
Four sources that have CO SLEDs in excess of a thermalized profile, where $L'_{\rm CO,a}/L'_{\rm CO,b} = 1$. All four have nearby sources at similar redshift. The {\it top panels} show the velocity-integrated fluxes of the spectral lines, with sources at more than 7 arcsec apart. The contours are drawn from $2\sigma$ for the left panel and from $3\sigma$ for the right panel. The white scale bar corresponds to 5 arcsec. The {\it bottom panels} show the CO SLEDs relative to the thermalized profile ({\it dashed black lines}) and the mean SLED with $\pm1\sigma$ standard deviation from \citet{Harrington2021}, similar to Fig.~\ref{fig:individualSLEDs}.
}
\label{fig:CUBS_SLED}
\end{figure*}

\begin{figure}
    \centering
    \includegraphics[width=\linewidth]{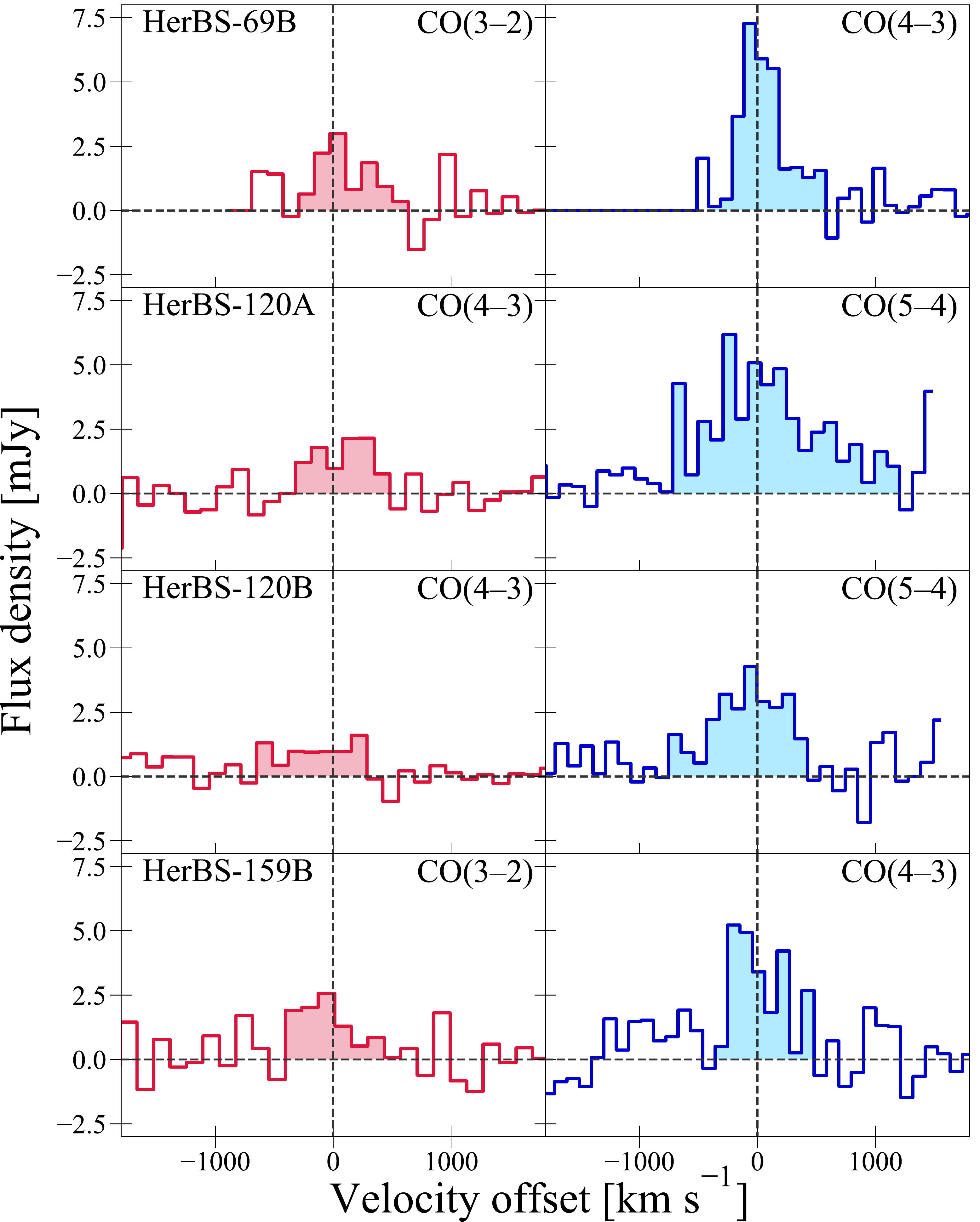}
    \caption{The spectra of four sources shown in Fig.~\ref{fig:CUBS_SLED}. The {\it left column} shows the lines in Band 3, and the {\it right column} shows the lines in Band 4 for each source. {\it Colour filled regions} correspond to the velocity range where we make the velocity-integrated intensity maps. The horizontal axis shows the velocity offset from the systemic velocity ($V=0~\mathrm{km~s^{-1}}$) obtained from the spectroscopic redshift. The spectral resolution is $0.05~\mathrm{GHz}$, which corresponds to $\sim150~\mathrm{km~s^{-1}}$ for Band 3 and $\sim100~\mathrm{km~s^{-1}}$ for Band 4.}
    \label{fig:CUBSSpectra}
\end{figure}

In Subsection~\ref{sec:COSLEDs}, we identified four sources with line ratios in excess of the thermalized profile ($L'_{\rm CO,a}/L'_{\rm CO,b} = 1$; eq. \ref{eq:JaJb}), namely HerBS-69B, -120A, -120B, and -159B. Interestingly, these are found only in fields with multiple sources where the companion galaxy has a slightly different redshift. 
The large angular separation between the components ($7$--$10~\mathrm{arcsecs}$) rules out galaxy-galaxy lensing, and the slightly different redshifts suggest they are not one source multiply imaged. 
In three out of four cases, they are the fainter component, i.e., B sources \footnote{For HerBS-69, the A source is 1.2 times brighter than the B source; for HerBS-120, the A source has a similar brightness as the B source; but for HerBS-159B, the A source is 3 times brighter than the B source.}. Figure~\ref{fig:CUBS_SLED} shows the velocity-integrated CO intensity maps with contours ({\it top panels}) and individual SLEDs ({\it bottom panels}) of these four sources. The CO SLEDs are steeper than the thermalized SLED by about $1$--$2\sigma$ (and of course, they are above the mean SLED from \citealt{Harrington2021}).
Figure~\ref{fig:CUBSSpectra} shows the Bands 3 (\textit{left} column) and 4 (\textit{right} column) spectra of these four sources extracted with the same aperture size. Colour filled regions indicate the velocity range across which we make the velocity integrated intensity maps (i.e., Fig.~\ref{fig:CUBS_SLED}). The vertical dashed line of each panel corresponds to the systemic velocity obtained from the spectroscopic redshift. As expected, we find that the line fluxes in Band~3 are lower than those of Band~4. 

\subsection{Are we confident that these SLEDs are real?}
We take several steps to ensure that the origins of these discrepant SLEDs are physical; we exclude statistical scatter in the ratios ($r_{a,b}$; eq. \ref{eq:JaJb}), calibration issues, line-fitting issues and issues associated with lensing.

Most of the galaxies in \cite{Urquhart2022} have ``normal'' line ratios (see e.g., Figure~\ref{fig:luminosityRatioStack}), while only six sources have values above unity, with two of these six sources consistent with a sub-thermal SLED. 
We measure the following luminosity ratios for these four sources: $r_{4,3} = 2.53\pm0.89$ for HerBS-69B; $r_{5,4} = 1.44\pm0.46$ for HerBS-120A; $r_{5,4} = 2.70\pm0.74$ for HerBS-120B; and $r_{4,3} = 1.69\pm0.44$ for HerBS-159B. 
As you can see, although these are consistent with thermalized profiles by $1$--$2.3\sigma$, we note a fundamental limitation in assessing the super-thermalized line profiles using our data. 
We optimized the velocity-integrated fluxes of the lines to maximally include signal at a moderate cost in signal-to-noise ratio. This results in large intrinsic uncertainties in the luminosity ratios, since we are taking the ratio of two roughly $5\sigma$ luminosities. This limits the maximum significance in line ratios of around $3.5\sigma (\approx 5 / \sqrt{2})$, as is seen in studies finding similar results \citep{Riechers2006QSO,Riechers2020,Weiss2007,Sharon2016}. Moreover, we would rarely achieve a $3.5 \sigma$ case, since we are identifying the super-thermalized cases by their line ratio being in excess of one (instead of zero). Although the luminosity ratios cannot exclude the explanation that these ratios are simple artefacts, the observed fluxes of the resolved lines (Figure~\ref{fig:CUBSSpectra}) show a more convincing case towards the veracity of the super-thermalized line profiles.

We estimate the potential for calibration issues by comparing the Bands 3 and 4 continuum emission from \referee{\cite{Bendo2023}}. The HerBS-120B source has $151~\mathrm{GHz}$ to $101~\mathrm{GHz}$ flux density ratio of $3.28$. This is not extraordinary compared to other sources. 
For the other two fields (HerBS-69 and -159), we do not detect the Band 3 continuum; however, we confirm that the brighter sources in the same field, i.e., A sources have normal (sub-thermal) CO SLEDs.
The photometric redshifts of the CUBS fields including ALMA Bands 3 and 4 agree with the spectroscopic solutions, and have similar uncertainties to other galaxies in the BEARS sample. The observational strategy in the BEARS programme images multiple galaxies using the same flux and phase calibrator, and here we note that HerBS-159 is observed in a different scheduling block than HerBS-120 and HerBS-69. The lack of any obvious calibration issues across the other sources in these scheduling blocks provides further confidence in the authenticity of these line ratios.

We also investigate potential issues with the line fitting. 
Based on Figure~\ref{fig:CUBSSpectra}, the CO emission lines are located at the band-edge of our tunings for the HerBS-120 and HerBS-69 fields. We therefore investigate the off-source root-mean square of the data cube as a function of frequency. We do not find any significant ($< 15$~per cent) variation in the noise at the position of the spectral lines. 
For HerBS-120A, the velocity width of both lines are different and thus we perhaps over-estimate (under-estimate) the line flux of Band~4 (Band~3).

We can also exclude differential magnification as an explanation, where the inhomogeneous magnification across the source causes flux ratios that are not representative of the entire source. The effects of differential magnification on CO SLEDs have been extensively simulated by \citet{serjeant2012}. Flux from spatially concentrated regions can be located close to a caustics (i.e., high-magnification region), leading to higher magnifications than for the rest of the system. However, this boosting cannot explain why high-$J$ CO transitions would appear to have super-thermal luminosities, because {\it all} transitions in the spatially-concentrated region would be similarly boosted. Differential magnification of thermalised or sub-thermalised SLEDs only generates thermalised or sub-thermalised SLEDs. \referee{We provide a more thorough investigation why this is the case in Appendix \ref{sec:whydifferentiallensingcannotproducesuperthermalizedlineprofiles}.}

\subsection{Physical interpretation}
The conditions of the ISM affect the CO line ratios of galaxies. Higher-$J$ CO lines trace denser gas components (with the CO line transitions having roughly $n_{\rm crit} \propto J^3$) and are often clumpy in nature, while the lower-$J$ CO lines can extend throughout and even beyond individual galaxies \citep[e.g.,][]{Cicone2021}. Basic \texttt{RADEX} \citep{vanderTak2007,ndRADEX} models suggest that high hydrogen density ($> 10^{4}~\mathrm{cm}^{-3}$) and high gas temperature ($T_{\rm kin} > 100~\mathrm{K}$) can indeed reproduce such super-thermalized luminosity ratios until $\sim1.7$ between CO~(4--3) and CO~(3--2) as well as CO~(5--4) and CO~(4--3), which shows that there is no simple model reproducing the line luminosity ratio of $\gtrsim2$. However, at least, these physical properties appear extreme when compared to previously observed gas conditions. \referee{Similar to differential lensing, a highly multi-phased ISM cannot reproduce the observed ratios, as can be seen by evaluating the discussion in Appendix~\ref{sec:whydifferentiallensingcannotproducesuperthermalizedlineprofiles} with the magnification, $\mu$, set to 1.}
To explain such extreme gas conditions we need strong heating sources, and thus we focus on dust-obscured AGN or galaxy mergers.

\citet{Riechers2006QSO} presented similar super-thermalized luminosity ratios of CO~(4--3) to CO~(2--1) or CO~(1--0) towards APM08279$+$5255, which is a lensed QSO at $z=3.91$. \citet{Weiss2007} suggested that the luminosity ratio indicates moderate opacities of low-$J$ CO transitions. Their large velocity gradient (LVG; \citealt{Sobolev1960}) models also show that they can well explain that with $n_\mathrm{H}=10^{4.2}~\mathrm{cm^{-3}}$ and $T_{\rm kin} = 220~\mathrm{K}$, which is consistent with our \texttt{RADEX} model. \citet{Sharon2016} reported on observations of a total of 14 known lensed and unlensed sources using the Very Large Array (VLA). They detected the CO~(1--0) emission in 13 sources (down to low significance) and reported one non-detection. They found four candidates that show super-thermalized luminosity ratios between CO~(3--2) and CO~(1--0). Three of these are lensed AGN host galaxies, and the other is a lensed merging system. \citet{Sharon2016} noted large uncertainties in the luminosity ratios and provide several hypotheses to the high ratios: 
(i) the emission is optically thin; 
(ii) the CO~(1--0) line is self-absorbed; and 
(iii) the source of optically thick emission has a temperature gradient. These hypotheses were also previously suggested by \cite{Bolatto2000,Bolatto2003}, in addition to varying beam-filling factors across the different CO transitions, although filling factor effects would instead result in sub-thermalized CO line ratios. This latter point is not expected to be an issue with our current marginally-resolved sources. Since our ratios do not include the CO~(1--0) emission line, the second option is also not a likely solution, but the optical depth effects or thermal gradients can possibly explain our result. In the local Universe, \citet{Meijerink2013} suggested a potential explanation through shocks around the AGN for a local super-thermalized candidate, NGC 6240. 
Here we note that these previous works found discrepant luminosity ratios based on CO~(1--0) -- typically found to be more extended -- so the reasons for the varying ratios might not be the same. Moreover, unlike other discrepant CO line ratio studies, we found these discrepant luminosity ratios using a single facility (ALMA), which makes the result less dependent on telescope-to-telescope variations.

AGN activity can heat gas and produce steep line intensity ratios out to (very) high CO line transitions \citep{Riechers2006QSO,Weiss2007}.
An AGN with bolometric luminosity $>L_*$ would be vanishingly unlikely to be a companion galaxy based on the number densities in \citet{Shen2020} ($1~\mathrm{arcmin}$ area and $\delta z=0.05$ spans 176~cMpc$^3$, compared to $\sim10^{-5}$~cMpc$^{-3}$ of $>L_*$ AGN) unless there is a causal factor in common with the DSFGs such as an environmental trigger or an interaction.

The super-thermalized ratios are observed solely for galaxies in systems of multiples. This could suggest that the presence of a nearby galaxy is important to produce the observed line ratios. Potentially, the interaction of merging galaxies could create the required high gas density and temperature conditions. The turbulent ISM resulting from galaxy mergers (or strong AGN winds) could affect the bulk gas (traced by the lower-$J$ transitions) differently relative to the dense star-forming clumps (traced by the higher-$J$ transitions). The larger line-widths of the bulk gas would push down the optical depth for a fixed column density, reducing the effective cross-section of the molecular gas clouds in the bulk gas and hence cause fainter emission from lower-$J$ CO transitions.
However, in our sample, the companion sources are at a projected distance of $50$ to $80~\mathrm{kpc}$, which is large on the typical distance scales of merging systems \citep{narayanan2015}.



\subsection{BEARS: a unique parent sample?}
Although we are unable to definitively exclude artefacts as the cause of these super-thermalized line profiles, 
we put these results into a larger cosmological perspective.
These four sources are selected directly from the large, homogeneous sample from \citet{Urquhart2022}, 
allowing us to infer the occurrence rate of DSFG properties. 
The typical lifetime of an DSFG (without excessive feeding; cf., \citealt{berta2021}) is on the order of the depletion time, typically $200~\mathrm{Myr}$ (e.g., \citealt{reuter20} and this work).
Our targets are four out of $46$ galaxies with multiple CO lines from \citet{Urquhart2022} (i.e., 9\%), thus we can jointly constrain the timescales and occurrence frequencies of these scenarios. For example, if AGN are universal in DSFG environments, then the AGN lifetimes must be around 9\% of $200~\mathrm{Myr}$, while rarer companion AGN must have longer lifetimes.

\section{Composite spectrum of all BEARS sources}
\label{sec:ch6}
The observations reported in \cite{Urquhart2022} directly detected CO~(2--1) to CO~(7--6) lines, as well as the two transitions of \textsc{[C\,i]} emission and the H$_2$O~(2$_{11}$--2$_{02}$) water line. In this section, we aim to statistically detect more emission lines by stacking the spectrum for each galaxy in their rest-frames.
This is similar to work on the SPT galaxies \citep[e.g.,][]{spilker2014,reuter20,Reuter2022}, as well as for \textit{Herschel} \citep{fudamoto2017}, LABOCA \citep{Birkin2021} and SCUBA-2 \citep{Chen2022} selected sources. In this Section, we aim to provide the archetypal spectrum of a hypothetical galaxy at a redshift of $2.5$ with an observed infrared luminosity of $3 \times{} 10^{13}$~$\mathrm{L}_{\odot}$. 

\subsection{Method}
We now re-extract the spectral data from the individual data cubes of the sources using an automated script. This evenly extracts the spectroscopic information for fair comparison of the bulk behaviour of all galaxies, although it prioritizes signal-to-noise over including all possible signal. 
Using the central positions of the continuum peak positions reported in \referee{\cite{Bendo2023}}, we extract the emission with a variable aperture between 1 to 3 times the radius of the beam to find the optimal balance between high signal-to-noise and full line extraction. 

\begin{figure}
    \centering
    \includegraphics[width=\linewidth]{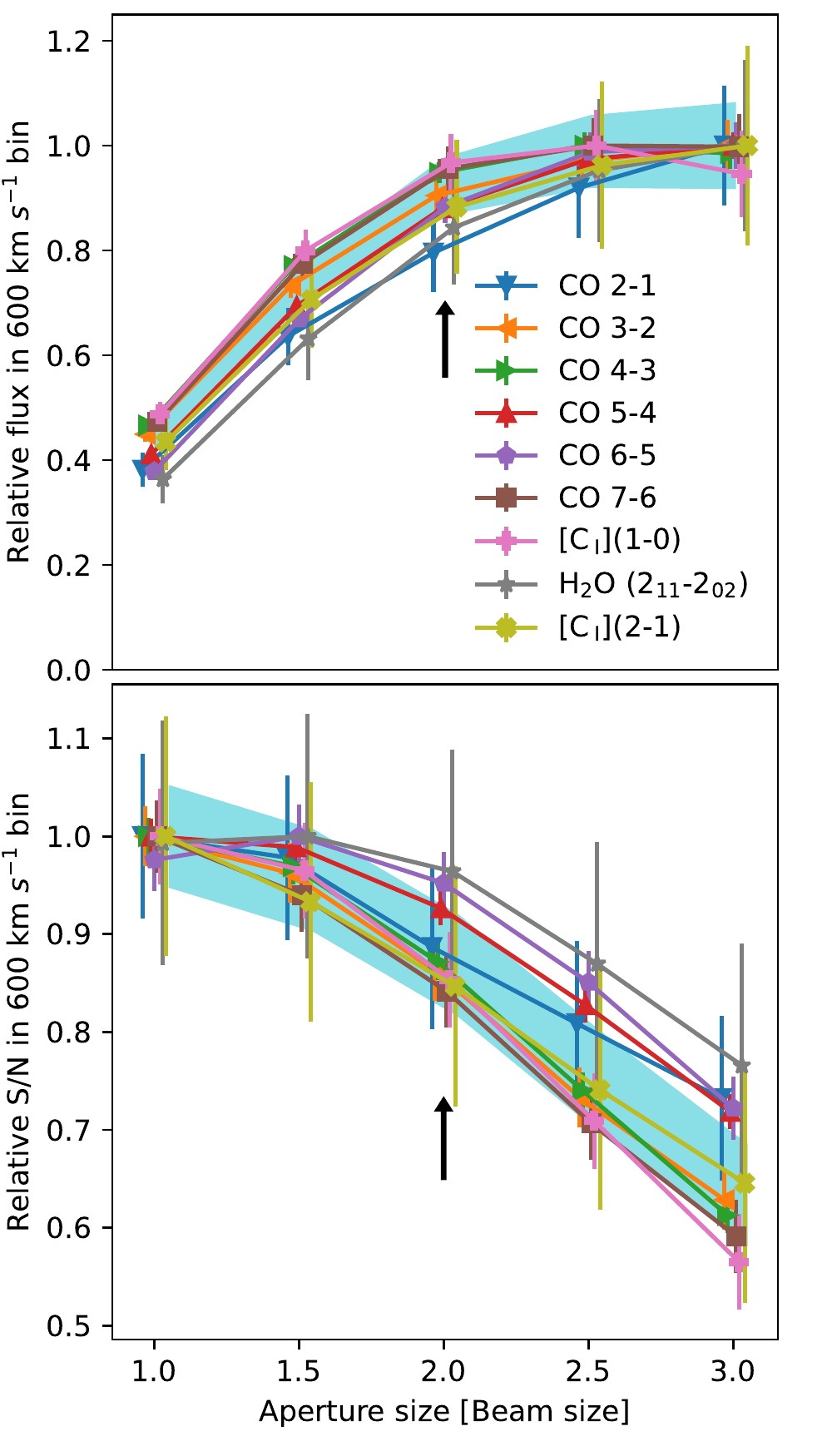}
    \caption{Spectral line fluxes ({\it top}) and SNRs ({\it bottom}) measured within $600~\mathrm{km~s^{-1}}$ bins versus the aperture size (in FWHM) plotted for multiple spectral lines. The {\it points} indicate individual spectral lines, and the {\it background fill} indicates the averaged relative flux ({\it top}) and SNR ({\it bottom}). We choose the extraction sizes of our apertures carefully, by balancing between a large enough aperture to include all the flux, while not degrading our signal-to-noise excessively. Hence, we decide to use an aperture with twice the radius as the beam ({\it arrow}). At this frequency, we are able to extract $92 \pm 6$\% of the flux density at a cost of $13 \pm 5$\% additional noise.}
    \label{fig:ApertureSizes}
\end{figure}

Figure~\ref{fig:ApertureSizes} shows the effective flux and signal-to-noise of the CO~(2--1) to CO~(7--6), [C\,{\sc i}]~($^3P_1$--$^3P_0$), [C\,{\sc i}]~($^3P_2$--$^3P_1$), and H$_2$O lines in an effort to find the optimum extraction area for the automated pipeline. The lines are evaluated within a $600~\mathrm{km~s}^{-1}$ bin; a bin width that was similarly chosen to include most of the signal while also achieving a high signal-to-noise ratio (see Section~\ref{sec:stackedResults}). We find the best extraction aperture to lie at twice the radius of the beam. With only marginal losses in signal-to-noise ratio ($\approx 13 \pm 5$\%), the majority of the flux is extracted across all the lines ($\approx 92 \pm 6$\%). Beyond this size, there appears a sharp down-turn in signal-to-noise ratio complicating the goal of the composite spectrum -- to reveal faint line emission. 
We correct the stacked spectrum by boosting the flux by 8\% ($= 1/0.92$), and accounting for 6\% extra noise in the extracted values, added in quadrature. 

We subtract the continuum emission directly from the spectra assuming a power-law dust continuum emission ($S_{\nu} = \rm A_i \nu^{(2 + \beta)}$), with $\beta_{\rm dust} = 2$. We separately fit the spectra from Bands 3 and 4, while masking out the data within $1500~\mathrm{km~s^{-1}}$ around the CO, [C\,{\sc i}] and H$_2$O lines. We compare to the 151~GHz continuum fluxes from \referee{\cite{Bendo2023}}, and we find a general agreement to their fluxes. On average, our flux estimates agree with the 151~GHz continuum fluxes, and we find a point-to-point standard deviation of $11$\%, although the scatter decreases for the brightest sources (to around $5$\%).

We decide to stack the spectrum of each galaxy with the goal of representing a single archetypal galaxy at $z = 2.5$ (approximately the mean of this sample; \citealt{Urquhart2022}) with an infrared luminosity of $3 \times{} 10^{13}~\mathrm{L}_{\odot}$. This involves normalizing all spectra to the same luminosity and to a common redshift. We use the same scaling factor as \cite{spilker2014}, which is derived from requiring a constant $L'$ across all redshifts in equation~\ref{eq:lprime}:
\begin{equation}
S_{\nu, \rm common} = S_{\rm \nu} \left(\frac{D_\mathrm{L}(z_{\rm source})}{D_\mathrm{L}(z_{\rm common})}\right)^2 \frac{1+z_{\rm common}}{1+z_{\rm source}}, \label{eq:commonRedshift}
\end{equation}
where $D_\mathrm{L}$ refers to the luminosity distance at redshift $z$, and $z_{\rm common}$ is set to 2.5. This factor accounts for the cosmological dimming for each spectral line, as well as for the redshift-dependence of the flux density unit.

We then normalize the luminosity of each galaxy to $3 \times{} 10^{13}~\mathrm{L}_{\odot}$, based on the luminosities calculated in Section~\ref{sec:SLcomparisonAndDustLuminosities}. Since source confusion could affect the \textit{Herschel} photometry, we take the $151$-$\mathrm{GHz}$ flux density that is detected for all sources, and assume a dust temperature  $T_\mathrm{d}$ of $35~\mathrm{K}$.\footnote{For a comparison to the line luminosities in \cite{spilker2014}, one can multiply our $L'$ values by 1.67 to compare our line luminosities to theirs. } Here we note several important caveats when creating a combined spectrum from sources across different frequencies, luminosities and redshifts.
Unlike previous methods, we provide a stacked spectrum normalized against the intrinsic properties of the observed galaxies. Previous methods provide their composite spectrum based on observed properties (e.g., \citealt{spilker2014} aim to provide the properties of a $z = 3$ SPT galaxy with $S_{\rm 1.4 mm} = 15~\mathrm{mJy}$). 
This is an important point since even at $1.4~\mathrm{mm}$, a source with constant flux-density undergoes a roughly $40$\% luminosity difference between $z =1$ and $z = 5$, despite the near-flat K-correction with redshift at $1.4~\mathrm{mm}$ (and in our case $2~\mathrm{mm}$). This results in a noticeable effect on a composite spectrum because each part of the rest-frame spectrum is sensitive to sources from different redshift regions. The luminosity variations with redshift cause an underestimate of the flux at higher redshifts compared to lower redshifts and produces an artificial SLED with a downward slope towards higher $J$, estimated in \cite{spilker2014} to be $\simeq15$\%. These choices make their stacked spectrum a representation of the average observed behaviour of an SPT galaxy. However, their line luminosities do not represent a single (hypothetical) galaxy across all frequencies and the CO line fluxes thus do not represent a typical CO SLED \citep{Reuter2022}.
We note that \cite{Birkin2021} and \cite{Chen2022} use a median or averaged spectrum based solely on the highest signal-to-noise emission. This provides an accurate picture of the average observational result for a galaxy in the sample, but does not reflect the average properties of any single existing or archetypal galaxy.

The stacked spectrum is created at several velocity resolutions by adding the rest-frequency spectra of each source, corrected for luminosity and redshift. Each spectrum is additionally weighted by the inverse variance in each channel. This final weighting step ensures a high signal-to-noise in the stacked spectrum, and removes much of the weighting by luminosity ($3 \times{} 10^{13}~\mathrm{L}_{\odot}$) and redshift ($z_{\rm common} = 2.5$) per each line (or any other weighting based on the properties of a galaxy, i.e., observed flux density or dust mass). In other words, the galaxy-based weighting step affects the ratios between lines, while the noise-based weighting ensures a high signal-to-noise across the spectrum.

\begin{figure}
    \centering
    \includegraphics[width=\linewidth]{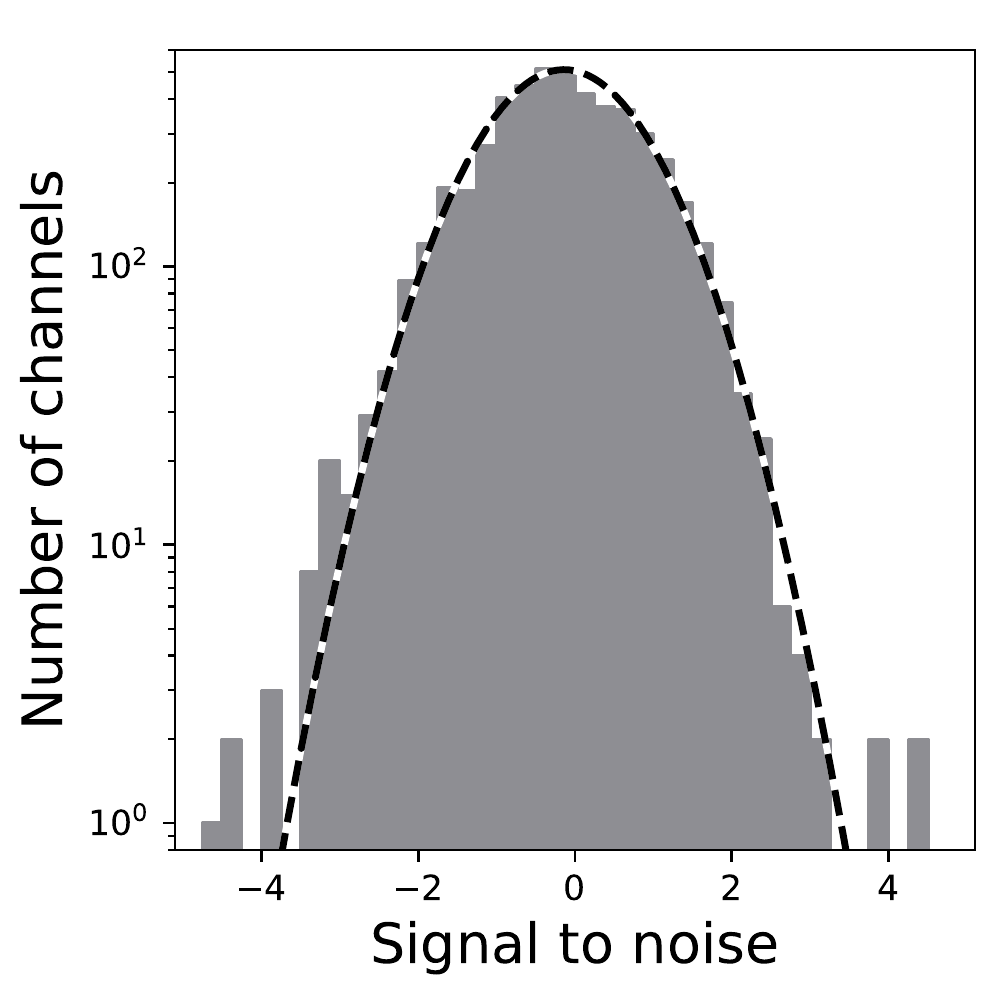}
    \caption{Histogram of 5000 signal-to-noise measurements in off-line $600~\mathrm{km~s}^{-1}$ bins of the stacked spectrum. The {\it dashed black-and-white line} indicates the unit-width Gaussian profile, which matches the bulk of the signal-to-noise distribution well. There exist some excess bins at both low and high signal-to-noise ratio, although there are very few of these.}
    \label{fig:gaussian_of_composite_spectrum}
\end{figure}

We estimate the noise properties of our stacked spectrum using the following procedure. We extract the signal-to-noise ratio in 5000 bins of $600~\mathrm{km~s^{-1}}$ width at random off-line frequencies. Figure~\ref{fig:gaussian_of_composite_spectrum} shows this signal-to-noise distribution, where the frequencies of the bins are chosen to not overlap with any known lines (see Appendix Table~\ref{tab:faintLines}). The {\it dashed black and white line} reflects a unit-width Gaussian profile, and accurately describes the histogram. The off-line stacked spectrum is thus well-described by a white noise spectrum, and importantly confirms there are no issues with coherent processes with our spectrum, such as imperfect continuum subtraction.

\subsection{Results}
\begin{figure*}
    \centering
    \includegraphics[width=0.95\textwidth]{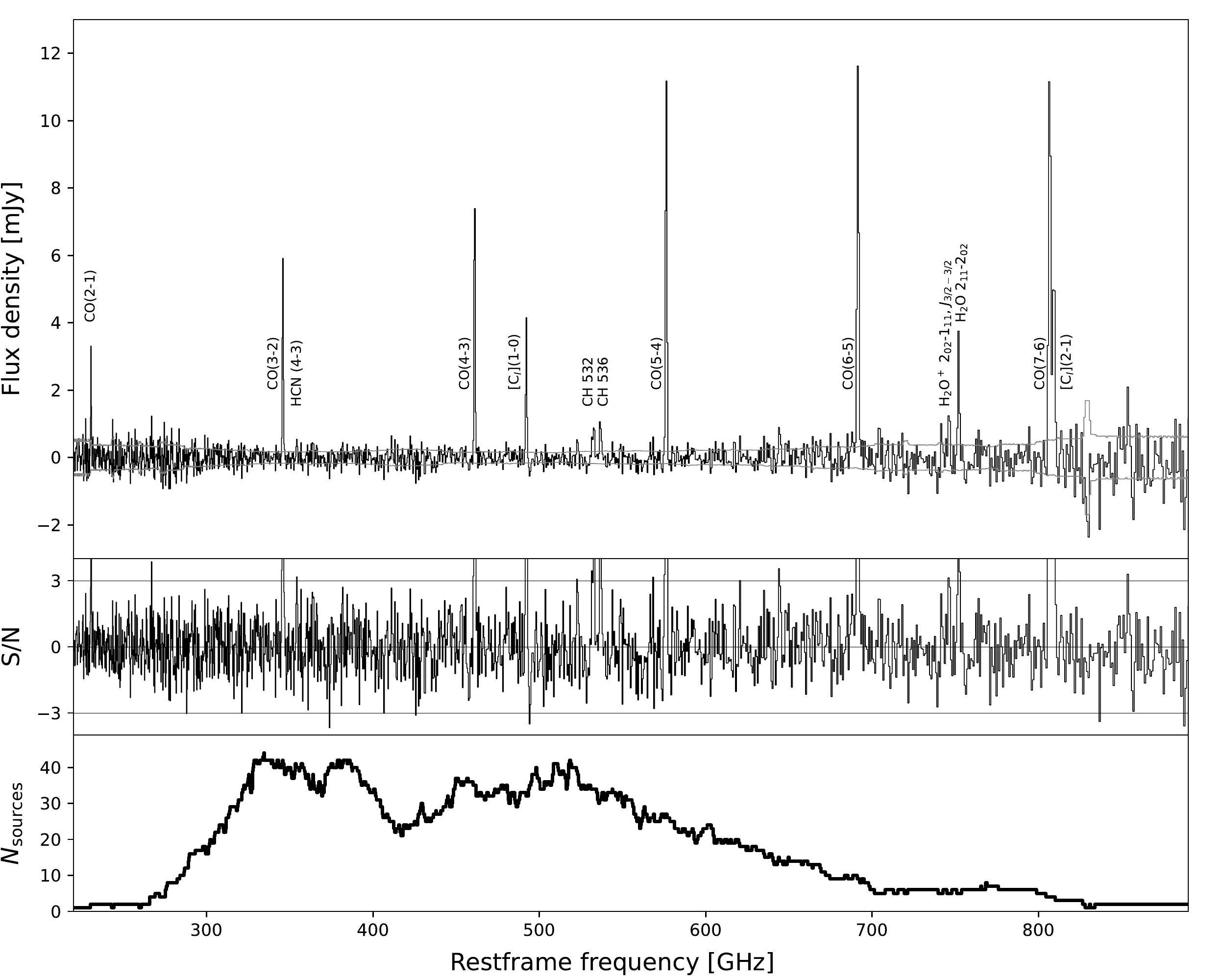}
    \caption{{\it Top panel:} Rest-frame composite spectrum between $220$ and $890~\mathrm{GHz}$ in $300~\mathrm{km~s}^{-1}$ bins, combining the data of the 71 galaxies in our sample. This spectrum is scaled such that it is representative of a {\it typical} $3 \times 10^{13}~\mathrm{L_\odot}$ BEARS galaxy at $z = 2.5$. We show the $\pm1\sigma$ standard deviation with a {\it thin grey} line. In addition to the CO, [C\,{\sc i}] and water lines, we find indications for CH emission in both the 532 and 536 transition, as well as H$_2$O$^+$ emission. Zoom-ins on these lines are shown in Fig.~\ref{fig:linesZoomIn}.
    {\it Middle panel:} Signal-to-noise ratios for lines across the entire composite spectrum. The {\it horizontal lines} show the plus- and minus $3\sigma$ confidence ranges.
    {\it Bottom panel:} Number of sources that contribute to the composite spectrum, as a function of restframe frequency. The low- and high-frequency tails of the spectrum rely on single-digit numbers of sources, with the central $300$ to $700~\mathrm{GHz}$ region combining between 10 and 45 galaxies in each $300~\mathrm{km~s}^{-1}$ bin.
    }
    \label{fig:compositeSpectrum}
\end{figure*}

\begin{figure*}
    \centering
    \includegraphics[width=0.95\textwidth]{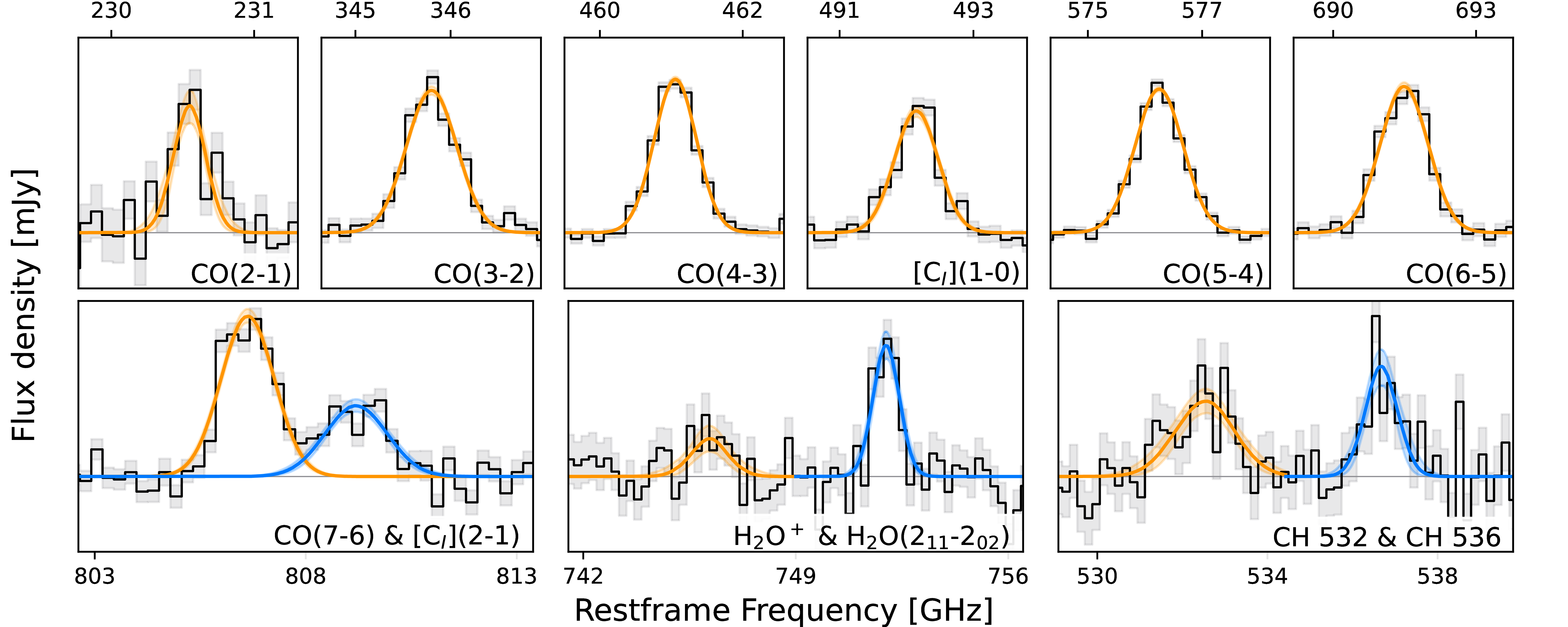}
    \caption{Stacking results for individual lines at the $70~\mathrm{km~s}^{-1}$ resolution. We fit the line profiles of each line with a Gaussian, and provide the fitted line properties in Table~\ref{tab:fittedSLlineproperties}. No obvious extended emission is seen in the combined profiles, similar to previous works that focused on stacking of DSFGs (e.g., \citealt{Birkin2021}).}
    \label{fig:linesZoomIn}
\end{figure*}

We show the composite spectrum in Figure~\ref{fig:compositeSpectrum}. The top panel shows the spectrum in bins of $300~\mathrm{km~s}^{-1}$ from $220$ to $890~\mathrm{GHz}$. The middle panel shows the signal-to-noise ratio of the stacked spectrum with the same binning. The bottom panel shows the number of sources contributing to the stacked spectrum as a function of frequency. Figure~\ref{fig:linesZoomIn} shows the zoomed-in spectra of the detected lines. The Gaussian fit parameters are shown in Table~\ref{tab:fittedSLlineproperties}.

\renewcommand{\arraystretch}{1.3}
\begin{table}
    \caption{Line properties fitted to the $70~\mathrm{km~s}^{-1}$ stacked spectrum.}
    \label{tab:fittedSLlineproperties}
    \centering
    \begin{tabular}{cccc} \hline
Line & $S\delta V$ & $\delta V$ & Luminosity \\ 
 & [$\mathrm{Jy~km~s}^{-1}$] & [$\mathrm{km~s}^{-1}$]  & [10$^7$ L$_{\odot}$] \\ \hline 
CO~($J=2$--1) &   $1.71^{+0.22}_{-0.22}$ &       $353^{+73}_{-63}$  & $5.12^{0.66}_{-0.66}$  \\
CO~($J=3$--2) &   $3.71^{+0.11}_{-0.11}$ &      $542^{+19}_{-18}$  & $16.65^{0.49}_{-0.49}$  \\
CO~($J=4$--3) &   $4.52^{+0.09}_{-0.09}$ &      $453^{+10}_{-10}$  & $27.05^{0.54}_{-0.54}$  \\
CO~($J=5$--4) &  $7.07^{+0.11}_{-0.10}$ &       $517^{+9}_{-9}$  & $52.89^{0.82}_{-0.75}$  \\
CO~($J=6$--5) &  $7.15^{+0.18}_{-0.18}$ &       $511^{+16}_{-15}$  & $64.18^{1.62}_{-1.62}$  \\
CO~($J=7$--6) &  $7.59^{+0.31}_{-0.31}$ &       $558^{+28}_{-26}$  & $79.5^{3.25}_{-3.25}$  \\
\textsc{[C\,i]}~($^3P_1$--$^3P_0$)  & $2.17^{+0.10}_{-0.09}$ &     $469^{+26}_{-25}$  & $13.86^{0.64}_{-0.57}$  \\
\textsc{[C\,i]}~($^3P_2$--$^3P_1$)  &  $3.90^{+0.32}_{-0.34}$ &    $645^{+67}_{-60}$  & $40.97^{3.36}_{-3.57}$  \\
CH~532 & $0.91^{+0.14}_{-0.14}$   & $928^{+209}_{-171}$ & $6.29^{0.97}_{-0.97}$  \\
CH~536 & $0.72^{+0.13}_{-0.12}$ &  $522^{+147}_{-119}$ & $5.02^{0.91}_{-0.84}$  \\
H$_2$O$^+$~746 & $0.81^{+0.51}_{-0.30}$ & $582^{+2764}_{-256}$ & $7.85^{4.94}_{-2.91}$  \\
H$_2$O~(2$_{11}$--2$_{02}$) & $1.83^{+0.21}_{-0.21}$ & $408^{+54}_{-48}$ & $17.86^{2.05}_{-2.05}$  \\
\hline
    \end{tabular}
\end{table}
\renewcommand{\arraystretch}{1.0}


\label{sec:stackedResults}
Similar to the work of \cite{spilker2014}, we use per-line bins to extract any potential line emission that is missed by the stacked spectrum. The composite spectrum could miss any emission by {\it smearing} flux across more than one spectral bin, especially since the bins are not necessarily centred on the positions of the line emission. Instead, we explore the optimum emission line by comparing the SNR of the highest-SNR line, CO~(5--4), across every velocity spacing between $200$ and $1000~\mathrm{km~s}^{-1}$ at $50~\mathrm{km~s}^{-1}$ increments. The highest SNR of the velocity-integrated flux is found at a bin-size of $600~\mathrm{km~s}^{-1}$. In a final attempt to optimally extract spectral lines, we stack all covered transitions of each line. We stack in $L'$ to correctly preserve the luminosity comparison of the ratios in $L'$ units (see equation~\ref{eq:commonRedshift}). The result is shown in Table~\ref{tab:stackOfLineTypes}, and zoom-ins on individual lines are shown in Figure~\ref{fig:linesZoomIn}.

Table~\ref{tab:faintLines} shows the resulting $L'$ estimates for 117 spectral lines within the $220$ to $890~\mathrm{GHz}$ composite spectrum. In order to optimally detect single species, we combine the bins of all transitions in \,$^{13}$CO, C$^{18}$O, HCN, HNC, HCO$^+$ and CH. The emission of several lines are potentially overlapping, particularly affecting the observed emission at the frequencies of CH~532 and HCN~(6--5), as well as CH~536 and HOC$^+$~(6--5). In this case, we follow the results from local ULIRGs by \cite{Rangwala2014}, and attribute the majority of the signal to the CH lines. 
All lines reported in the stacked spectrum, shown in Fig.~\ref{fig:compositeSpectrum}, are also seen in the bin-optimised extraction. 

The CO line ratios from the composite spectrum agree with the other CO SLEDs \citep[e.g.,][]{Harrington2021} and the individual galaxies, as seen in Figures~\ref{fig:togetherSLED} and \ref{fig:allCOscalingRelations}. The combined line emission of all CO lines adds to over $100 \sigma$. This allows us to look for large velocity tails in the emission of the spectral lines. Feeding and feedback are important components to the evolutionary track of high-redshift galaxies \citep{Peroux2020,berta2021}, and are sometimes revealed through wide velocity line profiles \citep{Ginolfi2020}. At high-redshift, outflow signatures are typically seen in \textsc{[C\,ii]} \citep{Fujimoto2019,Fujimoto2020,Ginolfi2020,Herrera-Camus2021,Izumi2021} and might not be visible in (higher-$J$ transitions of) CO lines, although \cite{Cicone2021} found an extended CO halo at $z \sim 2$ at velocities beyond $1000~\mathrm{km~s^{-1}}$. However, for our sample, both single-Gaussian fitting (Table~\ref{tab:fittedSLlineproperties}) and visual inspection of the stacked lines (Figure~\ref{fig:linesZoomIn}) do not show any large velocity components above $3\sigma$. Our findings are in line with the recent stacking work by \cite{Birkin2021}, which also fails to reveal any high-velocity tails, as well as in \cite{Meyer2022QSO} who emphasise the need for accurate continuum subtraction when comparing high-velocity tails.

$^{12}$CO lines can reach optically-thick column densities, but the isotopologues of CO often stay optically-thin. As such, the line ratios can be a useful probe of the optical depth. No individual CO isotopologues have been detected. Similarly, the combined stack of both $^{13}$CO and C$^{18}$O shows no significant emission. The resulting line ratio lower limits for $L'_{\rm ^{12}CO}/L'_{\rm ^{13}CO}$ and $L'_{\rm ^{12}CO}/L'_{\rm C^{18}O}$ are $> 35$.
The $L'_{\rm ^{12}CO}/L'_{\rm ^{13}CO}$ line ratio lower limit is around 2 times higher than observed for the SPT galaxies \citep{spilker2014}, although the relative errors are substantial. The $L'_{\rm ^{12}CO}/L'_{\rm C^{18}O}$ line ratio also remains undetected in the SPT survey, in line with our results. The transitions of the CO isotopologues are often observed together with the detected $^{12}$C$^{16}$O line transitions, which increases the robustness of our upper limits, because we are comparing like-for-like transitions for each galaxy in our isotopologue ratio estimates.

Our isotopologue results are within the typical range of other galaxies, although local molecular clouds typically have lower ratios. The Milky Way's molecular clouds have $L'_{\rm ^{12}CO}/L'_{\rm ^{13}CO}$ ratios between 5 and 10 \citep{Buckle2010,Buckle2012}, although the chemical abundances of the carbon element ($^{12}$C to $^{13}$C) evolves from 25 near the Galactic centre up to 100 in the solar vicinity \citep{Wilson1994,Wang2009}. \cite{Cao2017} and \cite{Cormier2018} report on spiral galaxies, with average ratios between 8 and 20. The local ULIRG Arp~220 has different isotopologue line ratios depending on the transition, ranging from 40 down to 8 \citep{Greve2009} for $^{13}$CO~(1--0) to (3--2), and a $L'_{\rm ^{12}CO}/L'_{\rm C^{18}O}$ ratio of 40 for C$^{18}$O (1--0). Similarly, \cite{Sliwa2017} find 60 to 200 for the $L'_{\rm ^{12}CO}/L'_{\rm ^{13}CO}$ and $L'_{\rm ^{12}CO}/L'_{\rm C^{18}O}$ ratios for the nearby ULIRG IRAS $13120-5453$. 

Higher-redshift galaxies also show relatively diverse isotopologue line ratios, with \cite{Mendez2020} finding $16$ and $40$ for $L'_{\rm ^{12}CO}/L'_{\rm ^{13}CO}$ and $L'_{\rm ^{12}CO}/L'_{\rm C^{18}O}$, respectively, for a sample of star-forming galaxies at $z \approx 0.02$--$0.2$. The DSFG Cosmic Eyelash (SMM~J2135$-$0102) has an $L'_{\rm ^{12}CO}/L'_{\rm ^{13}CO}$ ratio in excess of 60, with similarly-luminous $L'_{\rm C^{18}O}$. The Cloverleaf quasar \citep{Henkel2010} shows a high $^{13}$CO flux, with an associated ratio of 40. Individual observations of two SPT sources \citep{Bethermin2018} find a line ratio of around 26 for $L'_{\rm ^{12}CO}/L'_{\rm ^{13}CO}$, and \cite{Zhang2018} report $L'_{\rm ^{12}CO}/L'_{\rm ^{13}CO}$ ratios of 19--23 and $L'_{\rm ^{12}CO}/L'_{\rm C^{18}O}$ ratios between 25 and 33.
Our BEARS targets appear to agree with the more actively star-forming or AGN-dominated systems in the local and high-redshift Universe, which are suggestive of optically-thick emission, although the spatial variation of the isotopologue emission can cause filling factor effects \citep[e.g.,][]{Aalto1995}.

The relative ratios of the isotopologues to one another can further reveal the star-forming conditions of the system \citep{Davis2014,Jimenez2017}. The nucleosynthesis of the carbon and oxygen isotopes are both produced in the CNO cycle \citep{Maiolino2019}. The carbon isotope is formed through intermediate stars, on typical timescales of $>1~\mathrm{Gyr}$, while the oxygen isotope is produced through more massive stars \citep{Henkel1993,Wilson1994}. Both at low and high-redshift, low ratios of $L'_{\rm ^{13}CO}/L'_{\rm C^{18}O}$ can be interpreted as an effect of a variable IMF. For example, \cite{Sliwa2017} report ratios below 1 for $L'_{\rm ^{13}CO}/L'_{\rm C^{18}O}$ for the local ULIRG IRAS $13120-5453$, and \cite{Zhang2018} find that only a top-heavy IMF can produce the observed low ratios of $L'_{\rm ^{13}CO}/L'_{\rm C^{18}O}$. Our stacked observations are unable to detect this line ratio, although we are likely close to the detection limit of (one of the two) isotopologues given the existing detections in similar sources. These stacking observations argue towards the need for detailed and individual studies of isotopologues at high redshift, particularly of the brightest sources, instead of stacking studies across multiple sources.

The observed \textsc{[C\,i]}~($^3P_1$--$^3P_0$) and \textsc{[C\,i]}~($^3P_2$--$^3P_1$) line luminosities correlate with infrared luminosities for our sources, as well as local and other high-redshift galaxies over five orders of magnitude (Figure~\ref{fig:allCOscalingRelations}). The detection of neutral carbon and CO lines furthermore enables a comparison to PDR models and the line ratios of our stacked spectrum provide similar FUV radiation field and gas density values as the average sample, as seen in Figure~\ref{fig:PDRplot}.

These stacking attempts reveal HCN~(4--3), and tentatively show a small feature near the CH~532 associated with HCN~(6--5) in Figure~\ref{fig:linesZoomIn}, also noticeable in the larger velocity width of the CH~532 in Table~\ref{tab:fittedSLlineproperties} ($\delta V\approx 928~\mathrm{km~s^{-1}}$). These transitions suggest the presence of dense gas, since their critical density is about 100 to 1000 larger than those of CO lines ($10^{9 - 10}~\mathrm{cm}^{-3}$; \citealt{Jimenez-Donaire2019,Lizee2022}), and thus could be associated with dense star-forming regions \citep{Goldsmith2017} or with AGN \citep{Aalto2012,Lindberg2016,Cicone2020, Falstad2021}.
HCN is incidentally detected in the brightest high-redshift galaxies \citep{Riechers2006,riechers2010,Oteo2017, Canameras2020}. \referee{In contrast, \cite{Rybak2022} reported one detection of HCN~(1--0) emission line as a result of a deep survey with \textit{Karl G. Jansky} Very Large Array (VLA) towards six strongly lensed DSFGs. They suggest that in fact most DSFGs have low-dense gas fraction.} When stacking all the HCN lines across the entire sample, 140 line transitions are stacked, resulting in a $2.7\sigma$ tentative feature of $1.4 \times 10^9~\mathrm{K~km~s^{-1}~pc}^2$. \referee{We obtain the HCN/CO line luminosity ratio of $L^\prime_\mathrm{HCN}/L^\prime_\mathrm{CO}=0.022\pm0.008$ and an upper limit on the HCO$^+$/CO line luminosity ratio of $L^\prime_\mathrm{HCO^+}/L^\prime_\mathrm{CO} < 0.024$ based on our stacked spectrum. The deep VLA survey from \cite{Rybak2022} reports an upper limit in line luminosity ratio of $L^\prime_\mathrm{HCN}/L^\prime_\mathrm{CO} < 0.045$ and $L^\prime_\mathrm{HCO^+}/L^\prime_\mathrm{CO} < 0.043$, using solely the ground transitions. Since our results use a stack across multiple higher-order transitions, a direct comparison between these results is more difficult, however our results also suggest a dearth of dense gas in the BEARS sample.}
No other cyanide molecules nor the radical are detected. The individual and stacked lines agree with the observed line luminosities from \citet{spilker2014}, and suggest that dense star-forming regions are present across most DSFGs.

Five sources are observed at the rest-frame frequency of the H$_2$O~(2$_{11}$--2$_{02}$) line. The average line-to-total-infrared luminosity ratio is in line with the scaling relations of \cite{Jarugula2021} and \cite{Yang2016}, as well as the one fitted to our data. The scaling relation found in Section~\ref{sec:WaterSection} is instead due to the lower water line luminosity seen in infrared-fainter galaxies. 
The detection of H$_2$O$^+$ 746 allows us to make a rough estimation of the cosmic ray ionization rate. We find a luminosity ratio of H$_2$O$^+$ / H$_2$O $ \approx 0.4 \pm 0.2$, which is in agreement with \cite{Yang2016} (H$_2$O$^+$ / H$_2$O $ \approx 0.3 \pm 0.1$). The predicted ionization rate \citep{Meijerink2011} is around 10$^{-14} - 10^{-13}$~s$^{-1}$. 

The bright emission from the CH doublet at $532$ and $536~\mathrm{GHz}$ indicates the existence of high-density gas in X-ray-dominated regions (XDRs) associated with bright stars or AGN \citep{Meijerink2007,Rangwala2014} or strong irradiation through cosmic rays \citep{Benz2016}. Currently, the two scenarios are not easily distinguished \citep{Wolfire2022}, although enhanced excitation of high-$J$ CO lines through (resolved) observations could favour an XDR origin \citep{Vallini2019}, taking into account the effect of mechanical shock excitation of the high-$J$ CO components \citep{Meijerink2013, Falgarone2017}. Regardless of the origin of the CH~532 and 536 line emission, strong radiation sources and/or cosmic rays are necessary to explain the nature of these distant DSFGs.

\begin{table}
    \centering
    \caption{Stacks of line types.}
    \label{tab:stackOfLineTypes}
    \begin{tabular}{rccc} \hline
Line & N$_{\rm observations}$ & Line luminosity & SNR \\
&  &   [$10^9~\mathrm{K~km~s^{-1}~pc}^2$] \\ \hline
CO & 120 & $\boldsymbol{ 62.5 \pm 3.6}$ & ($103.1\sigma$) \\
$^{13}$CO & 122 & 0.5 $\pm$ 0.6 & (0.9$\sigma$) \\
C$^{18}$O & 126 & 0.9 $\pm$ 0.6 & (1.5$\sigma$) \\
\textsc{[C\,i]} & 36 & $\boldsymbol{ 24.8 \pm 1.7}$ & (25.6$\sigma$) \\
HCN & 140 & 1.4 $\pm$ 0.5 & (2.7$\sigma$) \\
HNC & 135 & 0.9 $\pm$ 0.6 & (1.6$\sigma$) \\
HCO$^+$ & 137 & 0.7 $\pm$ 0.5 & (1.3$\sigma$) \\
HOC$^+$ & 132 & 1.5 $\pm$ 0.6 & (2.6$\sigma$) \\
CH & 65 & $\boldsymbol{ 7.3 \pm 0.9}$ & (8.7$\sigma$) \\
H$_2$O$^+$ & 98 & 0.3 $\pm$ 0.4 & (0.7$\sigma$) \\
SiO & 273 & $-0.1$ $\pm$ 0.4 & ($-0.2\sigma$) \\
CS & 218 & 0.5 $\pm$ 0.5 & (1.0$\sigma$) \\
NH$_3$ & 152 & $-0.0$ $\pm$ 0.5 & ($-0.1\sigma$) \\
CCH & 133 & $-0.5$ $\pm$ 0.6 & ($-0.8\sigma$) \\
H21--28$\alpha$  & 220 & 0.1 $\pm$ 0.5 & (0.3$\sigma$) \\
\hline
\end{tabular}
\end{table}

\section{Conclusions}
\label{sec:ch7}
We have investigated the physical properties of the BEARS sample, which consists of 71 line-detected galaxies, based on 156 spectral line flux estimates including upper limits (the detections of 117 CO, 27 \textsc{[C\,i]}, and two H$_2$O lines, and the upper limits of a single CO, six \textsc{[C\,i]}, and three H$_2$O lines). We report the following conclusions:
\begin{itemize}
\renewcommand\labelitemi{\tiny \textbf{$\blacksquare{}$}}
    \item The average gas properties of our sample are similar to other DSFG samples. Especially, the CO SLEDs of most sources as well as the stacked CO SLED follows the mean CO SLED for \textit{Planck}-selected lensed DSFGs from \citet{Harrington2021}. 
    \item Our galaxies follow the relation between line luminosity and infrared luminosity over five orders of magnitude when compared to reference samples at low and high redshift. The Schmidt-Kennicutt relation, however, shows that our sources and other DSFGs are not located in same star formation phase as local and low-redshift gas-rich, normal star-forming systems. In addition, our sources seem to have slightly longer depletion times than other DSFGs from \citet{C.C.Chen2017}, although this effect could be explained by a difference in the size estimation of the star-forming regions.
    \item Most of our samples have dynamical masses between $10^{11}~\mathrm{M}_\odot$ and $10^{12}~\mathrm{M}_\odot$ and are found to be lower than molecular gas mass estimates. This means that we likely underestimate the dynamical mass within the BEARS systems. This could be caused by differential lensing or because these systems are dynamically-complex. 
    \item The dust-to-gas ratios of our sources do not vary with redshift, and the ratios are similar to that of the Milky Way. The low scatter and lack of trend with redshift suggest we are witnessing a single star-forming phase, and the high dust-to-gas ratio suggests that this phase occurred relatively recently. 
    \item The PDR conditions of the BEARS sources are similar to those of DSFGs and local U/LIRGs, with denser and more intense radiation environments than low-$z$ MS galaxies in line with previous studies \citep[e.g.,][]{Valentino2020CI}. We investigate these PDR conditions with \texttt{PDRT} and find that most of our galaxies are located within $10^4\leq n_\mathrm{H}\leq10^5~[\mathrm{cm}^{-3}]$ and $10^3\leq G_0\leq10^4~[\mathrm{Habing}]$.
    \item Our linear regression fitting of the H$_2$O~($2_{11}$--$2_{02}$) to infrared luminosity relation for low- and high-redshift samples favours a super-linear relation with $3\sigma$ significance, consistent with previous observations from \cite{Yang2016}. 
    \item We find four candidates (HerBS-69B, -120A, -120B, and -159B) that show ``super-thermalized'' CO line ratios. Although their ratios are consistent with thermalized one at $1$--$2.3\sigma$ due to the large uncertainty inherent in luminosity ratio estimates, especially, HerBS-69B and -120B stand out from the other 44 galaxies, suggesting a rare phase in the evolution of DSFGs. We note that we require follow-up observations to confirm their super-thermalized nature. 
    \item The deep stacked spectrum ($220$--$890~\mathrm{GHz}$) reveals an additional H$_2$O$^+$ line, as well as the dense gas tracer HCN~(4--3), and two tracers of XDR and/or cosmic-ray-dominated environments through CH~532 and 536. The total stack provides deep upper limits on the $^{13}$CO and C$^{18}$O isotopologues, in line with previous observations and line stacking experiments.
\end{itemize}

In the near future, we aim to expand upon the current studies with high-resolution imaging to reveal the morphological and kinematic properties of these galaxies. We place a particularly focus on the CUBS targets to confirm their super-thermalized line profiles, using tracers that reveal merging or AGN activity, as well as the highest-resolution imaging possible using the near-infrared James Webb Space Telescope. Further continuation of the study of the ISM of these galaxies is also necessary, notably towards water and carbon emission lines. Finally, an important goal is the redshift completion of the sample targeted in \cite{Urquhart2022}.

\section*{Acknowledgements}
\referee{We would like to thank the anonymous referee for their insightful comments and suggested additions.}
This work was supported by NAOJ ALMA Scientific Research Grant Nos. 2018-09B and JSPS KAKENHI No.~17H06130, 22H04939, and 22J21948.
SS was partly supported by the ESCAPE project; ESCAPE -- The European Science Cluster of Astronomy and Particle Physics ESFRI Research Infrastructures has received funding from the European Union’s Horizon 2020 research and innovation programme under Grant Agreement No.~824064. SS also thanks the Science and Technology Facilities Council for financial support under grant ST/P000584/1. SU would like to thank the Open University School of Physical Sciences for supporting this work. HU acknowledges support from JSPS KAKENHI Grant Number 20H01953. CY acknowledges support from ERC Advanced Grant 789410.

\section*{Data Availability}

The \textit{Herschel} SPIRE data can be downloaded from {\tt https://
www.h-atlas.org}, while the reduced, calibrated and science-ready
ALMA data are available from the ALMA Science Archive
at {\tt https://almascience.eso.org/asax/}~.
 



\bibliographystyle{mnras}
\bibliography{reference} 




\appendix

\section{Additional velocity-integrated line fluxes and upper limits}
We extracted three extra [C\,{\sc i}]~($^3P_1$--$^3P_0$), three extra [C\,{\sc i}]~($^3P_2$--$^3P_1$), and one extra H$_{\rm 2}$O~(2$_{\rm 11}$--2$_{\rm 02}$) lines on top of the lines reported in \cite{Urquhart2022}. Together with these, we list ten upper limits on line fluxes in Table~\ref{tab:upperlimitsAdditional}.

\begin{table}
    \caption{Additional velocity-integrated line fluxes and $3\sigma$ upper limits}
    \label{tab:upperlimitsAdditional}
    \centering
    \begin{tabular}{lcc}
        Source ID & Line & $S\delta V~[\mathrm{Jy~km~s^{-1}}]$ \\\hline
        HerBS-28 & H$_2$O~($2_{11}$--$2_{02}$) & $1.4\pm0.4$\\
        HerBS-41 & [C\,{\sc i}]~($^3P_1$--$^3P_0$) & $0.73^*$\\
           & [C\,{\sc i}]~($^3P_2$--$^3P_1$) & $1.3\pm0.4$\\
           & H$_2$O~($2_{11}$--$2_{02}$)  & $1.6^*$\\
        HerBS-77 & [C\,{\sc i}]~($^3P_1$--$^3P_0$) & $1.5^*$\\
        HerBS-80A & [C\,{\sc i}]~($^3P_1$--$^3P_0$) & $1.1^*$\\
        HerBS-90 & [C\,{\sc i}]~($^3P_2$--$^3P_1$) & $4.0\pm1.1$\\
        HerBS-117 & [C\,{\sc i}]~($^3P_2$--$^3P_1$) & $1.7\pm0.4$\\
        HerBS-121 & [C\,{\sc i}]~($^3P_1$--$^3P_0$) & $0.81^*$\\
            & H$_2$O~($2_{11}$--$2_{02}$) & $1.7^*$\\
        HerBS-131B & CO~($J=$3--2) & $1.1^*$\\
        HerBS-159A & [C\,{\sc i}]~($^3P_1$--$^3P_0$) & $1.0\pm0.2$\\
        HerBS-159B & [C\,{\sc i}]~($^3P_1$--$^3P_0$) & $0.71^*$\\
        HerBS-160 & [C\,{\sc i}]~($^3P_1$--$^3P_0$) & $1.2\pm0.3$\\
            & H$_2$O~($2_{11}$--$2_{02}$) & $1.4^*$\\
        HerBS-200 & [C\,{\sc i}]~($^3P_1$--$^3P_0$) & $2.2\pm0.6$\\
        HerBS-208B & [C\,{\sc i}]~($^3P_1$--$^3P_0$) & $1.2^*$\\ \hline
    \end{tabular}
    \raggedright \justify \vspace{-0.2cm}
    \textbf{Notes:} $^*$ $3\sigma$ upper limits.
\end{table}

\section{Scaling relations}
Table~\ref{tab:scalingrelations} shows the linear regression best-fit results for the lines shown in Figure~\ref{fig:allCOscalingRelations} according to the following prescription, 
\begin{equation}
    \log_{10} \left(L^\prime_{\rm line}~\left[\mathrm{K~km~s^{-1}~pc^2}\right]\right)=a \log_{10} \left(L_{\rm IR}~\left[\mathrm{L_\odot}\right]\right)+b.
\end{equation}

\begin{table}
\caption{Scaling relation fits to the lines shown in Fig.~\ref{fig:allCOscalingRelations}}
    \centering
    \begin{tabular}{lcc}
      Line   & $a$ & $b$ \\ \hline
        CO~($J=2$--1) & $0.79\pm0.04$ & $0.6\pm0.5$ \\
        CO~($J=3$--2) & $0.91\pm0.02$ & $-1.1\pm0.2$ \\
        CO~($J=4$--3) & $0.97\pm0.01$ & $-1.9\pm0.1$ \\
        CO~($J=5$--4) & $0.996\pm0.009$ & $-2.5\pm0.1$ \\
        CO~($J=6$--5) & $1.02\pm0.01$ & $-3.0\pm0.2$ \\
        CO~($J=7$--6) & $1.05\pm0.01$ & $-3.7\pm0.1$ \\
        \textsc{[C\,i]}~($^3P_1$--$^3P_0$) & $1.02\pm0.02$ & $-2.9\pm0.2$ \\
        \textsc{[C\,i]}~($^3P_2$--$^3P_1$) & $1.02\pm0.01$ & $-3.3\pm0.2$ \\ \hline
    \end{tabular}
    \label{tab:scalingrelations}
\end{table}

\section{Size estimation}
Table~\ref{tab:sizeestimation} shows the deconvolved sizes obtained from \texttt{CASA} task \texttt{IMFIT} to Band~4 continuum. \texttt{IMFIT} returns major axis ($\theta_\mathrm{maj}$) and minor axis ($\theta_\mathrm{min}$) and we use the radius as $r=\sqrt{\theta_\mathrm{maj}\theta_\mathrm{min}}$, which turn the surface area into $\pi r^2$. For the 21~sources that are not listed in Table~\ref{tab:sizeestimation}, we set the size to the minimum robust results found, equal to $0.14~\mathrm{arcseconds}$.

\begin{table}
\caption{The size estimation from \texttt{IMFIT} result}
    \centering
    \begin{tabular}{lccc}
      Source ID & Size [kpc] & Source ID & Size [kpc]\\ \hline
      HerBS-11 & $2.2\pm0.2$ & HerBS-93 & $3.8\pm0.5$ \\
      HerBS-14 & $4.17\pm0.06$ & HerBS-102 & $2.2\pm0.7$\\
      HerBS-18 & $3.7\pm0.2$ & HerBS-106 & $3.4\pm1.2$\\
      HerBS-21 & $3.9\pm0.2$ & HerBS-107 & $4.7\pm0.1$\\
      HerBS-22 & $3.8\pm0.2$ & HerBS-111 & $4.4\pm0.3$\\
      HerBS-24 & $1.1\pm0.8$ & HerBS-117 & $3.1\pm0.1$\\
      HerBS-25 & $6.9\pm0.9$ & HerBS-120B & $3.8\pm0.4$\\
      HerBS-27 & $2.8\pm0.5$ & HerBS-121 & $5.7\pm0.2$\\
      HerBS-28 & $5.0\pm0.1$ & HerBS-123 & $9.0\pm1.1$\\
      HerBS-37 & $5.6\pm0.3$ & HerBS-131B & $2.6\pm0.7$\\
      HerBS-39 & $6.8\pm0.4$ & HerBS-132 & $3.2\pm0.2$\\
      HerBS-41 & $2.0\pm0.6$ & HerBS-135A & $3.0\pm0.8$\\
      HerBS-42 & $2.3\pm0.4$ & HerBS-145 & $4.5\pm0.5$\\
      HerBS-47 & $5.2\pm0.2$ & HerBS-146B & $5.5\pm0.7$\\
      HerBS-49A & $6.8\pm0.2$ & HerBS-155 & $5.4\pm0.6$\\
      HerBS-55 & $5.1\pm0.2$ & HerBS-159A & $8.8\pm1.6$\\
      HerBS-57 & $3.8\pm0.4$ & HerBS-160 & $3.7\pm0.1$\\
      HerBS-60 & $5.0\pm0.3$ & HerBS-168 & $5.2\pm0.4$\\
      HerBS-63 & $2.5\pm0.7$ & HerBS-178A & $4.3\pm0.2$\\
      HerBS-69A & $2.2\pm1.1$ & HerBS-178B & $5.4\pm0.8$\\
      HerBS-73 & $2.0\pm0.5$ & HerBS-182 & $1.2\pm0.4$\\
      HerBS-77 & $3.1\pm3.2$ & HerBS-189 & $6.4\pm0.9$\\
      HerBS-80B & $4.7\pm0.3$ & HerBS-207 & $3.5\pm0.7$\\
      HerBS-86 & $3.0\pm0.5$ & HerBS-208A & $2.7\pm0.9$\\
      HerBS-90 & $4.6\pm1.0$ & HerBS-209 & $6.7\pm2.5$ \\ \hline
    \end{tabular}
    \label{tab:sizeestimation}
\end{table}



\section{Derived properties of the BEARS sample}
We list the galaxy properties derived througout this paper in Table~\ref{tab:extensiveTableOfAllProperties}. 

\begin{table}
    \centering
    \caption{Derived properties of the BEARS sample}
    \label{tab:extensiveTableOfAllProperties}
    \scalebox{0.86}{
    \begin{tabular}{lccccccccc}\hline
         ID & $\mu L_\mathrm{IR}$ & $\mu M_\mathrm{dust}$ & $\mu L^\prime_\mathrm{CO(1-0)}$ & $\mu M_\mathrm{H_2(CI)}$ & $\mu M_\mathrm{H_2(CO)}$ & $M_\mathrm{dyn}$ & $\Sigma_\mathrm{M_{H_2}(CI)}$ & $\Sigma_\mathrm{M_{H_2}(CO)}$ & $\Sigma_\mathrm{SFR}$ \\
        & $[10^{12}~\mathrm{L}_\odot]$ & $[10^{10}~\mathrm{M}_\odot]$ & $[10^{10}~\mathrm{K~km~s^{-1}~pc^2}]$ & $[10^{11}~\mathrm{M}_\odot]$ & $[10^{11}~\mathrm{M}_\odot]$ & $[10^{11}~\mathrm{M}_\odot]$ & $[10^4~\mathrm{M_\odot~pc^{-2}}]$ & $[10^4~\mathrm{M_\odot~pc^{-2}}]$ & $[10^2~\mathrm{M_\odot~yr^{-1}~kpc^{-2}}]$ \\\hline
        11&$66.4 $&$13.6 $&$58.9\pm10.9 $&$- $&$23.6\pm4.4 $&$1.5\pm0.3 $&$- $&$15.9 \pm 3.0 $&$6.6 \pm 1.7$\\
14&$97.8 $&$20.1 $&$52.9\pm14.5 $&$11.8\pm1.7 $&$21.2\pm5.8 $&$1.4\pm0.3 $&$2.2\pm0.3 $&$3.9 \pm 1.1 $&$2.6 \pm 0.7$\\
18&$58.0 $&$11.9 $&$40.7\pm11.5 $&$9.7\pm1.1 $&$16.3\pm4.6 $&$0.9\pm0.3 $&$2.4\pm0.3 $&$3.1 \pm 0.6 $&$2.1 \pm 0.5$\\
21&$45.8 $&$9.4 $&$76.1\pm43.5 $&$- $&$30.4\pm17.4 $&$4.9\pm2.2 $&$- $&$3.1 \pm 0.9 $&$1.4 \pm 0.4$\\
22&$50.8 $&$10.4 $&$61.3\pm25.4 $&$- $&$24.5\pm10.2 $&$7.5\pm3.7 $&$- $&$5.3 \pm 2.2 $&$1.6 \pm 0.4$\\
24&$56.7 $&$11.6 $&$20.9\pm4.0 $&$11.4\pm1.6 $&$8.3\pm1.6 $&$1.1 \pm 0.3 $&$27.7\pm3.9 $&$20.3 \pm 3.8 $&$20.2 \pm 5.1$\\
25&$59.8 $&$12.3 $&$78.2\pm33.1 $&$- $&$31.3\pm13.2 $&$1.3\pm0.5 $&$- $&$1.1 \pm 0.2 $&$0.6 \pm 0.1$\\
27&$98.6 $&$20.2 $&$79.7\pm32.7 $&$- $&$31.9\pm13.1 $&$2.7\pm1.0 $&$- $&$13.1 \pm 5.4 $&$6.0 \pm 1.5$\\
28&$71.8 $&$14.7 $&$85.2\pm51.5 $&$- $&$34.1\pm20.6 $&$7.5\pm2.7 $&$- $&$2.1 \pm 0.7 $&$1.4 \pm 0.3$\\
36&$77.5 $&$15.9 $&$23.6\pm6.9 $&$- $&$9.4\pm2.7 $&$1.2\pm0.3 $&$- $&$25.1 \pm 7.3 $&$30.3 \pm 7.6$\\
37&$29.7 $&$6.1 $&$34.7\pm16.7 $&$- $&$13.9\pm6.7 $&$4.8\pm2.1 $&$- $&$0.8 \pm 0.2 $&$0.4 \pm 0.1$\\
39&$46.7 $&$9.6 $&$33.7\pm9.7 $&$- $&$13.5\pm3.9 $&$9.3\pm5.9 $&$- $&$0.9 \pm 0.3 $&$0.5 \pm 0.1$\\
40&$22.0 $&$4.5 $&$16.2\pm4.5 $&$- $&$6.5\pm1.8 $&$2.6\pm0.6 $&$- $&$14.3 \pm 4.0 $&$7.1 \pm 1.8$\\
41&$47.1 $&$9.7 $&$18.9\pm6.0 $&$4.6^* $&$7.5\pm2.4 $&$2.6\pm0.7 $&$3.5^* $&$5.7 \pm 1.8 $&$5.3 \pm 1.3$\\
42&$27.8 $&$5.7 $&$38.9\pm23.9 $&$- $&$15.6\pm9.5 $&$4.0\pm0.9 $&$- $&$4.4 \pm 1.6 $&$2.4 \pm 0.6$\\
45&$13.2 $&$2.7 $&$10.8\pm2.7 $&$8.1\pm1.4 $&$4.3\pm1.1 $&$0.2\pm0.1 $&$19.1\pm3.2 $&$10.1 \pm 2.5 $&$4.5 \pm 1.1$\\
47&$25.0 $&$5.1 $&$21.6\pm8.4 $&$4.3\pm0.8 $&$8.6\pm3.4 $&$6.9\pm3.6 $&$0.5\pm0.1 $&$1.0 \pm 0.4 $&$0.4 \pm 0.1$\\
49A&$21.6 $&$4.4 $&$37.1\pm19.0 $&$- $&$14.8\pm7.6 $&$1.0\pm0.2 $&$- $&$0.6 \pm 0.2 $&$0.2 \pm 0.1$\\
49B&$8.8 $&$1.8 $&$8.4\pm3.7 $&$- $&$3.4\pm1.5 $&$1.1\pm0.5 $&$- $&$8.3 \pm 3.7 $&$3.2 \pm 0.8$\\
55&$19.3 $&$4.0 $&$16.3\pm3.6 $&$- $&$6.5\pm1.4 $&$1.8\pm0.6 $&$- $&$0.8 \pm 0.2 $&$0.3 \pm 0.1$\\
56C&$6.6 $&$1.4 $&$10.2\pm2.4 $&$- $&$4.1\pm0.9 $&$0.5\pm0.2 $&$- $&$9.8 \pm 2.3 $&$2.3 \pm 0.6$\\
57&$46.8 $&$9.6 $&$36.6\pm10.4 $&$- $&$14.7\pm4.2 $&$1.3\pm0.4 $&$- $&$3.2 \pm 0.9 $&$1.5 \pm 0.4$\\
60&$39.0 $&$8.0 $&$27.4\pm7.9 $&$- $&$11.0\pm3.2 $&$3.1\pm1.6 $&$- $&$1.4 \pm 0.4 $&$0.7 \pm 0.2$\\
63&$21.0 $&$4.3 $&$9.9\pm2.2 $&$5.4\pm0.8 $&$3.9\pm0.9 $&$1.7\pm0.5 $&$2.8\pm0.4 $&$2.0 \pm 0.5 $&$1.6 \pm 0.4$\\
68&$29.2 $&$6.0 $&$68.7\pm29.1 $&$- $&$27.5\pm11.6 $&$0.9\pm0.3 $&$- $&$36.4 \pm 7.7 $&$10.6 \pm 2.6$\\
69A&$15.2 $&$3.1 $&$19.5\pm4.0 $&$6.2\pm0.6 $&$7.8\pm1.6 $&$1.3\pm0.5 $&$4.2\pm0.4 $&$5.3 \pm 1.1 $&$1.5 \pm 0.4$\\
69B&$12.1 $&$2.5 $&$2.1\pm0.8 $&$1.4\pm0.4 $&$0.9\pm0.3 $&$0.08\pm0.01 $&$3.2\pm0.9 $&$1.9 \pm 0.7 $&$4.0 \pm 1.0$\\
73&$34.3 $&$7.0 $&$26.3\pm11.0 $&$- $&$10.5\pm4.4 $&$3.3\pm1.6 $&$- $&$8.7 \pm 3.6 $&$4.2 \pm 1.0$\\
77&$22.4 $&$4.6 $&$22.2\pm4.3 $&$3.6^* $&$8.9\pm1.7 $&$5.1\pm1.4 $&$1.2^* $&$3.0 \pm 0.6 $&$1.1 \pm 0.3$\\
80A&$6.3 $&$1.3 $&$14.5\pm3.5 $&$2.6^* $&$5.8\pm1.4 $&$2.1\pm0.2 $&$6.0^* $&$13.2 \pm 3.2 $&$2.1 \pm 0.5$\\
80B&$6.4 $&$1.3 $&$4.1\pm1.3 $&$- $&$1.6\pm0.5 $&$2.3\pm1.0 $&$- $&$0.2 \pm 0.1 $&$0.1 \pm 0.0$\\
81A&$11.4 $&$2.3 $&$27.7\pm8.6 $&$- $&$11.1\pm3.4 $&$2.1\pm1.4 $&$- $&$29.8 \pm 9.3 $&$4.5 \pm 1.1$\\
81B&$12.4 $&$2.5 $& $9.4\pm2.3 $&$- $&$3.7\pm0.9 $&$3.0\pm1.1 $&$- $&$9.0 \pm 2.2 $&$4.4 \pm 1.1$\\
86c&$29.5 $&$6.0 $&$28.7\pm5.6 $&$- $&$11.5\pm2.2 $&$3.2\pm1.0 $&$- $&$4.1 \pm 0.8 $&$1.5 \pm 0.4$\\
90&$41.3 $&$8.5 $&$26.1\pm8.1 $&$25.1\pm6.1 $&$10.4\pm3.2 $&$2.2\pm0.8 $&$3.8\pm0.9 $&$1.6 \pm 0.5 $&$0.9 \pm 0.2$\\
93&$27.6 $&$5.7 $&$18.8\pm4.6 $&$7.2\pm1.3 $&$7.5\pm1.8 $&$6.5\pm3.7 $&$1.6\pm0.3 $&$1.7 \pm 0.4 $&$0.9 \pm 0.2$\\
102&$21.1 $&$4.3 $&$21.4\pm6.7 $&$- $&$8.6\pm2.7 $&$1.1\pm0.6 $&$- $&$5.6 \pm 1.8 $&$2.0 \pm 0.5$\\
103&$23.6 $&$4.8 $&$30.9\pm12.7 $&$- $&$12.3\pm5.1 $&$1.6\pm0.8 $&$- $&$31.8 \pm 13.1 $&$8.9 \pm 2.2$\\
106&$29.7 $&$6.1 $&$18.8\pm4.0 $&$4.4\pm0.8 $&$7.5\pm1.6 $&$3.6\pm2.9 $&$1.2\pm0.2 $&$2.1 \pm 0.4 $&$1.2 \pm 0.3$\\
107&$18.9 $&$3.9 $&$16.8\pm3.6 $&$- $&$6.7\pm1.4 $&$0.8\pm0.3 $&$- $&$1.0 \pm 0.2 $&$0.4 \pm 0.1$\\
111&$22.7 $&$4.6 $&$29.6\pm6.0 $&$7.5\pm0.8 $&$11.9\pm2.4 $&$6.7\pm2.9 $&$1.2\pm0.1 $&$2.0 \pm 0.4 $&$0.6 \pm 0.1$\\
117&$39.8 $&$8.2 $&$25.9\pm11.0 $&$- $&$10.4\pm4.4 $&$2.5\pm0.8 $&$- $&$3.3 \pm 1.4 $&$1.9 \pm 0.5$\\
120A&$18.5 $&$3.8 $&$10.7\pm4.3 $&$- $&$4.3\pm1.7 $&$2.0\pm0.8 $&$- $&$11.4 \pm 4.6 $&$7.3 \pm 1.8$\\
120B&$18.2 $&$3.7 $&$6.7\pm3.1 $&$- $&$2.7\pm1.2 $&$8.9\pm5.1 $&$- $&$0.4 \pm 0.1 $&$0.6 \pm 0.1$\\
121&$27.0 $&$5.5 $&$32.8\pm10.1 $&$4.5^* $&$13.1\pm4.1 $&$3.3\pm1.0 $&$0.4^* $&$1.3 \pm 0.4 $&$0.4 \pm 0.1$\\
122A&$6.3 $&$1.3 $&$13.9\pm5.8 $&$- $&$5.5\pm2.3 $&$0.6\pm0.3 $&$- $&$14.1 \pm 5.9 $&$2.4 \pm 0.6$\\
123&$22.4 $&$4.6 $&$22.7\pm4.6 $&$10.3\pm1.3 $&$9.1\pm1.8 $&$3.5\pm1.3 $&$0.4\pm0.1 $&$0.4 \pm 0.1 $&$0.13 \pm 0.03$\\
131B&$15.7 $&$3.2 $&$5.9\pm1.8 $&$10.3\pm0.9 $&$2.4\pm0.7 $&$6.2\pm2.3 $&$4.8\pm0.4 $&$0.8 \pm 0.2 $&$1.1 \pm 0.3$\\
132&$16.7 $&$3.4 $&$19.3\pm4.1 $&$3.4\pm0.6 $&$7.7\pm1.7 $&$2.8\pm1.4 $&$1.1\pm0.2 $&$2.5 \pm 0.5 $&$0.8 \pm 0.2$\\
135A&$17.1 $&$3.5 $&$12.4\pm3.1 $&$2.9\pm0.5 $&$5.0\pm1.3 $&$1.8\pm0.6 $&$1.1\pm0.2 $&$1.8 \pm 0.5 $&$0.9 \pm 0.2$\\
138B&$23.6 $&$4.8 $&$3.7\pm0.7 $&$- $&$1.5\pm0.3 $&$0.07\pm0.02 $&$- $&$3.2 \pm 0.6 $&$7.5 \pm 1.9$\\
141&$16.0 $&$3.3 $&$20.8\pm5.1 $&$10.6\pm1.7 $&$8.3\pm2.0 $&$0.6\pm0.2 $&$23.6\pm3.7 $&$18.5 \pm 4.5 $&$5.2 \pm 1.3$\\
145&$13.8 $&$2.8 $&$19.9\pm4.5 $&$- $&$8.0\pm1.8 $&$8.2\pm2.0 $&$- $&$1.2 \pm 0.3 $&$0.3 \pm 0.1$\\
146B&$8.8 $&$1.8 $&$10.7\pm3.1 $&$- $&$4.3\pm1.3 $&$3.7\pm1.6 $&$- $&$0.5 \pm 0.1 $&$0.14 \pm 0.03$\\
155&$34.7 $&$7.1 $&$21.0\pm8.6 $&$- $&$8.4\pm3.5 $&$0.7\pm0.4 $&$- $&$0.9 \pm 0.4 $&$0.5 \pm 0.1$\\
159A&$14.0 $&$2.9 $&$19.1\pm3.7 $&$2.3\pm0.5 $&$7.6\pm1.5 $&$1.6\pm0.7 $&$0.10\pm0.02 $&$0.3 \pm 0.1 $&$0.08 \pm 0.02$\\
159B&$4.8 $&$1.0 $&$3.2\pm0.9 $&$1.7^* $&$1.3\pm0.3 $&$0.6\pm0.2 $&$3.8^* $&$2.9 \pm 0.8 $&$1.6 \pm 0.4$\\
160&$50.5 $&$10.4 $&$19.4\pm6.1 $&$7.2\pm1.8 $&$7.8\pm2.4 $&$2.9\pm0.7 $&$1.7\pm0.4 $&$1.8 \pm 0.6 $&$1.7 \pm 0.4$\\
163A&$6.5 $&$1.3 $&$4.3\pm1.6 $&$- $&$1.7\pm0.6 $&$0.7\pm0.3 $&$- $&$4.6 \pm 1.7 $&$2.6 \pm 0.6$\\
168&$49.2 $&$10.1 $&$17.1\pm3.6 $&$- $&$6.9\pm1.5 $&$11.1\pm3.0 $&$- $&$0.8 \pm 0.2 $&$0.9 \pm 0.2$\\
178A&$14.1 $&$2.9 $&$16.4\pm3.3 $&$- $&$6.5\pm1.3 $&$5.0\pm2.2 $&$- $&$1.1 \pm 0.2 $&$0.4 \pm 0.1$\\
178B&$9.9 $&$2.0 $&$14.2\pm3.0 $&$- $&$5.7\pm1.2 $&$8.2\pm2.7 $&$- $&$0.6 \pm 0.1 $&$0.16 \pm 0.04$\\
178C&$6.2 $&$1.3 $&$7.6\pm2.1 $&$- $&$3.0\pm0.8 $&$0.6\pm0.3 $&$- $&$7.4 \pm 2.1 $&$2.2 \pm 0.6$\\
182&$21.8 $&$4.5 $&$15.7\pm3.9 $&$7.2\pm1.2 $&$6.3\pm1.6 $&$3.8\pm1.1 $&$15.7\pm2.5 $&$13.7 \pm 3.4 $&$7.0 \pm 1.7$\\
184&$28.6 $&$5.9 $&$21.6\pm6.1 $&$6.0\pm0.9 $&$8.7\pm2.5 $&$0.5\pm0.2 $&$14.3\pm2.0 $&$20.6 \pm 5.8 $&$10.0 \pm 2.5$\\
189&$29.0 $&$6.0 $&$27.4\pm7.9 $&$- $&$10.9\pm3.2 $&$2.6\pm0.9 $&$- $&$0.8 \pm 0.2 $&$0.3 \pm 0.1$\\
200&$16.7 $&$3.4 $&$11.4\pm2.7 $&$4.8\pm1.3 $&$4.6\pm1.1 $&$0.8\pm0.4 $&$10.8\pm3.0 $&$10.2 \pm 2.5 $&$5.5 \pm 1.4$\\
207&$22.5 $&$4.6 $&$26.5\pm3.1 $&$- $&$10.6\pm1.2 $&$3.2\pm1.0 $&$- $&$2.7 \pm 0.3 $&$0.8 \pm 0.2$\\
208A&$15.7 $&$3.2 $&$17.0\pm4.5 $&$5.6\pm0.8 $&$6.8\pm1.8 $&$6.9\pm2.7 $&$2.4\pm0.4 $&$2.9 \pm 0.8 $&$1.0 \pm 0.2$\\
208B&$10.8 $&$2.2 $&$11.7\pm2.8 $&$3.4^* $&$4.7\pm1.1 $&$2.5\pm1.3 $&$8.0^* $&$11.0 \pm 2.7 $&$3.8 \pm 0.9$\\
209&$13.6 $&$2.8 $&$10.8\pm2.5 $&$- $&$4.3\pm1.0 $&$6.6\pm0.8 $&$- $&$0.3\pm 0.1 $&$0.14 \pm0.04$\\\hline
    \end{tabular}
    }
\raggedright \justify \vspace{-0.2cm}
\textbf{Notes:} $^*$ Calculated from $3\sigma$ upper limits shown in Table~\ref{tab:upperlimitsAdditional}
\end{table}

\section{Why differential lensing or a multi-phase ISM cannot produce super-thermalized line profiles}
\label{sec:whydifferentiallensingcannotproducesuperthermalizedlineprofiles}
\referee{Here, to see why differential magnification nor multi-phase ISMs cannot create super-thermalised lines, we consider the following toy model. 
We suppose the case that there are two CO transitions, $J=4$--3 and $J=3$--2. For thermalised lines, we can write $L^\prime_\mathrm{CO(4-3)}=L^\prime_\mathrm{CO(3-2)}$. For thermalised or sub-thermalised lines in a single-phase ISM region, we can generalise this to $L^\prime_\mathrm{CO(4-3)}=kL^\prime_\mathrm{CO(3-2)}$ where the constant $0<k\leq1$, with $k=1$ meaning it is thermalised.}

\referee{Now suppose that there is some variation in $k$ across the galaxy, perhaps due to the multi-phase ISM, so we split the galaxy into $N$ discrete regions, labelled $i=1$ to $i=N$. Each region has a thermalisation coefficient, $k_i$. In other words, for each region
\begin{equation}
L_i^\prime{}_\mathrm{CO(4-3)}=k_i L_i^\prime{}_\mathrm{CO(3-2)}.
\end{equation}
}
\referee{Now we assume that this galaxy also has differential magnification, with each region being magnified by a factor of $|\mu_i|$. We can set every $|\mu_i|$ to be $|\mu_i|\geq 1$ (though formally our toy model only requires having at least one non-zero $|\mu_i|$). The total observed CO~($J=4$--3) line luminosity will be
\begin{equation}
    \begin{split}
        \label{eq:useful}
        L_\mathrm{obs}^\prime{}_\mathrm{CO(4-3)} =& \sum_{i=1}^N |\mu_i| L_i^\prime{}_\mathrm{CO(4-3)}\\
        =& \sum_{i=1}^N |\mu_i| k_i L_i^\prime{}_\mathrm{CO(3-2)}.
    \end{split} 
\end{equation}
Meanwhile, the total observed CO~($J=3$--2) will be
\begin{equation}
    L_\mathrm{obs}^\prime{}_\mathrm{CO(3-2)} = \sum_{i=1}^N |\mu_i| L_i^\prime{}_\mathrm{CO(3-2)}.
\end{equation}
If we take the ratio of these two observed line luminosities, we obtain an ``effective'' observed total $k$ for the differentially-magnified galaxy:
\begin{equation}
    \begin{split}
        k_\mathrm{obs} =&~\frac{L_\mathrm{obs}^\prime{}_\mathrm{CO(4-3)}}{L_\mathrm{obs}^\prime{}_\mathrm{CO(3-2)}}\\
    =&~\frac{\sum_{i=1}^N |\mu_i| k_i L_i^\prime{}_\mathrm{CO(3-2)}}{\sum_{i=1}^N |\mu_i| L_i^\prime{}_\mathrm{CO(3-2)}}
    \end{split}
\end{equation}
where we have used equation~\ref{eq:useful}. However, every $k_i\leq1$, which means that every term in the numerator will be smaller than or equal to the corresponding term in the denominator. Therefore $k_\mathrm{obs}$ is necessarily $k_\mathrm{obs}\leq 1$. Therefore, no amount of differential magnification nor combination of regions with different ISM phases can create a super-thermalised line ratio ($k_\mathrm{obs}>1$) from individual thermalised or sub-thermalised components with each individual $k_i\leq 1$. }

\section{Stacked spectrum line luminosities}
We report the line luminosities of the stacked spectrum in Table~\ref{tab:faintLines}.

\begin{table}
    \caption{Spectral line luminosities from the stacked spectrum.}
    \label{tab:faintLines}
\centering
    \begin{tabular}{lccccclcccc}\hline 
    Line & Frequency & Sources & Line luminosity & S/N &  & Line & Frequency & Sources & Line luminosity & S/N  \\ 
     & [GHz] &  & [$10^9~\mathrm{K~km~s^{-1}~pc}^2$] & & & & [GHz] &  & [$10^9~\mathrm{K~km~s^{-1}~pc}^2$] \\ \hline
CO~(2--1) & 230.5380 & 2 & $\boldsymbol{ 129.3 \pm 14.9}$ & (9.9$\sigma$)  & & H$_2$O$^+$~607 & 607.2273 & 20 & $-0.1$ $\pm$ 1.1 & ($-0.1\sigma$) \\
CO~(3--2) & 345.7960 & 42 & $\boldsymbol{ 98.1 \pm 6.1}$ & (37.9$\sigma$)  & & H$_2$O$^+$~631 & 631.7241 & 17 & $-0.6$ $\pm$ 1.0 & ($-0.5\sigma$) \\
CO~(4--3) & 461.0408 & 35 & $\boldsymbol{ 74.2 \pm 4.4}$ & (53.8$\sigma$)  & & H$_2$O$^+$~634 & 634.2729 & 17 & $-0.8$ $\pm$ 1.1 & ($-0.7\sigma$) \\
CO~(5--4) & 576.2679 & 27 & $\boldsymbol{ 69.4 \pm 4.0}$ & (68.0$\sigma$)  & & H$_2$O$^+$~721 & 721.9274 & 6 & $-2.5$ $\pm$ 1.5 & ($-1.7\sigma$) \\
CO~(6--5) & 691.4731 & 10 & $\boldsymbol{ 50.5 \pm 3.1}$ & (40.1$\sigma$)  & & H$_2$O$^+$~742 & 742.1090 & 5 & 1.4 $\pm$ 1.5 & (1.0$\sigma$) \\
CO~(7--6) & 806.8984 & 4 & $\boldsymbol{ 38.2 \pm 2.6}$ & (25.6$\sigma$)  & & H$_2$O$^+$~746 & 746.5417 & 5 & $\boldsymbol{ 5.0 \pm 1.4}$ & (3.8$\sigma$) \\
$^{13}$CO~(2--1) & 220.3987 & 1 & 7.3 $\pm$ 18.4 & (0.4$\sigma$)  & & H$_2$O$^+$~761 & 761.8188 & 6 & 0.2 $\pm$ 1.2 & (0.2$\sigma$) \\
$^{13}$CO~(3--2) & 330.5880 & 41 & 4.3 $\pm$ 3.2 & (1.4$\sigma$)  & & LiH~(1--0) & 443.9529 & 30 & $-2.4$ $\pm$ 1.7 & ($-1.5\sigma$) \\
$^{13}$CO~(4--3) & 440.7652 & 28 & 2.2 $\pm$ 2.0 & (1.1$\sigma$)  & & CH~532 & 532.7239 & 34 & $\boldsymbol{ 6.8 \pm 1.2}$ & (5.8$\sigma$) \\
$^{13}$CO~(5--4) & 550.9263 & 29 & 1.6 $\pm$ 1.2 & (1.4$\sigma$)  & & CH~536 & 536.7614 & 31 & $\boldsymbol{ 7.8 \pm 1.3}$ & (6.5$\sigma$) \\
$^{13}$CO~(6--5) & 661.0673 & 14 & 1.2 $\pm$ 1.2 & (1.0$\sigma$)  & & OH~425 & 425.0363 & 25 & 0.4 $\pm$ 2.5 & (0.2$\sigma$) \\
$^{13}$CO~(7--6) & 771.1841 & 7 & $-0.9$ $\pm$ 1.0 & ($-0.9\sigma$)  & & OH~446 & 446.2910 & 31 & 2.5 $\pm$ 1.5 & (1.6$\sigma$) \\
$^{13}$CO~(8--7) & 881.2728 & 2 & $-1.2$ $\pm$ 1.5 & ($-0.8\sigma$)  & & CN~(N = 3--2) & 340.2478 & 41 & 2.1 $\pm$ 2.8 & (0.8$\sigma$) \\
C$^{18}$O~(2--1) & 219.5603 & 1 & $-24.3$ $\pm$ 19.8 & ($-1.2\sigma$)  & & CN~(N = 4--3) & 453.6067 & 36 & 2.2 $\pm$ 1.4 & (1.6$\sigma$) \\
C$^{18}$O~(3--2) & 329.3305 & 42 & 2.7 $\pm$ 3.2 & (0.9$\sigma$)  & & CN~(N = 5--4) & 566.9470 & 26 & 0.8 $\pm$ 1.1 & (0.7$\sigma$) \\
C$^{18}$O~(4--3) & 439.0888 & 27 & $-0.2$ $\pm$ 2.0 & ($-0.1\sigma$)  & & CN~(N = 6--5) & 680.2641 & 9 & 0.4 $\pm$ 1.3 & (0.3$\sigma$) \\
C$^{18}$O~(5--4) & 548.8310 & 33 & 2.0 $\pm$ 1.1 & (1.8$\sigma$)  & & SiO~(7--6) & 303.9270 & 19 & 5.3 $\pm$ 4.9 & (1.1$\sigma$) \\
C$^{18}$O~(6--5) & 658.5533 & 14 & 1.0 $\pm$ 1.2 & (0.9$\sigma$)  & & SiO~(8--7) & 347.3306 & 38 & $-1.7$ $\pm$ 2.6 & ($-0.7\sigma$) \\
C$^{18}$O~(7--6) & 768.2515 & 7 & 0.8 $\pm$ 1.1 & (0.7$\sigma$)  & & SiO~(9--8) & 390.7284 & 40 & 0.3 $\pm$ 2.3 & (0.1$\sigma$) \\
C$^{18}$O~(8--7) & 877.9219 & 2 & $-1.0$ $\pm$ 1.6 & ($-0.7\sigma$)  & & SiO~(10--9) & 434.1196 & 25 & 1.2 $\pm$ 2.2 & (0.6$\sigma$) \\
\textsc{[C~i]}~($^3P_1$--$^3P_0$) & 492.1606 & 32 & $\boldsymbol{ 29.0 \pm 2.1}$ & (23.0$\sigma$)  & & SiO~(11--10) & 477.5031 & 35 & $-1.0$ $\pm$ 1.3 & ($-0.8\sigma$) \\
\textsc{[C~i]}~($^3P_2$--$^3P_1$) & 809.3419 & 4 & $\boldsymbol{ 18.8 \pm 1.8}$ & (12.4$\sigma$)  & & SiO~(12--11) & 520.8782 & 40 & 1.2 $\pm$ 1.2 & (1.0$\sigma$) \\
HCN~(3--2) & 265.8864 & 3 & 0.2 $\pm$ 9.2 & (0.0$\sigma$)  & & SiO~(13--12) & 564.2440 & 27 & $-0.8$ $\pm$ 1.1 & ($-0.7\sigma$) \\
HCN~(4--3) & 354.5055 & 40 & $\boldsymbol{ 7.6 \pm 2.6}$ & (3.0$\sigma$)  & & SiO~(14--13) & 607.5994 & 20 & $-0.5$ $\pm$ 1.1 & ($-0.4\sigma$) \\
HCN~(5--4) & 443.1161 & 29 & 3.0 $\pm$ 1.7 & (1.8$\sigma$)  & & SiO~(15--14) & 650.9436 & 14 & $-0.4$ $\pm$ 1.1 & ($-0.4\sigma$) \\
HCN~(6--5) & 531.7164 & 34 & ${ 4.4 \pm 1.2}$ & (3.8$\sigma$)  & & SiO~(16--15) & 694.2754 & 9 & 0.3 $\pm$ 1.3 & (0.2$\sigma$) \\
HCN~(7--6) & 620.3040 & 20 & 1.4 $\pm$ 1.1 & (1.4$\sigma$)  & & SiO~(17--16) & 737.5939 & 6 & 0.1 $\pm$ 1.3 & (0.1$\sigma$) \\
HCN~(8--7) & 708.8770 & 6 & 0.8 $\pm$ 1.4 & (0.5$\sigma$)  & & CS~(6--5) & 293.9122    & 17 & $-10.7$ $\pm$ 5.8 & ($-1.9\sigma$) \\
HCN~(9--8) & 797.4333 & 6 & $-1.0$ $\pm$ 1.2 & ($-0.9\sigma$)  & & CS~(7--6) & 342.8830    & 42 & 3.5 $\pm$ 2.7 & (1.3$\sigma$) \\
HCN~(10--9) & 885.9707 & 2 & $-2.2$ $\pm$ 1.5 & ($-1.5\sigma$)  & & CS~(8--7) & 391.8470    & 39 & $-0.2$ $\pm$ 2.3 & ($-0.1\sigma$) \\
HNC~(3--2) & 271.9811 & 4 & 8.7 $\pm$ 11.2 & (0.8$\sigma$)  & & CS~(10--9) & 489.7510 & 33 & 1.0 $\pm$ 1.3 & (0.8$\sigma$) \\
HNC~(4--3) & 362.6303 & 35 & 0.3 $\pm$ 3.0 & (0.1$\sigma$)  & & CS~(11--10) & 538.6888 & 33 & $-0.2$ $\pm$ 1.2 & ($-0.1\sigma$) \\
HNC~(5--4) & 453.2699 & 36 & 3.4 $\pm$ 1.4 & (2.4$\sigma$)  & & CS~(12--11) & 587.6162 & 23 & 1.0 $\pm$ 1.1 & (0.9$\sigma$) \\
HNC~(6--5) & 543.8976 & 34 & 1.8 $\pm$ 1.2 & (1.6$\sigma$)  & & CS~(13--12) & 636.5318 & 15 & $-0.6$ $\pm$ 1.2 & ($-0.5\sigma$) \\
HNC~(7--6) & 634.5108 & 17 & 0.4 $\pm$ 1.1 & (0.3$\sigma$)  & & CS~(14--13) & 685.4348 & 10 & 1.0 $\pm$ 1.2 & (0.8$\sigma$) \\
HNC~(8--7) & 725.1073 & 6 & $-0.3$ $\pm$ 1.3 & ($-0.2\sigma$)  & & CS~(15--14) & 734.3240 & 6 & 0.8 $\pm$ 1.3 & (0.6$\sigma$) \\
HNC~(9--8) & 815.6847 & 3 & $-1.1$ $\pm$ 1.6 & ($-0.7\sigma$)  & & NH$_3$~(1$_0$--0$_0$) & 572.4982 & 25 & 1.0 $\pm$ 1.0 & (1.0$\sigma$) \\
HCO$^+$~(3--2) & 267.5576 & 4 & 5.1 $\pm$ 8.3 & (0.6$\sigma$)  & & N$_2$H$^+$~(3--2) & 279.5117 & 8 & 1.4 $\pm$ 9.0 & (0.2$\sigma$) \\
HCO$^+$~(4--3) & 356.7342 & 40 & 3.4 $\pm$ 2.5 & (1.4$\sigma$)  & & N$_2$H$^+$~(4--3) & 372.6725    & 39 & $-1.6$ $\pm$ 2.3 & ($-0.7\sigma$) \\
HCO$^+$~(5--4) & 445.9029 & 31 & 4.5 $\pm$ 1.5 & (3.0$\sigma$)  & & N$_2$H$^+$~(5--4) & 465.8250 & 32 & 0.1 $\pm$ 1.5 & (0.1$\sigma$) \\
HCO$^+$~(6--5) & 535.0616 & 33 & 1.6 $\pm$ 1.2 & (1.3$\sigma$)  & & N$_2$H$^+$~(6--5) & 558.9667    & 27 & $-2.1$ $\pm$ 1.3 & ($-1.7\sigma$) \\
HCO$^+$~(7--6) & 624.2085 & 17 & 0.6 $\pm$ 1.1 & (0.5$\sigma$)  & & N$_2$H$^+$~(7--6) & 652.0959    & 15 & $-0.8$ $\pm$ 1.1 & ($-0.7\sigma$) \\
HCO$^+$~(8--7) & 713.3414 & 5 & 0.1 $\pm$ 1.4 & (0.0$\sigma$)  & & N$_2$H$^+$~(8--7) & 745.2103    & 6 & 1.9 $\pm$ 1.3 & (1.5$\sigma$) \\
HCO$^+$~(9--8) & 802.4582 & 5 & $-2.5$ $\pm$ 1.4 & ($-1.9\sigma$)  & & CCH~(3--2) & 262.0042    & 2 & 1.0 $\pm$ 9.4 & (0.1$\sigma$) \\
HCO$^+$~(10--9) & 891.5573 & 2 & $-0.6$ $\pm$ 1.5 & ($-0.4\sigma$)  & & CCH~(4--3) & 349.3387    & 40 & 1.1 $\pm$ 2.6 & (0.4$\sigma$) \\
HOC$^+$~(3--2) & 268.4510 & 4 & $-1.4$ $\pm$ 8.5 & ($-0.2\sigma$)  & & CCH~(5--4) & 436.6604    & 27 & $-4.0$ $\pm$ 2.1 & ($-1.9\sigma$) \\
HOC$^+$~(4--3) & 357.9219 & 40 & $-5.0$ $\pm$ 2.4 & ($-2.1\sigma$)  & & CCH~(6--5) & 523.9704    & 37 & $-0.9$ $\pm$ 1.2 & ($-0.7\sigma$) \\
HOC$^+$~(5--4) & 447.3818 & 30 & $-0.7$ $\pm$ 1.5 & ($-0.5\sigma$)  & & CCH~(7--6) & 611.2650 & 20 & $-0.1$ $\pm$ 1.1 & ($-0.1\sigma$) \\
HOC$^+$~(6--5) & 536.8279 & 31 & ${ 7.6 \pm 1.3}$ & (6.4$\sigma$)  & & CCH~(8--7) & 698.5416    & 7 & 0.3 $\pm$ 1.4 & (0.2$\sigma$) \\
HOC$^+$~(7--6) & 626.2575 & 17 & $-0.0$ $\pm$ 1.1 & ($-0.0\sigma$)  & & H21$\alpha$ & 662.4042    & 13 & 0.7 $\pm$ 1.2 & (0.6$\sigma$) \\
HOC$^+$~(8--7) & 715.6679 & 6 & 0.3 $\pm$ 1.4 & (0.2$\sigma$)  & & H22$\alpha$ & 577.8964    & 27 & 0.3 $\pm$ 1.0 & (0.3$\sigma$) \\
HOC$^+$~(9--8) & 805.0563 & 4 & 0.7 $\pm$ 1.5 & (0.5$\sigma$)  & & H23$\alpha$ & 507.1755    & 35 & $-0.8$ $\pm$ 1.1 & ($-0.8\sigma$) \\
H$_2$O~(5$_{1, 5}$--4$_{2, 2}$) & 325.1529 & 38 & $-5.8$ $\pm$ 3.4 & ($-1.7\sigma$)  & & H24$\alpha$ & 447.5403    & 30 & 0.1 $\pm$ 1.5 & (0.0$\sigma$) \\
H$_2$O~(4$_{1, 4}$--3$_{2, 1}$) & 380.1974 & 42 & 0.7 $\pm$ 2.2 & (0.3$\sigma$)  & & H25$\alpha$ & 396.9008 & 35 & $-1.6$ $\pm$ 2.4 & ($-0.7\sigma$) \\
H$_2$O~(4$_{2, 3}$--3$_{3, 0}$) & 448.0011 & 31 & 1.8 $\pm$ 1.5 & (1.2$\sigma$)  & & H26$\alpha$ & 353.6227 & 41 & 5.2 $\pm$ 2.5 & (2.1$\sigma$) \\
H$_2$O~(5$_{3, 3}$--4$_{4, 0}$) & 474.6891 & 33 & $-2.2$ $\pm$ 1.3 & ($-1.6\sigma$)  & & H27$\alpha$ & 316.4154 & 29 & $-4.7$ $\pm$ 4.0 & ($-1.2\sigma$) \\
H$_2$O~(1$_{1, 0}$--1$_{0, 1}$) & 556.9360 & 30 & $-1.0$ $\pm$ 1.2 & ($-0.8\sigma$)  & & H28$\alpha$ & 284.2506 & 10 & 9.0 $\pm$ 8.0 & (1.1$\sigma$) \\
H$_2$O~(2$_{1, 1}$--2$_{0, 2}$) & 752.0331 & 5 & $\boldsymbol{10.6 \pm 1.4}$ & (8.5$\sigma$)  & & CH$^+$ & 835.0789 & 2 & $-0.5$ $\pm$ 1.7 & ($-0.3\sigma$) \\
H$_2$O$^+$~604 & 604.6786 & 22 & 0.6 $\pm$ 1.1 & (0.6$\sigma$) \\
\hline
\end{tabular}
\raggedright \justify \vspace{-0.2cm}
\textbf{Notes:} Lines in bold-face indicate $> 3\sigma$ detections. Note that the apparent HCN~(6--5) and HOC$^+$~(6--5) ``detections'' are due to line confusion with CH~532 and 536.
\end{table}

\section{Diversity in the line profiles}
\begin{figure*}
    \centering
    \includegraphics[width=\linewidth]{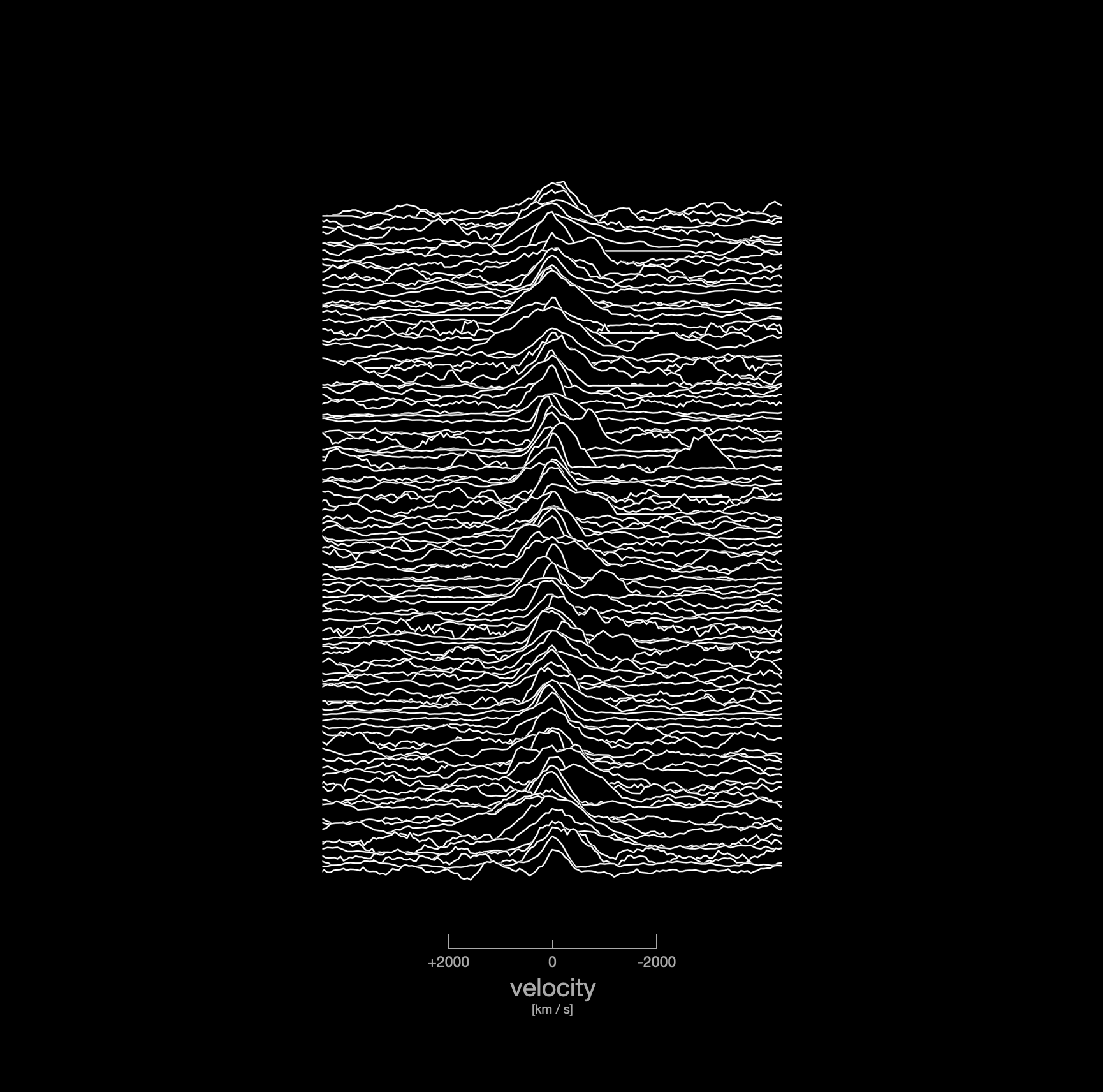}
    \caption{Demonstration of the diversity of the CO and \textsc{[C\,i]}~($^3P_1$--$^3P_0$) line profiles in the BEARS data set, visualised by superposing them with slight offsets, i.e., the vertical axis is linear in flux, scaled to the peak flux density value of each line. The horizontal axis shows the line velocity. Note the striking variation in widths and shapes, which we have found is not as obviously apparent in other visualisations of the data. For plotting purposes, we exclude any sources within $4000~\mathrm{km~s^{-1}}$ of the edge of the spectral windows. The line profiles are stacked in no particular order.}
    \label{fig:JD_graph}
\end{figure*}

Figure~\ref{fig:JD_graph} graphically demonstrates the diversity in line profiles across the BEARS systems.

{\noindent
\small
\textit{
$^{1}$Department of physics, Graduate School of Science, Nagoya University, Nagoya, Aichi 464-8602, Japan\\
$^{2}$National Astronomical Observatory of Japan, 2-21-1, Osawa, Mitaka, Tokyo 181-8588, Japan\\
$^{3}$School of Physical Sciences, The Open University, Milton Keynes, MK7 6AA, UK\\
$^{4}$UK ALMA Regional Centre Node, Jodrell Bank Centre for Astrophysics, Department of Physics and Astronomy, University of Manchester, Oxford Road, Manchester M13 9PL, UK\\
$^{5}$School of Physics and Astronomy, Cardiff University, The Parade, Cardiff, CF24 3AA, UK\\
$^{6}$European Southern Observatory, Alonso de Córdova 3107, Vitacura, Casilla 19001, Santiago de Chile, Chile\\
$^{7}$Institute for Advanced Research, Nagoya University, Furocho, Chikusa, Nagoya 464-8602, Japan\\
$^{8}$Institut de Radioastronomie Millimétrique (IRAM), 300 Rue de la Piscine, 38400 Saint-Martin-d’H\`{e}res, France\\
$^{9}$Department of Physics \& Astronomy, University of California, Irvine, 4129 Reines Hall, Irvine, CA 92697, USA\\
$^{10}$Institut d’Astrophysique de Paris, Sorbonne Universit\'{e},UPMC Universit\'{e} Paris 6 and CNRS, UMR 7095, 98 bis boulevard Arago, F-75014 Paris, France\\
$^{11}$INAF – Osservatorio Astronomico di Padova, Vicolo dell'Osservatorio 5, I-35122 Padova, Italy\\
$^{12}$Centre de Recherche Astrophysique de Lyon, ENS de Lyon, Universit\'{e} Lyon 1, CNRS, UMR5574, 69230 Saint-Genis-Laval, France\\
$^{13}$I. Physikalisches Institut, Universit\"at zu K\"oln, Z\"ulpicher Strasse 77, D-50937 K\"oln, Germany\\
$^{14}$Department of Physics and Astronomy, University of British Columbia, 6224 Agricultural Road, Vancouver, BC V6T 1Z1, Canada\\
$^{15}$Astrophysics Branch, NASA—Ames Research Center, MS 245-6, Moffett Field, CA 94035, USA\\
$^{16}$Leiden Observatory, Leiden University, PO Box 9513, NL-2300 RA Leiden, Netherlands\\
$^{17}$Department of Earth and Space Sciences, Chalmers University of Technology, Onsala Observatory, 439 94 Onsala, Sweden\\
$^{18}$Centre for Extragalactic Astronomy, Durham University, Department of Physics, South Road, Durham DH1 3LE, UK\\
$^{19}$European Southern Observatory, Karl Schwarzschild Strasse 2, D-85748 Garching, Germany\\
$^{20}$Department of Physics and Astronomy, Rutgers, the State University of New Jersey, 136 Frelinghuysen Road, Piscataway, NJ 08854-8019, USA\\
$^{21}$Department of Physics and Astronomy, University of the Western Cape, Robert Sobukwe Road, Bellville 7535, South Africa\\
$^{22}$Institut d’Astrophysique Spatiale (IAS), CNRS et Universit\'{e} Paris Sud, Orsay, France\\
$^{23}$Dipartimento di Fisica e Astronomia “G. Galilei”, Universit’di Padova, Vicolo dell’Osservatorio 3, I-35122, Padova, Italy\\
$^{24}$Aix-Mar seille Univer sit\`{e}, CNRS and CNES, Laboratoire d’Astrophysique de Marseille, 38, rue Fr\`{e}d\`{e}ric Joliot-Curie, F-13388 Marseille, France\\
$^{25}$Instituto Astrof\'{i}sica de Canarias (IAC), E-38205 La Laguna, Tenerife, Spain\\
$^{26}$Dpto. Astrof\'{i}sica, Universidad de la , E-38206 La Laguna, Tenerife, Spain\\
$^{27}$School of Physics \& Astronomy, University of Nottingham, University Park, Nottingham NG7 2RD, UK\\
$^{28}$University of Bologna—Department of Physics and Astronomy “Augusto Righi” (DIFA), Via Gobetti 93/2, I-40129, Bologna, Italy\\
$^{29}$INAF—Osservatorio di Astrofisica e Scienza dello Spazio, Via Gobetti 93/3, I-40129, Bologna, Italy\\
$^{30}$CAS Key Laboratory for Research in Galaxies and Cosmology, Department of Astronomy, University of Science and Technology of China, Hefei 230026, People’s Republic of China\\
$^{31}$School of Astronomy and Space Sciences, University of Science and Technology of China, Hefei, Anhui 230026, People’s Republic of China\\
$^{32}$Institute of Deep Space Sciences, Deep Space Exploration Laboratory, Hefei 230026, People’s Republic of China\\
$^{33}$Institute of Astronomy, University of Cambridge, Madingley Road, Cambridge CB30HA, UK\\
$^{34}$Departamento de Fisica, Universidad de Oviedo, C. Federico Garcia Lorca 18, E-33007 Oviedo, Spain\\
$^{35}$Instituto Universitario de Ciencias y Tecnologas Espaciales de Asturias (ICTEA), C. Independencia 13, E-33004 Oviedo, Spain\\
$^{36}$Department of Astronomy, University of Maryland, College Park, MD 20742, USA\\
$^{37}$Instituto Nacional de Astrof\'{ı}sica, \'{O}ptica y Electr\'{o}nica, Tonantzintla, 72000 Puebla, M\'{e}xico\\
\referee{$^{38}$}Institute of Astronomy, Graduate School of Science, The University of Tokyo, 2-21-1 Osawa, Mitaka, Tokyo 181-0015, Japan\\
\referee{$^{39}$}Research Center for the Early Universe, Graduate School of Science, The University of Tokyo, 7-3-1 Hongo, Bunkyo-ku, Tokyo 113-0033, Japan\\
\referee{$^{40}$}Department of Astronomy, University of Cape Town, Private Bag X3, Rondebosch 7701, Cape Town, South Africa\\
\referee{$^{41}$}INAF, Instituto di Radioastronomia-Italian ARC, Via Piero Gobetti 101, I-40129 Bologna, Italy\\
\referee{$^{42}$}SISSA, Via Bonomea 265, 34136 Trieste, Italy\\
\referee{$^{43}$}Joint ALMA Observatory, Alonso de Córdova 3107, Vitacura 763-0355, Santiago de Chile, Chile\\
\referee{$^{44}$}Ikerbasque Foundation, University of the Basque Country, DIPC Donostia, Spain\\
\referee{$^{45}$}Sub-department of Astrophysics, University of Oxford, Denys Wilkinson Building, Keble Road, Oxford, OX1 3RH, UK\\
\referee{$^{46}$}National Radio Astronomy Observatory, 520 Edgemont Road, Charlottesville, VA 22903, USA\\
\referee{$^{47}$}Max-Planck-Institut für Radioastronomie, Auf dem Hügel 69, 53121 Bonn, Germany\\
}
}
\bsp	
\label{lastpage}
\end{document}